%% file: main.tex
\documentclass[runningheads,envcountsame]{llncs}

\makeatletter
\RequirePackage[bookmarks,unicode,colorlinks=true]{hyperref}%
   \def\@citecolor{blue}%
   \def\@urlcolor{blue}%
   \def\@linkcolor{blue}%

\def\orcidID#1{\href{http://orcid.org/#1}{\smash{\protect\raisebox{-1.25pt}{\protect\includegraphics{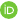}}}}}
\makeatother

\usepackage{xcolor,colortbl}
\usepackage{mathtools}
\usepackage{multirow}
\usepackage{stmaryrd}
\usepackage{cancel}
\usepackage{amsmath,amssymb}
\usepackage{thmtools, thm-restate}
\usepackage[shortlabels]{enumitem}
\usepackage{microtype}
\usepackage{wrapfig}
\usepackage{pbox}
\usepackage{marvosym}
\usepackage{listings}
\usepackage{ifthen}
\usepackage[ruled, vlined, linesnumbered]{algorithm2e}
\hypersetup{
	pdftitle={Relating Different Evaluation Strategies of Probabilistic Term Rewriting for Multiple Notions of Termination}, colorlinks=true, linkcolor=blue, citecolor=olive, filecolor=magenta, urlcolor=cyan
}
\usepackage{todonotes}
\usepackage{placeins}
\usepackage{float}
\usepackage{tikz}
\usepackage{caption}
\usepackage{subcaption}
\usepackage{mymatrix}
\usepackage{nicefrac,xfrac}
\usepackage{centernot}
\usepackage{array,longtable}
\usetikzlibrary{shapes,calc,arrows,automata,decorations.pathmorphing,backgrounds,arrows.meta,shapes.geometric,shapes.multipart,shapes.misc}
\pgfdeclarelayer{edgelayer}
\pgfdeclarelayer{nodelayer}
\pgfsetlayers{background,edgelayer,nodelayer,main}
\tikzstyle{none}=[inner sep=0mm]
\tikzstyle{moveBlock}=[fill=white, draw=black, shape=rectangle]
\tikzstyle{target}=[fill=white, draw=black, shape=circle]

\tikzstyle{dotHead}=[dotted, ->]
\tikzstyle{dotWithoutHead}=[dotted, -]
\tikzstyle{dashHead}=[dashed,->]
\tikzstyle{dashWithoutHead}=[dashed,-]
\tikzstyle{arrow}=[->]

\RequirePackage{makecell}

\usepackage[capitalize,nameinlink]{cleveref}
\usepackage{IEEEtrantools}

\newenvironment{myproof}{
	\noindent{\it Proof.}
}{\qed
	\medskip
}
\newenvironment{myproofsketch}{
	\noindent{\it Proof Sketch.}
}{\qed
	\medskip
}
\newenvironment{counterexample}[1][]{\refstepcounter{example}\par\medskip
   \noindent \textit{Counterexample~\theexample 
   \ifthenelse{\equal{#1}{}}{.}{\, (#1).}} \rmfamily}{\medskip}
\input{commands}

\SetKwInOut{Input}{input}
\SetKwInOut{Output}{output}

\definecolor{Gray}{gray}{0.85}
\definecolor{LightCyan}{rgb}{0.88,1,1}

\newcolumntype{a}{>{\columncolor{Gray}}c}
\newcolumntype{b}{>{\columncolor{white}}c}

 \newcommand{\makeproof}[2]{}
 \newcommand{\paper}[1]{}
 \newcommand{\report}[1]{#1}

 \report{
  \setlength{\textwidth}{125mm}
   \setlength{\textheight}{198mm}
 }

 \newcommand{\notnew}[1]{}

\newcounter{auxctr}
\newcounter{iASTtofASTOne}
\newcounter{iASTtofASTTwo}
\newcounter{iASTtofASTThree}
\newcounter{iASTtofASTFour}
\newcounter{wASTtofAST}
\newcounter{liASTtoiAST}
 
\usepackage{cite}
\AtBeginDocument{\renewcommand{\doi}[1]{\url{https://doi.org/#1}}}

\title{From Innermost to Full Almost-Sure Termination of Probabilistic Term
  Rewriting\thanks{funded by the Deutsche Forschungsgemeinschaft (DFG, German Research
    Foundation) - 235950644 (Project GI 274/6-2) and the DFG Research Training Group 2236 UnRAVeL}}
\titlerunning{From Innermost to Full AST of PTRSs}
\author{Jan-Christoph Kassing\paper{$^{(\mbox{\Letter})}$}\orcidID{0009-0001-9972-2470} \and Florian Frohn\paper{$^{(\mbox{\Letter})}$}\orcidID{0000-0003-0902-1994}
  \and Jürgen Giesl\paper{$^{(\mbox{\Letter})}$}\orcidID{0000-0003-0283-8520}}
\institute{LuFG Informatik 2, RWTH Aachen University, Aachen, Germany}
\authorrunning{J.-C.\ Kassing, F.\ Frohn, J.\ Giesl}

\begin{document}
\allowdisplaybreaks

\maketitle \input{abstract}
\input{introduction}
\input{pre}

\input{probtrs}
\input{relating}
\input{improve}
\input{evaluation}
\bibliographystyle{splncs04}
\newcommand{\noopsort}[1]{}
\bibliography{biblio}
\report{

\appendix
\renewcommand{\theHsection}{appendix.\Alph{section}}
\input{appendix}

\renewcommand{\theHtheorem}{pastsast.\arabic{theorem}}
\input{pastandsast}
}
\end{document}

%% file: commands.tex
\makeatletter \newcommand*\bigcdot{\mathpalette\bigcdot@{.4}}
\newcommand*\bigcdot@[2]{\mathbin{\vcenter{\hbox{\scalebox{#2}{$\m@th#1\bullet$}}}}}
\makeatother

\renewcommand{\emptyset}{\varnothing}

\newcommand{\disabledcomment}[1]{}
\newcommand{\oldcomment}[1]{}

\newcommand{\dontprint}[1]{}

\renewcommand{\epsilon}{\varepsilon}

\newcommand{\IN}{\mathbb{N}}
\newcommand{\IR}{\mathbb{R}}

\newcommand{\R}{\mathcal{R}}

\newcommand{\SSS}{\mathcal{S}}

\newcommand{\aprove}{\textsf{AProVE}}
\newcommand{\muterm}{\textsf{MuTerm}}
\newcommand{\natt}{\textsf{NaTT}}
\newcommand{\ttttwo}{\textsf{TTT2}}
\newcommand{\ceta}{\textsf{CeTA}}

\crefname{definition}{Def.}{Def.}
\crefname{example}{Ex.}{Ex.}
\crefname{counterexample}{Counterex.}{Counterex.}
\crefname{appendix}{App.}{App.}
\crefname{ex}{Ex.}{Ex.}
\crefname{theorem}{Thm.}{Thm.}
\crefname{lemma}{Lemma}{Lemmas}
\crefname{remark}{Rem.}{Rem.}
\crefname{section}{Sect.}{Sect.}
\crefname{subsection}{Sect.}{Sect.}
\crefname{subsubsection}{Sect.}{Sect.}
\crefname{line}{Line}{Lines}
\crefname{corollary}{Cor.}{Cor.}
\crefname{figure}{Fig.}{Fig.}
\crefname{enumi}{}{}
\crefname{algorithm}{Alg.}{Alg.}

\renewcommand{\emptyset}{\varnothing}

\newcommand{\F}[1]{\mathfrak{#1}}
\newcommand{\C}[1]{\mathcal{#1}}

\makeatletter
\def\moverlay{\mathpalette\mov@rlay}
\def\mov@rlay#1#2{\leavevmode\vtop{%
   \baselineskip\z@skip \lineskiplimit-\maxdimen
   \ialign{\hfil$\m@th#1##$\hfil\cr#2\crcr}}}
\newcommand{\charfusion}[3][\mathord]{
    #1{\ifx#1\mathop\vphantom{#2}\fi
        \mathpalette\mov@rlay{#2\cr#3}
      }
    \ifx#1\mathop\expandafter\displaylimits\fi}
\makeatother

\newcommand{\Var}{\mathcal{V}}

\newcommand{\TSet}[2]{\mathcal{T}\left(#1,#2\right)}

\newcommand{\VSet}{\mathcal{V}}

\renewcommand{\O}{\mathcal{O}}

\newcommand{\FDist}{\operatorname{FDist}}
\newcommand{\Supp}{\operatorname{Supp}}

\newcommand{\rootsym}{\operatorname{root}}

\newcommand{\edh}{\operatorname{edh}}
\newcommand{\edl}{\operatorname{edl}}

\renewcommand{\ts}{\mathsf{s}}
\renewcommand{\O}{\mathcal{O}}
\newcommand{\tf}{\mathsf{f}}
\newcommand{\tg}{\mathsf{g}}
\renewcommand{\th}{\mathsf{h}}
\newcommand{\ta}{\mathsf{a}}
\newcommand{\tb}{\mathsf{b}}
\newcommand{\tc}{\mathsf{c}}
\newcommand{\td}{\mathsf{d}}

\newcommand{\trw}{\mathsf{rw}}
\newcommand{\tcons}{\mathsf{cons}}

\newcommand{\tenc}{\mathsf{enc}}
\newcommand{\targenc}{\mathsf{argenc}}

\newcommand{\cv}{\operatorname{cv}}
\newcommand{\bv}{\operatorname{bv}}
\newcommand{\dv}{\operatorname{dv}}

\newcommand{\ctleaf}{\operatorname{Leaf}}

\makeatletter
\NewDocumentCommand{\dparrow}{+O{} +O{0.5cm}}{%
\begin{tikzpicture}[baseline=-0.5ex] {
\node[inner sep=0](@1) at (0,0) {};
\node[inner sep=0](@2) at (#2,0) {};
\draw [arrows={-Triangle[open]},shorten >= 1pt,shorten <= 1pt](@1)--(@2) node[pos=.5,above,inner sep=1pt] {\ensuremath{#1}};}
\end{tikzpicture}\xspace
}

\NewDocumentCommand{\myto}{+O{} +O{0.5cm}}{%
\begin{tikzpicture}[baseline=-0.5ex] {
\node[inner sep=0](@1) at (0,0) {};
\node[inner sep=0](@2) at (#2,0) {};
\draw [arrows={-to}](@1)--(@2) node[pos=.5,above,inner sep=1pt] {\ensuremath{#1}};}
\end{tikzpicture}\xspace
}

\NewDocumentCommand{\paraarrow}{+O{} +O{0.4cm}}{%
\begin{tikzpicture}[baseline=-0.5ex] {
\node[inner sep=0](@1) at (0,0) {};
\node[inner sep=0](@2) at (#2,0) {};
\node[inner sep=0](@3) at (0.07,0) {};
\draw [arrows={-to}](@1)--(@2) node[pos=.5,above,inner sep=1pt] {\ensuremath{#1}};
\draw [arrows={-to}](@1)--(@3);}
\end{tikzpicture}\xspace
}

\NewDocumentCommand{\paradparrow}{+O{} +O{0.4cm}}{%
\begin{tikzpicture}[baseline=-0.5ex] {
\node[inner sep=0](@1) at (0,0) {};
\node[inner sep=0](@2) at (#2,0) {};
\node[inner sep=0](@3) at (0.07,0) {};
\draw [arrows={-Triangle[open]}](@1)--(@2) node[pos=.5,above,inner sep=1pt] {\ensuremath{#1}};
\draw [arrows={-to}](@1)--(@3);}
\end{tikzpicture}\xspace
}

\newcommand{\oset}[2]{%
  {\mathop{#2}\limits^{\vbox to 1\ex@{\kern-\tw@\ex@
   \hbox{\scriptsize #1}\vss}}}}

\newcommand{\osetthree}[2]{%
  {\mathop{#2}\limits^{\vbox to 3\ex@{\kern-\tw@\ex@
   \hbox{\scriptsize #1}\vss}}}}

\newcommand{\osetfive}[2]{%
  {\mathop{#2}\limits^{\vbox to 5\ex@{\kern-\tw@\ex@
   \hbox{\scriptsize #1}\vss}}}}

\newcommand{\osetminus}[2]{%
  {\mathop{#2}\limits^{\vbox to -2\ex@{\kern-\tw@\ex@
   \hbox{\scriptsize #1}\vss}}}}
\makeatother

\newcommand{\ito}{\mathrel{\smash{\stackrel{\raisebox{3.4pt}{\tiny $\mathsf{i}\:$}}{\smash{\rightarrow}}}}}

\newcommand{\lito}{\mathrel{\smash{\stackrel{\raisebox{3.4pt}{\tiny $\mathsf{li}\:$}}{\smash{\rightarrow}}}}}

\newcommand{\topara}{\mathrel{\paraarrow}}
\newcommand{\toparas}{\mathrel{\paraarrow}_{\SSS}}

\newcommand{\itopara}{\mathrel{\smash{\stackrel{\raisebox{3.4pt}{\tiny
   $\mathsf{i}\:$}}{\smash{\paraarrow}}}}}
\newcommand{\itoparas}{\mathrel{\smash{\stackrel{\raisebox{3.4pt}{\tiny
   $\mathsf{i}\:$}}{\smash{\paraarrow}}}}_{\SSS}}

\newcommand{\iparaarrow}{\mathrel{\smash{\stackrel{\raisebox{3.4pt}{\tiny $\mathsf{i}\:$}}{\smash{\paraarrow}}}}}

\newcommand{\leftR}{\mathrel{\prescript{}{\R}{\leftarrow}}}
\newcommand{\leftROne}{\mathrel{\prescript{}{\R_1}{\leftarrow}}}

\newcommand{\itor}{\mathrel{\ito_{\R}}}
\newcommand{\litor}{\mathrel{\lito_{\R}}}
\newcommand{\litos}{\mathrel{\lito_{\SSS}}}

\newcommand{\itorExOne}{\mathrel{\ito_{\R_1}}}

\newcommand{\itoRd}{\mathrel{\ito_{\R_{\td}}}}

\newcommand{\itos}{\mathrel{\ito_{\SSS}}}

\newcommand{\xliftto}[1]{\mathrel{\smash{\stackrel{\raisebox{3.4pt}{\tiny $#1\:$}}{\smash{\liftto}}}}}
\newcommand{\liftto}{\rightrightarrows}
\newcommand{\iliftto}{\xliftto{\mathsf{i}}}

\newcommand{\liliftto}{\xliftto{\mathsf{li}}}

\newcommand{\paraliftto}{\mathrel{\ooalign{\raisebox{2pt}{$\paraarrow$}\cr\hfil\raisebox{-2pt}{$\paraarrow$}\hfil}}}
\newcommand{\iparaliftto}{\mathrel{\smash{\stackrel{\raisebox{3.4pt}{\tiny $\mathsf{i}\:$}}{\smash{\paraliftto}}}}}
\newcommand{\iparalifttos}{\mathrel{\smash{\stackrel{\raisebox{3.4pt}{\tiny $\mathsf{i}\:$}}{\smash{\paraliftto}}}}_{\SSS}}

\newcommand{\ilifttoRrw}{\mathrel{\iliftto_{\SSS_{\trw}}}}
\newcommand{\lifttoRrw}{\mathrel{\liftto_{\SSS_{\trw}}}}

\newcommand{\lilifttoRrw}{\mathrel{\liliftto_{\SSS_{\trw}}}}

%% file: abstract.tex
\begin{abstract}
There are many evaluation strategies for term rewrite systems, 
but proving termination automatically is usually easiest for innermost rewriting.
Several syntactic criteria exist when innermost termination implies full
 termination.
  We adapt these criteria to the probabilistic setting, e.g., we show when it suffices to
  analyze almost-sure termination (AST) w.r.t.\ innermost rewriting to prove
  full AST of probabilistic term rewrite systems.
 These criteria also apply to other notions of termination like positive AST.
 We implemented and evaluated our new contributions in the tool \aprove.
\end{abstract}

%% file: introduction.tex
\section{Introduction}\label{sec-introduction}

Termination analysis is one of the main tasks in program verification, and 
techniques and tools to analyze termination of
term rewrite systems (TRSs) automatically
have been studied for decades. 
While a direct application of classical reduction orderings is often too weak,
these orderings can be used successfully within the \emph{dependency pair} (DP) framework
\cite{arts2000termination,giesl2006mechanizing}. 
This framework  allows for modular termination proofs
by decomposing the original termination problem into sub-problems whose termination can
then be analyzed independently using different techniques.
Thus, DPs are used in essentially all current termination tools for TRSs (e.g.,
\aprove{} \cite{JAR-AProVE2017}, \muterm{} \cite{gutierrez_mu-term_2020}, \natt{} \cite{natt_sys_2014}, \ttttwo{} \cite{ttt2_sys}).
To allow certification of termination proofs with DPs, they have been formalized in
several proof assistants and there exist several corresponding certification tools
for termination proofs with DPs (e.g.,
\ceta{} \cite{ceta_sys}).

On the other hand, \emph{probabilistic} programs are used to describe randomized
algorithms and probability distributions, with applications in many areas, see, e.g., \cite{Gordon14}.
To use TRSs also for such programs, \emph{probabilistic term rewrite systems} (PTRSs) were introduced in \cite{BournezRTA02,bournez2005proving,avanzini2020probabilistic}.
In the probabilistic setting, there are several notions of ``termination''.
In this paper, we mostly focus on analyzing \emph{almost-sure termination} (AST), i.e., we
want to prove automatically that the probability \pagebreak[2] for termination is $1$.

While there exist many automatic approaches to prove (P)AST of imperative programs
on numbers (e.g.,
\cite{kaminski2018weakest,mciver2017new,TACAS21,lexrsm,FoundationsTerminationMartingale2020,FoundationsExpectedRuntime2020,rsm,cade19,dblp:journals/pacmpl/huang0cg19,amber,ecoimp,absynth}),
there are only  few automatic approaches for programs with complex non-tail recursive
structure
\cite{beutner2021probabilistic,Dallago2017ProbSizedTyping,lago_intersection_2021}.
 The
approaches that are also suitable for algorithms on recursive data structures
\cite{wang2020autoexpcost,LeutgebCAV2022amor,KatoenPOPL23}
are mostly specialized for specific data structures and cannot easily be adjusted to
other (possibly user-defined) ones, or are not yet fully automated.

For innermost AST (i.e., AST restricted to rewrite sequences where one only
evaluates at innermost positions), we recently presented an adaption of the DP framework
which allows us to benefit from a similar modularity
as in the non-probabilistic setting \cite{kassinggiesl2023iAST,FLOPS2024}.
Unfortunately, there is no such modular powerful approach
available for \emph{full} AST (i.e., AST when
considering arbitrary rewrite sequences).
Up to now, full AST of PTRSs can only be proved via a direct
application of orderings \cite{avanzini2020probabilistic,kassinggiesl2023iAST}, but  there is no
corresponding adaption of dependency pairs. (As explained in \cite{kassinggiesl2023iAST},
a DP framework to analyze full instead of innermost AST would be
``considerably more involved''.)
Indeed, also in the non-probabilistic setting, 
innermost termination is usually substantially easier to prove than full termination,
see, e.g., \cite{arts2000termination,giesl2006mechanizing}. 
To lift innermost termination proofs to full rewriting,
in the non-probabilistic setting, there exist several sufficient criteria  which ensure that innermost termination implies full termination \cite{Gramlich1995AbstractRB}.

Up to now no such results were known in the probabilistic setting. 
Our paper presents the first sufficient criteria for PTRSs which ensure that AST
coincides for full and innermost rewriting, and we also show similar results for
other rewrite strategies
like \emph{leftmost-innermost} rewriting.
We focus on criteria that can be checked automatically, so we can combine our results
with the DP framework for proving innermost AST of PTRSs
\cite{kassinggiesl2023iAST,FLOPS2024}.  
In this way, we obtain a modular powerful technique that can also
prove AST for \emph{full} rewriting automatically.

We will also consider the stronger notion of \emph{positive almost-sure
termination} (PAST) \cite{DBLP:conf/mfcs/Saheb-Djahromi78,bournez2005proving}, which
requires that the expected runtime is finite, and show that our criteria for the
relationship between full and innermost probabilistic rewriting hold for PAST as well.
In contrast to AST, PAST is not modular, i.e., the sequence of two programs that are PAST
may yield a program that is not PAST (see, e.g., \cite{kaminski2018weakest}).
Therefore, up to now there is no variant of DPs that allows to prove PAST of PTRSs, but
there only exist techniques
to apply polynomial or matrix orderings directly \cite{avanzini2020probabilistic}.

We start with preliminaries on term rewriting in \cref{Preliminaries}.
Then we recapitulate PTRSs based on
\cite{bournez2005proving,avanzini2020probabilistic,Faggian2019ProbabilisticRN,kassinggiesl2023iAST,diazcaro_confluence_2018} in
\cref{Probabilistic Term Rewriting}.
In \cref{Relating AST and its Restricted Forms} we show that the properties of \cite{Gramlich1995AbstractRB}
that ensure equivalence of innermost and full termination do not suffice in the probabilistic setting and extend them accordingly.
In particular, we show that innermost and full AST coincide for PTRSs that are
non-overlapping
and linear.
This result also holds for PAST, as well as for strategies like
leftmost-innermost evaluation.
In \cref{Improving Applicability} we show how to weaken the linearity
requirement in order to prove
full AST for larger classes of PTRSs.
The implementation of our criteria in the tool \aprove{} is evaluated in \cref{Evaluation}.
We refer to 
\paper{\cite{REPORT} for all proofs.}\report{App.\ \ref{appendix} for all proofs regarding AST and to App.\ \ref{PASTandSAST} for all
proofs regarding PAST.}

%% file: pre.tex
\section{Preliminaries}\label{Preliminaries}

We assume 
familiarity with term rewriting \cite{baader_nipkow_1999} and regard (possibly infinite) TRSs
over a 
(possibly infinite) signature $\Sigma$ 
and a set of variables $\VSet$.
Consider the TRS $\R_{\td}$ that doubles a natural number (represented by the terms $\ts$ and $\O$) with the rewrite rules $\td(\ts(x)) \to \ts(\ts(\td(x)))$ and $\td(\O) \to \O$ as an example.
A TRS $\R$ induces a \emph{rewrite relation} ${\to_{\R}} \subseteq \TSet{\Sigma}{\VSet}
\times \TSet{\Sigma}{\VSet}$ on terms where $s \to_{\R} t$ holds if there is a position $\pi$, a rule $\ell \to r \in \R$, and a substitution $\sigma$ such that $s|_{\pi}=\ell\sigma$ and $t = s[r\sigma]_{\pi}$.
A rewrite step $s \to_{\R} t$ is an \emph{innermost} rewrite step (denoted $s \itor t$) if
all proper subterms of the used redex $\ell\sigma$ are in normal form w.r.t.\ $\R$ (i.e.,
they do not contain redexes themselves and thus, they cannot be reduced with $\to_\R$).
For example, we have $\td(\ts(\td(\ts(\O)))) \itoRd \td(\ts(\ts(\ts(\td(\O)))))$.

Let $<$ be the prefix ordering on positions and let $\leq$ be its reflexive closure.
Then for two
parallel positions $\tau$ and $\pi$ we define
$\tau \prec \pi$ if we have $i < j$ for the unique $i,j$ such that $\chi.i \leq \tau$ and $\chi.j \leq \pi$,
where $\chi$ is the longest common prefix of $\tau$ and $\pi$.
An innermost rewrite step $s \itor t$ at position $\pi$ is
\emph{leftmost} (denoted $s \litor t$) if there exists no redex at a position
$\tau$ with $\tau \prec \pi$.

We call a TRS $\R$ \emph{strongly (innermost/leftmost innermost) normalizing} (SN / iSN / liSN) if
$\to_{\R}$ ($\itor$ / $\litor$) is well founded.
SN is also called ``\emph{terminating}'' and iSN/liSN are called
``\emph{innermost/leftmost innermost terminating}''.
If every term $t \in \TSet{\Sigma}{\VSet}$ has a normal form
(i.e., we have $t \to_{\R}^* t'$ where $t'$ is in normal form)
then we call $\R$ \emph{weakly normalizing} (WN).
Two terms $s,t$ are \emph{joinable} via $\R$ (denoted $s \downarrow_{\R} t$) if there exists a term $w$ such that $s \to_{\R}^* w \leftarrow_{\R}^* t$.
Two rules $\ell_1 \to r_1, \ell_2 \to r_2 \in \R$ with renamed variables such that
$\VSet(\ell_1) \cap \VSet(\ell_2) = \emptyset$ are \emph{overlapping}
if there exists a non-variable position $\pi$ of $\ell_1$ such that $\ell_1|_{\pi}$ and
$\ell_2$ are unifiable with a mgu $\sigma$.
If $(\ell_1 \to r_1)
= (\ell_2 \to r_2)$, then we require that $\pi \neq \varepsilon$.
$\R$ is \emph{non-overlapping} (NO) if it has no overlapping rules.
As an example, the TRS $\R_{\td}$ is non-overlapping.
A TRS is \emph{left-linear} (LL) (\emph{right-linear}, RL) if
every variable occurs at most once in the left-hand side  (right-hand side) of a rule.
A TRS is \emph{linear} if it is both left- and right-linear.
A TRS is \emph{non-erasing} (NE)
if in every rule,
all variables of the left-hand side also occur
in the right-hand side.

Next, we recapitulate the relations
between iSN, SN, liSN, and WN
in the non-probabilistic setting.
We start with the relation between iSN and SN.

\begin{counterexample}[Toyama's Counterexample \cite{Toyama87}]\label{example:diff-SN-vs-iSN-toyama}
    The TRS $\R_1$ with the rules $\tf(\ta,\tb,x) \to \tf(x,x,x)$, $\tg \to \ta$, and $\tg \to \tb$
    is not SN since we have $\tf(\ta, \tb, \tg) \to_{\R_1} \tf(\tg, \tg, \tg) \to_{\R_1}
    \tf(\ta, \tg, \tg) \to_{\R_1} \tf(\ta, \tb, \tg) \to_{\R_1} \ldots$
    But the only innermost rewrite sequences starting with $\tf(\ta, \tb, \tg)$ are
    $\tf(\ta, \tb, \tg) \itorExOne \tf(\ta, \tb, \ta) \itorExOne \tf(\ta, \ta, \ta)$ and
    $\tf(\ta, \tb, \tg) \itorExOne \tf(\ta, \tb, \tb) \itorExOne \tf(\tb, \tb, \tb)$,
    i.e., both reach normal forms in the end.
    Thus, $\R_1$ is iSN as
    we have to rewrite the inner $\tg$ before we can use the $\tf$-rule. 
\end{counterexample}

The first property known to ensure equivalence of SN and iSN is orthogonality.
A TRS is \emph{orthogonal} if it is non-overlapping and left-linear.

\begin{restatable}[From iSN to SN (1), \cite{ODonnell77}]{theorem}{SNvsiSN1}\label{properties-eq-SN-iSN-1}
    If a TRS $\R$ is orthogonal, then $\R$ is SN iff $\R$ is iSN.
\end{restatable}

Then, in \cite{Gramlich1995AbstractRB} it was shown that one can remove the left-linearity requirement.

\begin{restatable}[From iSN to SN (2), \cite{Gramlich1995AbstractRB}]{theorem}{SNvsiSN2}\label{properties-eq-SN-iSN-2}
    If a TRS $\R$ is non-overlapping, then $\R$ is SN iff $\R$ is iSN.
\end{restatable}

Finally, \cite{Gramlich1995AbstractRB} also refined
\Cref{properties-eq-SN-iSN-2} further.
A TRS $\R$ is an \emph{overlay system} (OS) if
its rules may only overlap at the root position, i.e.,
$\pi = \varepsilon$.
For \Cref{example:diff-SN-vs-iSN-toyama} one can see that the
only overlap occurs at the root position, i.e., $\R_1$ is an overlay system.
Furthermore, a TRS is \emph{locally confluent} (or \emph{weakly Church-Rosser}, abbreviated WCR) if for all terms $s,t_1,t_2$ such that $t_1 \leftR s \rightarrow_{\R} t_2$ the terms $t_1$ and $t_2$ are joinable.
So $\R_1$ is \emph{not} WCR, as we have $\tf(\ta,\tb,\ta) \leftROne \tf(\ta,\tb,\tg) \rightarrow_{\R_1} \tf(\ta,\tb,\tb)$, but $\tf(\ta,\tb,\ta) \centernot\downarrow_{\R_1} \tf(\ta,\tb,\tb)$.
If a TRS has both of these properties, then iSN and SN are again equivalent.

\begin{restatable}[From iSN to SN (3), \cite{Gramlich1995AbstractRB}]{theorem}{SNvsiSN3}\label{properties-eq-SN-iSN-3}
    If a TRS $\R$ is a locally confluent overlay system, then $\R$ is SN iff $\R$ is iSN.
\end{restatable}

\Cref{properties-eq-SN-iSN-3} is stronger than \Cref{properties-eq-SN-iSN-2} 
as every non-overlapping TRS is a locally confluent overlay system.
We recapitulate the relation between WN and SN next.

\begin{counterexample}\label{example:diff-SN-vs-WN}
    Consider the TRS $\R_2$ with the rules $\tf(x) \to \tb$ and $\ta \to \tf(\ta)$.
    This TRS is not SN since we can always rewrite the inner $\ta$ to get $\ta \to_{\R_2} \tf(\ta) \to_{\R_2} \tf(\tf(\ta)) \to_{\R_2} \ldots$, but it is WN since we can also rewrite the outer $\tf(\ldots)$ 
    before we use the $\ta$-rule twice, resulting in the term $\tb$, which is a normal form.
    For the TRS $\R_3$ with the rules $\tf(\ta) \to \tb$ and $\ta \to \tf(\ta)$, the situation is similar.
\end{counterexample}

The TRS $\R_2$ from \Cref{example:diff-SN-vs-WN} is erasing and
$\R_3$ is overlapping.
For TRSs with neither of those two properties, SN and WN are equivalent.

\begin{restatable}[From WN to SN \cite{Gramlich1995AbstractRB}]{theorem}{SNvsWN}\label{properties-SN-vs-WN}
    If a TRS $\R$ is non-overlapping and non-erasing, then $\R$ is SN iff $\R$ is WN.
\end{restatable}

Finally, we look at the difference between rewrite strategies that use an ordering for parallel redexes like leftmost innermost rewriting compared to just innermost rewriting.
It turns out that such an ordering does not interfere with termination at all.

\begin{restatable}[From liSN to iSN \cite{KRISHNARAO2000141}]{theorem}{iSNvsliSN}\label{properties-iSN-vs-liSN}
    For all TRSs $\R$ we have that $\R$ is iSN iff $\R$ is liSN.
\end{restatable}

The relations between the different properties for non-probabilistic TRSs (given in
 \Cref{properties-eq-SN-iSN-3}, \ref{properties-SN-vs-WN},
and \ref{properties-iSN-vs-liSN}) are summarized below.
\begin{center}
	\begin{tikzpicture}
        \tikzstyle{adam}=[rectangle,thick,draw=black!100,fill=white!100,minimum size=4mm]
        \tikzstyle{empty}=[rectangle,thick,minimum size=4mm]
        
        \node[empty] at (0, -1)  (ast) {\textbf{SN}};
        \node[empty] at (-2, -1)  (iast) {\textbf{iSN}};
        \node[empty] at (2, -1)  (wast) {\textbf{WN}};
        \node[empty] at (-4, -1)  (liast) {\textbf{liSN}};

        \draw (iast) edge[implies-implies,double equal sign distance] (liast);
        \draw (ast) edge[-implies,double equal sign distance, bend left] (iast);
        \draw (iast) edge[-implies,double equal sign distance, bend left] node[sloped, anchor=center,above] {\scriptsize \textbf{OS} + \textbf{WCR}} (ast);
        \draw (ast) edge[-implies,double equal sign distance, bend right] (wast);
        \draw (wast) edge[-implies,double equal sign distance, bend right] node[sloped, anchor=center,above] {\scriptsize \textbf{NO}+\textbf{NE}} (ast);
        \draw (iast.south) edge[-implies,double equal sign distance, bend right] (wast.south);

    \end{tikzpicture}
\end{center}

%% file: probtrs.tex
\section{Probabilistic Term Rewriting}\label{Probabilistic Term Rewriting}

In this section, we recapitulate \emph{probabilistic TRSs}
\cite{avanzini2020probabilistic,bournez2005proving, kassinggiesl2023iAST}.
In contrast to TRSs, a PTRS has finite multi-distributions\footnote{The restriction to
finite multi-distributions allows us to simplify the handling of PTRSs in the proofs.}
on the right-hand sides of its rewrite rules.\footnote{A different form of probabilistic rewrite rules was proposed in \textsf{PMaude} \cite{pmaude_2006}, where numerical extra variables in right-hand sides of rules are instantiated according to a probability distribution.}
A finite \emph{multi-distribution} $\mu$ on a set $A \neq \emptyset$ is a  finite multiset of pairs $(p:a)$, where $0 < p \leq 1$ is a probability and $a \in A$, such that $\sum _{(p:a) \in \mu}p = 1$.
$\FDist(A)$ is the set of all finite multi-distributions on $A$.
For $\mu\in\FDist(A)$, its \emph{support} is the multiset $\Supp(\mu)=\{a \mid (p:a)\in\mu$ for some $p\}$.
A \emph{probabilistic rewrite rule} is a pair $\ell \to \mu \in \TSet{\Sigma}{ \VSet} \times \FDist(\TSet{\Sigma}{\VSet})$ such that $\ell \not\in \VSet$ and $\VSet(r) \subseteq \VSet(\ell)$ for every $r \in \Supp(\mu)$.
A \emph{probabilistic TRS} (PTRS) is a (possibly infinite) set $\SSS$ of probabilistic rewrite rules.
Similar to TRSs, the PTRS $\SSS$ induces a \emph{rewrite relation} ${\to_{\SSS}} \subseteq \TSet{\Sigma}{\VSet} \times \FDist(\TSet{\Sigma}{\VSet})$ where $s \to_{\SSS} \{p_1:t_1, \ldots, p_k:t_k\}$  if there is a position $\pi$, a rule $\ell \to \{p_1:r_1, \ldots, p_k:r_k\} \in \SSS$, and a substitution $\sigma$ such that $s|_{\pi}=\ell\sigma$ and $t_j = s[r_j\sigma]_{\pi}$ for all $1 \leq j \leq k$.
We call $s \to_{\SSS} \mu$ an \emph{innermost} rewrite step (denoted $s \itos \mu$)
if all proper subterms of the used redex $\ell\sigma$ are in normal form
w.r.t.\ $\SSS$.
We have $s \litos \mu$ if the rewrite step $s \itos \mu$ at position $\pi$ is
leftmost (i.e., there is no redex at a position $\tau$ with $\tau \prec \pi$).
For example, the PTRS $\SSS_{\trw}$ with the only rule $\tg \to
\{\nicefrac{1}{2}:\tc(\tg,\tg), \; \nicefrac{1}{2}:\bot\}$ corresponds to a symmetric
random walk on the number of $\tg$-symbols in a term.

As in \cite{avanzini2020probabilistic,kassinggiesl2023iAST,Faggian2019ProbabilisticRN,diazcaro_confluence_2018}, we
\emph{lift} $\to_{\SSS}$ to a rewrite relation between multi-distributions in order to track all probabilistic rewrite sequences (up to non-determinism) at once.
For any $0 < p \leq 1$ and any $\mu \in \FDist(A)$, let $p \cdot \mu = \{ (p\cdot q:a) \mid (q:a) \in \mu \}$.

\begin{definition}[Lifting]\label{def:lifting}
    The \emph{lifting} ${\liftto} \subseteq \FDist(\TSet{\Sigma}{\VSet}) \times \FDist(\TSet{\Sigma}{\VSet})$ of a relation ${\to} \subseteq \TSet{\Sigma}{\VSet} \times \FDist(\TSet{\Sigma}{\VSet})$ is the smallest relation with:
	\begin{itemize}
	    \item[$\bullet$] If $t \in \TSet{\Sigma}{\VSet}$ is in normal form w.r.t.~$\rightarrow$, then $\{1: t\} \liftto \{1:t\}$.
		\item[$\bullet$] If $t \to \mu$, then $\{1: t\} \liftto \mu$.
		\item[$\bullet$] If  for all $1 \leq j \leq k$ there are $\mu_j, \nu_j \in
                  \FDist(\TSet{\Sigma}{\VSet})$ with $\mu_j \liftto \nu_j$ and $0 < p_j
                  \leq 1$ with $\sum_{1 \leq j \leq k} p_j = 1$, then $\bigcup_{1 \leq j
                    \leq k} p_j \cdot \mu_j \liftto \bigcup_{1 \leq j \leq k} p_j \cdot
                  \nu_j$. 
	\end{itemize}
\end{definition}
For a PTRS $\SSS$, we write $\liftto_\SSS$, $\iliftto_{\SSS}$, and $\liliftto_{\SSS}$ for
the liftings of $\to_{\SSS}$, $\itos$, and $\litos$, respectively.

\begin{example} \label{example:PTRS-random-walk-lifting-sequence}
  For example, we obtain the following $\lifttoRrw$-rewrite sequence (which is also a
  $\ilifttoRrw$-sequence, but not a $\lilifttoRrw$-sequence).
    {\small
	\[
        \begin{array}{ll}
            &\{1:\tg\}\\
	        \lifttoRrw&\{\nicefrac{1}{2}:\tc(\tg,\tg), \nicefrac{1}{2}:\bot\}\\
            \lifttoRrw& \{\nicefrac{1}{4}:\tc(\tc(\tg,\tg),\tg),\nicefrac{1}{4}:\tc(\bot,\tg), \nicefrac{1}{2}:\bot \}\\
            \lifttoRrw& \{ \nicefrac{1}{8}:\tc(\tc(\tg,\tg),\tc(\tg,\tg)), \nicefrac{1}{8}:\tc(\tc(\tg,\tg),\bot), \nicefrac{1}{8}:\tc(\bot,\tc(\tg,\tg)), \nicefrac{1}{8}:\tc(\bot,\bot), \nicefrac{1}{2}:\bot \}
        \end{array}
    \]}
\end{example}
\pagebreak[2]

To express the concept of almost-sure termination, one has to determine the probability for normal forms in a multi-distribution.

\begin{definition}[$|\mu|_{\SSS}$] \label{def:prob-abs-value}
    For a PTRS $\SSS$, $\mathtt{NF}_{\SSS} \subseteq \TSet{\Sigma}{\VSet}$ denotes the
    set of all \emph{normal forms} w.r.t.\ $\SSS$.
    For any $\mu \in \FDist(\TSet{\Sigma}{\VSet})$, let $|\mu|_{\SSS} = \sum_{(p:t) \in \mu, t \in \mathtt{NF}_{\SSS}} p$.
\end{definition}

\begin{example}\label{example:prob-abs-value}
	Consider $\{ \nicefrac{1}{8}:\tc(\tc(\tg,\tg),\tc(\tg,\tg)), \nicefrac{1}{8}:\tc(\tc(\tg,\tg),\bot), \nicefrac{1}{8}:\tc(\bot,\tc(\tg,\tg)),\linebreak \nicefrac{1}{8}:\tc(\bot,\bot), \nicefrac{1}{2}:\bot \} = \mu$ from \Cref{example:PTRS-random-walk-lifting-sequence}.
	Then $|\mu|_{\SSS_{\trw}} = \nicefrac{1}{8} + \nicefrac{1}{2}  = \nicefrac{5}{8}$,
        since $\tc(\bot,\bot)$ and $\bot$ are both normal forms w.r.t.\ $\SSS_{\trw}$.
\end{example}

\begin{definition}[AST]\label{def:ptrs-innermost-term-innermost-AST}
	Let $\SSS$ be a PTRS and $\vec{\mu} = (\mu_n)_{n \in \IN}$ be an infinite $\liftto_{\SSS}$-rewrite
    sequence, i.e., $\mu_n \liftto_{\SSS} \mu_{n+1}$ for all $n \in \IN$.  We say that
    $\vec{\mu}$ \emph{converges with probability} $\lim\limits_{n \to \infty}|\mu_n|_{\SSS}$.
    $\SSS$ is \emph{almost-surely terminating (AST)} (\emph{innermost AST (iAST)} / \emph{leftmost innermost AST (liAST)}) if $\lim\limits_{n \to \infty} |\mu_n|_{\SSS} = 1$
    holds for every infinite $\liftto_{\SSS}$- ($\iliftto_{\SSS}$- / $\liliftto_{\SSS}$-) rewrite sequence $(\mu_n)_{n \in \IN}$.
        To highlight the consideration of AST for \emph{full} (instead of innermost) rewriting, we
    also speak of \emph{full AST} (fAST) instead of ``AST''.    
    We say that $\SSS$ is \emph{weakly AST} (wAST) if for every term $t$ there exists an infinite 
    $\liftto_{\SSS}$-rewrite sequence $(\mu_n)_{n \in \IN}$ with $\lim\limits_{n \to \infty} |\mu_n|_{\SSS} = 1$ 
    and $\mu_0 = \{1:t\}$.
 \end{definition}

\begin{example}\label{example:rw-is-AST} 
    For every infinite extension $(\mu_n)_{n \in \IN}$ of the  $\lifttoRrw$-rewrite sequence in
    \cref{example:PTRS-random-walk-lifting-sequence}, we have $\lim\limits_{n \to \infty}
    |\mu_n|_{\SSS_{\trw}} = 1$. Indeed, $\SSS_{\trw}$ is fAST and thus also iAST, liAST, and wAST.
\end{example}

Next, we define \emph{positive} almost-sure termination 
that considers the \emph{expected derivation length} $\edl(\vec{\mu})$ of a rewrite sequence $\vec{\mu}$,
i.e., the expected number of steps until one reaches a normal form.
For PAST, we require that the expected derivation lengths of all possible rewrite sequences are finite.
In the following definition, $(1 - |\mu_n|_{\SSS})$ is the probability of terms that are \emph{not} in normal form w.r.t.\ $\SSS$ after the $n$-th step.

\begin{definition}[$\edl$, PAST]\label{def:expected-derivation-length-PAST}
	Let $\SSS$ be a PTRS and $\vec{\mu} = (\mu_n)_{n \in \IN}$ be an infinite $\liftto_{\SSS}$-rewrite sequence.  
    By $\edl(\vec{\mu}) = \sum_{n = 0}^{\infty} (1 - |\mu_n|_{\SSS})$
    we denote the \emph{expected derivation length} of $\vec{\mu}$.
	$\SSS$ is \emph{positively almost-surely terminating (PAST)} 
    (\emph{innermost PAST (iPAST)} / \emph{leftmost innermost PAST (liPAST)}) if $\edl(\vec{\mu})$ is finite
    for every infinite $\liftto_{\SSS}$- 
    ($\iliftto_{\SSS}$- / $\liliftto_{\SSS}$-) rewrite sequence $\vec{\mu} = (\mu_n)_{n \in \IN}$.\footnote{This
    definition is from \cite{avanzini2020probabilistic},
    where it is also explained why this definition of PAST
    is equivalent to the one of, e.g., \cite{bournez2005proving}.}
    Again, we also speak of \emph{full PAST} (fPAST) when considering PAST for the \emph{full}
    rewrite relation $\liftto_{\SSS}$.    
    We say that $\SSS$ is \emph{weakly PAST} (wPAST) if for every term $t$ there exists an infinite $\liftto_{\SSS}$-rewrite
    sequence $\vec{\mu} = (\mu_n)_{n \in \IN}$ such that $\edl(\vec{\mu})$ is finite and $\mu_0 = \{1:t\}$.
\end{definition}

It is well known that PAST implies AST, but not vice versa.

\begin{example}
    For every infinite extension $\vec{\mu} = (\mu_n)_{n \in \IN}$ of the $\lifttoRrw$-rewrite sequence in
    \cref{example:PTRS-random-walk-lifting-sequence}, the expected derivation length
    $\edl(\vec{\mu})$ is infinite, hence $\SSS_{\trw}$ is not PAST w.r.t.\ any of the
    strategies regarded in this paper.
\end{example}

In \cite{dblp:conf/vmcai/fuc19,avanzini2020probabilistic},
PAST was strengthened further to
\emph{bounded} or \emph{strong almost-sure termination} (SAST).
Indeed, our results on PAST can also be adapted to SAST
\paper{(see \cite{REPORT}).}
\report{(see App.\ \ref{PASTandSAST}).}

Many properties of TRSs from \Cref{Preliminaries} can be lifted to PTRSs in a
straightforward way:
A PTRS $\SSS$ is right-linear (non-erasing) iff the TRS
$\{\ell \to r \mid \ell \to \mu \in \SSS, r \in \Supp(\mu)\}$
has the respective property.
Moreover, all properties that just consider the left-hand sides, e.g., left-linearity,
being
non-overlapping, orthogonality,  and being an overlay system,
can be lifted to PTRSs directly as well, since their rules again only have a single
left-hand side.

%% file: relating.tex
\section{Relating Variants of AST}\label{Relating AST and its Restricted Forms}

Our goal is to relate AST of full rewriting
to restrictions of fAST, i.e., to iAST (\Cref{subsection-iast-fast}), wAST (\Cref{subsection-wast-fast}), and liAST (\Cref{subsection-liast-fast}).
More precisely, we want to find properties of PTRSs which are suitable for automated
checking and which guarantee that two variants of AST are equivalent.
Then for example, we can use existing tools that analyze iAST in order to prove fAST.
Clearly, we have to impose at least the same requirements as in the
non-probabilistic setting, as every TRS $\R$ can be transformed into a PTRS $\SSS$
by replacing every rule $\ell \to r$ with $\ell \to \{1:r\}$.
Then $\R$ is SN / iSN / liSN iff $\SSS$ is fAST / iAST / liAST.
While we mostly focus on AST, all results and counterexamples in this section also hold for PAST.

\subsection{From iAST to fAST}\label{subsection-iast-fast}

Again, we start by analyzing the relation between iAST and fAST.
The following example shows that \Cref{properties-eq-SN-iSN-1} does not carry over to the probabilistic setting,
i.e., orthogonality is not sufficient to ensure that iAST implies fAST.

\begin{counterexample}[Orthogonality Does Not Suffice]\label{example:diff-AST-vs-iAST-dup}
    Consider the orthogonal PTRS $\SSS_1$ with the two rules:

    \vspace*{-.6cm}
    \begin{minipage}[t]{5cm}
        \begin{align*}
            \tg &\to \{\nicefrac{3}{4}:\td(\tg), \nicefrac{1}{4}:\bot\}
        \end{align*}
    \end{minipage}
    \hspace*{.5cm}
    \begin{minipage}[t]{5cm}
        \begin{align*}
            \td(x) &\to \{1:\tc(x,x)\}
        \end{align*}
    \end{minipage}
    \vspace*{.2cm}

    \noindent
    This PTRS is not fAST (and thus, also not fPAST), as we have
    $\{1:\tg\} \liftto_{\SSS_1}^2 \{\nicefrac{3}{4}:\tc(\tg,\tg), \nicefrac{1}{4}:\bot\}$,
    which corresponds to a random walk biased towards non-termination (since $\tfrac{3}{4} > \tfrac{1}{2}$).

    However, the $\td$-rule can only duplicate normal forms in innermost evaluations.
    To see that $\SSS_1$ is iPAST (and thus, also iAST), consider the following rewrite sequence $\vec{\mu}$:
    {\small
    \[
        \{1:\tg\} \iliftto_{\SSS_1} \{\nicefrac{3}{4}:\td(\tg), \nicefrac{1}{4}:\bot\} \iliftto_{\SSS_1} \{(\nicefrac{3}{4})^2:\td(\td(\tg)),\nicefrac{1}{4} \cdot \nicefrac{3}{4}:\td(\bot), \nicefrac{1}{4}:\bot\} \iliftto_{\SSS_1} \ldots
    \]}
    We can also view this rewrite sequence as a tree:

\medskip
    
    \begin{center}
        \begin{tikzpicture}
            \tikzstyle{adam}=[thick,draw=black!100,fill=white!100,minimum size=4mm, shape=rectangle split, rectangle split parts=2,rectangle split horizontal]
            \tikzstyle{empty}=[rectangle,thick,minimum size=4mm]

            \node[empty] at (-7, 2)  (a) {$\mu_0:$};
            \node[adam] at (0, 2)  (1) {$1$ \nodepart{two} $\tg$};

            \node[empty] at (-7, 1)  (b) {$\mu_1:$};
            \node[adam] at (-1.5, 1)  (11) {$\nicefrac{3}{4}$\nodepart{two}$\td(\tg)$};
            \node[adam] at (1.5, 1)  (12) {$\nicefrac{1}{4}$\nodepart{two}$\bot$};

            \node[empty] at (-7, 0)  (c) {$\mu_2:$};
            \node[adam] at (-3, 0)  (111) {$(\nicefrac{3}{4})^2$\nodepart{two}$\td(\td(\tg))$};
            \node[adam] at (0, 0)  (112) {$\nicefrac{1}{4} \cdot \nicefrac{3}{4}$\nodepart{two}$\td(\bot)$};

            \node[empty] at (-7, -1)  (d) {$\mu_3:$};
            \node[adam] at (-4.8, -1)  (1111) {$(\nicefrac{3}{4})^3$\nodepart{two}$\td(\td(\td(\tg)))$};
            \node[adam] at (-1.2, -1)  (1112) {$\nicefrac{1}{4} \cdot (\nicefrac{3}{4})^2$\nodepart{two}$\td(\td(\bot))$};
            \node[empty] at (1.5, -1)  (1121) {$\ldots$};

            \node[empty] at (-4.8, -2)  (11111) {$\ldots$};
            \node[empty] at (0.3, -2)  (11121) {$\ldots$};

            \draw (1) edge[->] (11);
            \draw (1) edge[->] (12);
            \draw (11) edge[->] (111);
            \draw (11) edge[->] (112);
            \draw (111) edge[->] (1111);
            \draw (111) edge[->] (1112);
            \draw (112) edge[->] (1121);
            \draw (1112) edge[->] (11121);
            \draw (1111) edge[->] (11111);
            \draw (1112) edge[->] (11121);
        \end{tikzpicture}
    \end{center}
    The branch to the right that starts with $\bot$ stops after $0$ innermost steps, the branch that starts with $\td(\bot)$ stops after $1$ innermost steps, the branch that starts with $\td(\td(\bot))$ stops after $2$ innermost steps, and so on.
    So if we start with the term $\td^n(\bot)$, then we reach a normal form after
    $n$ steps, and we reach $\td^n(\bot)$ after $n+1$ steps
    from the initial term $\tg$, where $\td^n(\bot) = \underbrace{\td(\ldots(\td}_{n\text{-times}}(\bot))\ldots)$.
    Hence, for every $k \in \IN$ we have
    $|\mu_{2\cdot k + 1}|_{\SSS_1} = |\mu_{2 \cdot k + 2}|_{\SSS_1} = \sum_{n=0}^{k} \nicefrac{1}{4} \cdot (\nicefrac{3}{4})^n$
    and thus   
    \[ 
        \begin{array}{rclcl}
            \edl(\vec{\mu}) &=& \sum_{n = 0}^{\infty} (1 - |\mu_n|_{\SSS_1}) 
            &=&
            1 + 2 \cdot  \sum_{k \in \IN} (1 - |\mu_{2\cdot k + 1}|_{\SSS_1}) \\
            &=&
            1 + 2 \cdot  \sum_{k \in \IN} (1 - \sum_{n=0}^{k} \nicefrac{1}{4} \cdot (\nicefrac{3}{4})^n) 
            &=&
            1 + 2 \cdot  \sum_{k \in \IN} (\nicefrac{3}{4})^{k+1}\\
            &=&
            (2 \cdot  \sum_{k \in \IN} (\nicefrac{3}{4})^{k}) -1 
            &=& 
            7
        \end{array}
        \]
    Analogously, in all other innermost rewrite sequences,  the $\td$-rule can also only duplicate normal
    forms. Thus, all possible innermost rewrite sequences have finite expected derivation
    length. Therefore, $\SSS_1$ is iPAST and thus, also iAST.
    The latter can also be proved automatically by our implementation of the probabilistic DP
    framework for iAST \cite{kassinggiesl2023iAST} in \aprove{}.
\end{counterexample}

To construct a counterexample for AST of $\SSS_1$, we exploited the fact that $\SSS_1$ is not right-linear.
Indeed, requiring right-linearity yields our desired result.
\paper{For reasons of space, here we only give a proof sketch. 
As mentioned, all full proofs can be found in \cite{REPORT}.}
\report{As mentioned, all proofs regarding AST can be found in
  App.\ \ref{appendix}  and  all proofs regarding PAST can be found in
   App.\ \ref{PASTandSAST}.}

\setcounter{iASTtofASTOne}{\value{theorem}}

\begin{restatable}[From iAST/iPAST to fAST/fPAST (1)]{theorem}{ASTAndIASTPropertyOne}\label{properties-eq-AST-iAST-1}
    If a PTRS $\SSS$ is orthogonal and right-linear (i.e., non-overlapping and linear), then:
    \begin{align*}
        \SSS \text{ is fAST} &\Longleftrightarrow \SSS \text{ is iAST}\\
        \SSS \text{ is fPAST} &\Longleftrightarrow \SSS \text{ is iPAST}\!
    \end{align*}
\end{restatable}

\begin{myproofsketch}
    We only have to prove the non-trivial direction ``$\Longleftarrow$''.
    The proofs for all theorems in this section (for both AST and PAST) follow a similar structure.
    We always iteratively replace rewrite steps by steps that use the desired strategy and
    ensure that this does not increase
    the probability of termination (resp.\ the expected derivation
    length).
    For this replacement, we lift the corresponding construction from the non-probabilistic to the
    probabilistic setting. However, this cannot be done directly but instead, we have to
    regard the ``limit'' of a sequence of transformation steps.

    We first consider fAST and iAST.
    Let $\SSS$ be a PTRS that is non-overlapping, linear, and not fAST.
    Thus,  there exists an infinite rewrite sequence $\vec{\mu} = (\mu_n)_{n \in \IN}$ such that $\lim_{n \to \infty} |\mu_n|_{\SSS} = c$ for some $c \in \IR$ with $0 \leq c < 1$.
    Our goal is  to transform this sequence into an innermost sequence that converges at most with
    probability $c$. If the sequence is not yet an innermost one, then in 
    $(\mu_n)_{n \in \IN}$ at least one rewrite step is performed with a redex that is not an innermost redex.
    Since $\SSS$ is non-overlapping, we can replace a first such non-innermost rewrite step with an innermost rewrite step using
    a similar construction as in the non-probabilistic setting.
    In this way, we result in a rewrite sequence $\vec{\mu}^{(1)} = (\mu^{(1)}_n)_{n \in \IN}$ with 
    $\lim_{n \to \infty} |\mu^{(1)}_n|_{\SSS} = \lim_{n \to \infty} |\mu_n|_{\SSS} = c$.
    Here, linearity is needed to ensure that the probability of termination
    does not increase during this replacement.
    We can then repeat this replacement for every non-innermost rewrite step, i.e., we again
    replace a first non-innermost rewrite step in $(\mu^{(1)}_n)_{n \in \IN}$ to obtain
    $(\mu^{(2)}_n)_{n \in \IN}$ with  the same termination probability, etc.
    In the end, the limit of all these rewrite sequences $\lim_{i \to \infty}
    (\mu^{(i)}_n)_{n \in \IN}$ is an innermost rewrite sequence that converges with
    probability at most
    $c < 1$, and hence, the PTRS $\SSS$ is not innermost AST.

    For fPAST and iPAST,
    we start with an infinite rewrite sequence $\vec{\mu}$ such that $\edl(\vec{\mu}) = \infty$.
    Again, we replace the first non-innermost rewrite step with an innermost rewrite step
    using exactly the same construction as before to obtain $\vec{\mu}^{(1)}$, etc., since
    $\vec{\mu}^{(1)}$ does not only have 
    the same termination probability as
    $\vec{\mu}$, but we also have 
    $\edl(\vec{\mu}^{(1)}) \geq \edl(\vec{\mu})$.
    In the end, the limit of all these rewrite sequences 
    $\lim_{i \to \infty}\vec{\mu}^{(i)}$ is an innermost rewrite sequence 
    such that $\edl(\lim_{i \to \infty}\vec{\mu}^{(i)}) \geq \edl(\vec{\mu}) = \infty$, 
    and hence, the PTRS $\SSS$ is not innermost PAST.
\end{myproofsketch}

One may wonder whether we can remove the left-linearity requirement from \Cref{properties-eq-AST-iAST-1}, 
as in the non-probabilistic setting.
It turns out that this is not possible.

\begin{counterexample}[Left-Linearity Cannot be Removed]\label{example:diff-AST-vs-iAST-left-lin}
    Consider the PTRS $\SSS_2$ with the rules:
    
    \vspace*{-.5cm}
    \begin{minipage}[t]{5cm}
        \begin{align*}
            \tf(x,x) &\to \{1:\tf(\ta,\ta)\}
        \end{align*}
    \end{minipage}
    \hspace*{.5cm}
    \begin{minipage}[t]{5cm}
        \begin{align*}
            \ta &\to \{\nicefrac{1}{2}:\tb, \nicefrac{1}{2}:\tc\}
        \end{align*}
    \end{minipage}
    \vspace*{.2cm}

    \noindent
    $\SSS_2$ is not fAST (hence also not fPAST), since $\{1:\tf(\ta,\ta)\} \liftto_{\SSS_2} \{1:\tf(\ta,\ta)\} \liftto_{\SSS_2} \ldots$ is an infinite rewrite sequence that converges with probability $0$.
    However, it is iPAST (and hence, iAST) since the corresponding innermost sequence has the form $\{1:\tf(\ta,\ta)\} \iliftto_{\SSS_2} \{\tfrac{1}{2}:\tf(\tb,\ta),\tfrac{1}{2}:\tf(\tc,\ta)\} \iliftto_{\SSS_2} \{\tfrac{1}{4}:\tf(\tb,\tb), \tfrac{1}{4}:\tf(\tb,\tc), \tfrac{1}{4}:\tf(\tc,\tb), \tfrac{1}{4}:\tf(\tc,\tc)\}$.
    Here, the last distribution contains two normal forms $\tf(\tb,\tc)$ and $\tf(\tc,\tb)$ that did not occur in the previous rewrite sequence.
    Since all innermost rewrite sequences keep on adding such normal forms after a certain
    number of steps for each start term, 
    they always have finite expected derivation length and thus,
    converge with probability $1$ (again, iAST can be shown automatically by \aprove{}).
    Note that adding the requirement of being non-erasing would not help to get rid of the left-linearity either, 
    as shown by the PTRS $\SSS_3$ which results from $\SSS_2$ by replacing the
    $\tf$-rule with $\tf(x,x) \to \{1:\td(\tf(\ta,\ta), x)\}$.
\end{counterexample}

The problem here is that although we rewrite both occurrences of $\ta$ with the same rewrite rule,
the two $\ta$-symbols are replaced by two different terms (each with a probability $> 0$).
This is impossible in the non-probabilistic setting.

Next, one could try to adapt \cref{properties-eq-SN-iSN-3} to the probabilistic setting
(when requiring linearity in addition).
So one could investigate whether iAST implies fAST for PTRSs that are linear locally confluent
overlay systems. A PTRS $\SSS$ is \emph{locally confluent} if 
for all multi-distributions $\mu, \mu_1, \mu_2$ such that $\mu_1 \leftleftarrows_{\SSS} \mu
\liftto_{\SSS} \mu_2$, there exists a multi-distribution $\mu'$ such that 
 $\mu_1 \liftto_{\SSS}^* \mu'
\leftleftarrows_{\SSS}^* \mu_2$, see \cite{diazcaro_confluence_2018}. Note that in
contrast to the non-probabilistic setting, there are non-overlapping PTRSs that are not
locally confluent
(e.g., the variant $\SSS_2'$ of $\SSS_2$ that consists of the rules
 $\tf(x,x) \to \{1:\td\}$ and $\ta \to \{\nicefrac{1}{2}:\tb, \nicefrac{1}{2}:\tc\}$,
since we have 
 $\{1:\td\} \leftleftarrows_{\SSS_2'} \{1:\tf(\ta,\ta)\}
\liftto_{\SSS_2'} \{\nicefrac{1}{2}:\tf(\tb,\ta), \nicefrac{1}{2}:\tf(\tc,\ta)\}$ and the two
resulting multi-distributions are not joinable).
Since $\SSS_2'$ is not linear, it does not show that such an adaption of \cref{properties-eq-SN-iSN-3} would not
subsume \Cref{properties-eq-AST-iAST-1}.

In contrast to the proof of \cref{properties-eq-SN-iSN-1},
the proof of \cref{properties-eq-SN-iSN-3} relies on a minimality requirement for
the used redex.
In the non-probabilistic setting,
whenever a term $t$ starts an infinite rewrite sequence, then there exists a position
$\pi$ of $t$ such that there is an infinite rewrite sequence of $t$ starting with the redex
$t|_\pi$, but no  infinite rewrite sequence of $t$ starting with a redex at a position
$\tau > \pi$ which is strictly below $\pi$.
In other words, if $t$ starts an infinite rewrite sequence, then
there is a ``minimal'' infinite rewrite sequence starting in $t$, i.e., as soon as one
reduces a proper subterm of one of the redexes in the sequence, then one obtains a term
which is terminating.
However, such minimal infinite sequences do not always exist in the probabilistic setting.

\begin{example}[No Minimal Infinite Rewrite Sequence for AST]\label{example:minimality-in-positions-AST}
    Reconsider the PTRS $\SSS_1$ from \cref{example:diff-AST-vs-iAST-dup}, which is not fAST.
    However, there is no ``minimal'' rewrite sequence with convergence probability $< 1$
    such that one rewrite step at a
    proper subterm of a redex would modify the multi-distribution in such a way that now
    only rewrite sequences with convergence probability $1$ are possible. We have
    $\{1:\tg\} \liftto_{\SSS_1}
    \{\nicefrac{3}{4}:\td(\tg), \nicefrac{1}{4}:\bot\}$. In
    \Cref{example:diff-AST-vs-iAST-dup}, we now alternated between the
    $\td$- and the $\tg$-rule, resulting in a biased random walk, i.e., we obtained
    $\{\nicefrac{3}{4}:\td(\tg), \nicefrac{1}{4}:\bot\}  \liftto_{\SSS_1}
    \{\nicefrac{3}{4}:\tc(\tg,\tg), \nicefrac{1}{4}:\bot\}  \liftto_{\SSS_1}
    \{(\nicefrac{3}{4})^2:\tc(\td(\tg),\tg), \nicefrac{3}{4} \cdot \nicefrac{1}{4}:\tc(\bot,\tg), \nicefrac{1}{4}:\bot\}  \liftto_{\SSS_1}
    \ldots$\
    The steps with the $\td$-rule use redexes that have $\tg$ as a proper subterm.

    However, there does not exist any ``minimal'' non-fAST sequence.
    If we rewrite the proper subterm $\tg$ of a redex $\td(\tg)$, then this still yields a
    multi-distribution that is not fAST, i.e., it can still start a rewrite sequence with
    convergence probability $< 1$. For example, we have  $\{\nicefrac{3}{4}:\td(\tg),
    \nicefrac{1}{4}:\bot\}  \liftto_{\SSS_1}
    \{(\nicefrac{3}{4})^2:\td(\td(\tg)),\nicefrac{1}{4} \cdot \nicefrac{3}{4}:\td(\bot),
    \nicefrac{1}{4}:\bot\}$, but the obtained multi-distribution still contains the subterm
    $\tg$,
    and thus, one can still continue the rewrite sequence in such a way that
    its convergence probability is $< 1$.
    Again, the same example also shows that there is no 
    ``minimal'' non-fPAST sequence.
\end{example}

It remains open whether one can also adapt
\cref{properties-eq-SN-iSN-3} to the probabilistic setting (e.g., if one can replace
non-overlappingness in \Cref{properties-eq-AST-iAST-1}
by the requirement of locally confluent overlay systems). 
There are two main difficulties when trying to adapt the proof of this theorem to PTRSs.
First, the minimality requirement cannot be imposed in the probabilistic setting, as
discussed above.
In the non-probabilistic setting, this requirement is needed  to ensure that rewriting
below a position that was reduced in the original (minimal) infinite rewrite sequence leads to a strongly normalizing rewrite sequence.
Second, the original proof of \cref{properties-eq-SN-iSN-3} uses Newman's Lemma
\cite{Newman42}  which states
that local confluence implies confluence for strongly normalizing terms $t$, and thus
it implies that $t$ has a unique normal form.
Local confluence and adaptions of the unique normal form property for the probabilistic
setting have
been studied in \cite{Faggian2019ProbabilisticRN,diazcaro_confluence_2018}, which concluded that obtaining an
analogous statement to Newman's Lemma for PTRSs that are AST (or PAST) would be very difficult.
The reason is that one cannot use well-founded induction on the length of a rewrite
sequence of a PTRS that is AST (or PAST), since these 
rewrite sequences may be infinite.

\subsection{From wAST to fAST}\label{subsection-wast-fast}

Next, we investigate wAST. Since iAST implies wAST, 
we essentially have the same problems as for innermost AST, i.e.,
in addition to non-overlappingness, we need linearity, as seen in
\Cref{example:diff-AST-vs-iAST-dup,example:diff-AST-vs-iAST-left-lin},
as $\SSS_1$ and $\SSS_3$ are iAST (and hence wAST) but not fAST,
while they are non-overlapping and non-erasing, but not linear.
Furthermore, we need non-erasingness as we did in the non-probabilistic setting for the
same reasons, see
\Cref{example:diff-SN-vs-WN}.

\setcounter{wASTtofAST}{\value{theorem}}

\begin{restatable}[From wAST/wPAST to fAST/fPAST]{theorem}{ASTvswAST}\label{properties-AST-vs-wAST}
  If a PTRS $\SSS$ is non-overlapping, linear, and non-erasing, then

\vspace*{-.5cm}
  
    \begin{align*}
        \SSS \text{ is fAST} &\Longleftrightarrow \SSS \text{ is wAST}\\
        \SSS \text{ is fPAST} &\Longleftrightarrow \SSS \text{ is wPAST}\!
    \end{align*}
\end{restatable}

\subsection{From liAST to fAST}\label{subsection-liast-fast}

Finally, we look at leftmost-innermost AST as an example for a rewrite strategy that uses an ordering for parallel redexes.
In contrast to the non-probabilistic setting, it turns out that liAST and iAST are not equivalent in general.
The counterexample is similar to \Cref{example:diff-AST-vs-iAST-left-lin}, which illustrated that fAST and iAST are not equivalent without left-linearity.

\begin{counterexample}\label{example:liAST-vs-iAST}
    Consider the PTRS $\SSS_4$ with the five rules:
    
    \vspace*{-.5cm}
    \begin{minipage}[t]{4cm}
        \vspace*{.2cm}
        \begin{align*}
            \ta &\to \{1:\tc_1\}\\
            \ta &\to \{1:\tc_2\}
        \end{align*}
    \end{minipage}
    \hspace*{.5cm}
    \begin{minipage}[t]{7cm}
        \begin{align*}
            \tb &\to \{\nicefrac{1}{2}:\td_1, \nicefrac{1}{2}:\td_2\}\\
            \tf(\tc_1,\td_1) &\to \{1:\tf(\ta,\tb)\}\\
            \tf(\tc_2,\td_2) &\to \{1:\tf(\ta,\tb)\}
        \end{align*}
    \end{minipage}
    \vspace*{.1cm}

    \noindent
    This PTRS is not iAST (and hence not iPAST) since there exists the infinite rewrite sequence
    $\{1:\tf(\ta,\tb)\} \iliftto_{\SSS_4} \{\nicefrac{1}{2}:\tf(\ta,\td_1),
    \nicefrac{1}{2}:\tf(\ta,\td_2)\} \iliftto_{\SSS_4} \{\nicefrac{1}{2}:\tf(\tc_1,\td_1),
    \nicefrac{1}{2}:\tf(\tc_2,\td_2)\} \iliftto_{\SSS_4} \{\nicefrac{1}{2}:\tf(\ta,\tb),
    \nicefrac{1}{2}:\tf(\ta,\tb)\} \iliftto_{\SSS_4} \ldots$, which converges with probability $0$.
    It first ``splits'' the term $\tf(\ta,\tb)$ with the $\tb$-rule, and then applies one of the two different $\ta$-rules to each of the resulting terms.
    In contrast, when applying a leftmost innermost rewrite strategy, we have to decide which $\ta$-rule to use.
    For example, we have $\{1:\tf(\ta,\tb)\} \liliftto_{\SSS_4} \{1:\tf(\tc_1,\tb)\} \liliftto_{\SSS_4} \{\nicefrac{1}{2}:\tf(\tc_1,\td_1), \nicefrac{1}{2}:\tf(\tc_1,\td_2)\}$.
    Here, the second term $\tf(\tc_1,\td_2)$ is a normal form.
    Since all leftmost innermost rewrite sequences keep on adding such normal forms after
    a certain number of steps for each start term, the PTRS is liAST (and also liPAST).
\end{counterexample}

The counterexample above can easily be adapted to variants of innermost rewriting that impose different orders on parallel redexes like, e.g., \emph{rightmost} innermost rewriting.

However, liAST and iAST are again equivalent for non-overlapping TRSs.
For such TRSs, at most one rule can be used to rewrite at a given position, which prevents the problem illustrated in \Cref{example:liAST-vs-iAST}.

\setcounter{liASTtoiAST}{\value{theorem}}

\begin{restatable}[From liAST/liPAST to iAST/iPAST]{theorem}{iASTvsliAST}\label{properties-iAST-vs-liAST}
  If a PTRS $\SSS$ is non-overlapping, then

  \vspace*{-.8cm}
  
    \begin{align*}
        \SSS \text{ is iAST} &\Longleftrightarrow \SSS \text{ is liAST}\\
        \SSS \text{ is iPAST} &\Longleftrightarrow \SSS \text{ is liPAST}\!
    \end{align*}
\end{restatable}

The relations between the different properties for AST of PTRSs (given in
\Cref{properties-eq-AST-iAST-1}, \ref{properties-AST-vs-wAST}, and \ref{properties-iAST-vs-liAST})
are summarized below.
An analogous figure also holds for PAST.

\begin{center}
	\begin{tikzpicture}
        \tikzstyle{adam}=[rectangle,thick,draw=black!100,fill=white!100,minimum size=4mm]
        \tikzstyle{empty}=[rectangle,thick,minimum size=4mm]

        \node[empty] at (0, -1)  (ast) {\textbf{fAST}};
        \node[empty] at (-3, -1)  (iast) {\textbf{iAST}};
        \node[empty] at (3, -1)  (weak-AST) {\textbf{wAST}};
        \node[empty] at (-6, -1)  (liast) {\textbf{liAST}};

        \draw (iast) edge[-implies,double equal sign distance, bend left] (liast);
        \draw (liast) edge[-implies,double equal sign distance, bend left] node[sloped, anchor=center,above] {\scriptsize \textbf{NO}} (iast);
        \draw (ast) edge[-implies,double equal sign distance, bend left] (iast);
        \draw (iast) edge[-implies,double equal sign distance, bend left] node[sloped, anchor=center,above] {\scriptsize \textbf{NO}+\textbf{LL}+\textbf{RL}} (ast);
        \draw (ast) edge[-implies,double equal sign distance, bend right] (weak-AST);
        \draw (weak-AST) edge[-implies,double equal sign distance, bend right] node[sloped, anchor=center,above] {\scriptsize \textbf{NO}+\textbf{LL}+\textbf{RL}+\textbf{NE}} (ast);
        \draw (iast) edge[-implies,double equal sign distance, bend right, out=310, in=235, looseness=0.5] (weak-AST);
        \draw (liast) edge[-implies,double equal sign distance, bend right, out=310,
          in=250, looseness=0.5] (weak-AST.south);
    \end{tikzpicture}
\end{center}

\vspace*{-.5cm}

%% file: improve.tex
\section{Improving Applicability}\label{Improving Applicability}

In this section, we improve the applicability of \cref{properties-eq-AST-iAST-1},  
which relates fAST and iAST.\report{\footnote{This section contains some mistakes, which are fixed in the journal version~\cite{kassing2025LMCS}, see \url{https://jckassing.com/publication/2024-04}.}}
The results of \Cref{sec:simultaneous rewriting} allow us to remove the requirement of
left-linearity by modifying the rewrite relation to \emph{simultaneous rewriting}.
Then in \Cref{sec:spareness} we show that the requirement of right-linearity can be
weakened to \emph{spareness} if one only considers rewrite sequences that start with \emph{basic terms}.

\subsection{Removing Left-Linearity by Simultaneous Rewriting}\label{sec:simultaneous rewriting}

First, we will see that we do not need to require left-linearity if we allow the
simultaneous reduction of several copies of identical redexes. For a PTRS $\SSS$, this
results in the notion of
\emph{simultaneous rewriting},
denoted $\toparas$.
While $\itoparas$ over-approximates $\itos$, existing techniques for proving iAST
\cite{kassinggiesl2023iAST,FLOPS2024} (except for the rewriting
processor\footnote{This processor is an
optional transformation technique which was added in \cite{FLOPS2024} when
improving the DP framework 
further since it 
sometimes helps
to increase power, but all other (major) DP processors do not distinguish between  $\itos$ and
$\itoparas$.})
do not distinguish between both notions of rewriting, 
i.e., these techniques even prove
that every rewrite sequence
with the lifting ${\iparalifttos}$ of 
$\itoparas$
converges with probability 1.
So for non-overlapping and right-linear PTRSs, these techniques
can be used to prove innermost almost-sure termination
w.r.t.\ $\toparas$, which then implies fAST. 
The following example illustrates our approach for handling non-left-linear PTRSs by
applying the same rewrite rule at parallel positions simultaneously. 

\begin{example}[Simultaneous Rewriting]\label{example:simultaneous-simultaneous-rewriting}
    Reconsider the PTRS $\SSS_2$ from \Cref{example:diff-AST-vs-iAST-left-lin} with the rules $\tf(x,x) \to \{1:\tf(\ta,\ta)\}$ and $\ta \to \{\nicefrac{1}{2}:\tb, \nicefrac{1}{2}:\tc\}$ which is iAST, but not fAST.
    Our new rewrite relation ${\paraliftto_{\SSS_2}}$ allows us to reduce several copies
    of the same redex simultaneously, so that we get $\{1:\tf(\ta,\ta)\}
    \iparaliftto_{\SSS_2} \{\tfrac{1}{2}:\tf(\tb,\tb),\tfrac{1}{2}:\tf(\tc,\tc)\}
    \iparaliftto_{\SSS_2} \{\nicefrac{1}{2}:\tf(\ta,\ta),
    \nicefrac{1}{2}:\tf(\ta,\ta)\}$,
    i.e.,
    this $\,\iparaliftto_{\SSS_2}$-sequence converges with probability 0 and thus,
    $\SSS_2$ is \emph{not} iAST w.r.t.\ $\topara_{\SSS_2}$. 
    Note that we simultaneously reduced both occurrences of $\ta$ in the first step.
\end{example}

\begin{definition}[Simultaneous Rewriting]\label{def:PTRS}
	Let $\SSS$ be a PTRS\@.
    A term $s$ rewrites \emph{simultaneously} to a multi-distribution $\mu = \{p_1:t_1, \ldots,
    p_k:t_k\}$ (denoted $s \topara_{\SSS} \mu$) if there is a non-empty set of
    parallel positions $\Pi = \{\pi_1, \ldots, \pi_m\}$, a rule $\ell \to \{p_1:r_1, \ldots, p_k:r_k\} \in \SSS$, and a substitution $\sigma$ such that $s|_{\pi}=\ell\sigma$ for every position $\pi \in \Pi$ and $t_j = s[r_j\sigma]_{\pi_1} \ldots [r_j\sigma]_{\pi_m}$ for all $1 \leq j \leq k$.
	We call $s \topara_{\SSS} \mu$ an \emph{innermost simultaneous} rewrite step (denoted $s \itopara_\SSS \mu$) if all proper subterms of the redex $\ell\sigma$ are in normal form w.r.t.\ $\SSS$.
\end{definition}

Clearly, if the set of positions $\Pi$ from \Cref{def:PTRS} is a singleton, then the
resulting simultaneous rewrite step is an ``ordinary'' probabilistic rewrite step, i.e.,
${\to_{\SSS}} \subseteq {\topara_{\SSS}}$ and
${\itos} \subseteq {\itoparas}$.

\begin{restatable}[From $\paraarrow_{\SSS}$ to $\to_{\SSS}$]{corollary}{rewriteRelationRelation}\label{rewrite-relation-relation-cor}
  If $\SSS$ is fAST (iAST) w.r.t.\ $\paraarrow_{\SSS}$, 
  i.e., every infinite $\paraliftto_\SSS$- (resp.\
  $\iparaliftto_\SSS$-) rewrite sequence converges with probability $1$,
  then $\SSS$ is fAST (iAST). Analogously, if $\SSS$ is fPAST (iPAST) w.r.t.\ $\paraarrow_{\SSS}$, 
  i.e., every infinite $\paraliftto_\SSS$- (resp.\
  $\iparaliftto_\SSS$-) rewrite sequence has finite expected derivation length,  
  then $\SSS$ is fPAST (iPAST).
\end{restatable}

However, the converse of \cref{rewrite-relation-relation-cor} does not hold.
\cref{example:simultaneous-simultaneous-rewriting} shows that
 $\itoparas$ allows for rewrite sequences that are not possible with
$\itos$, and
the following example shows the same for 
$\toparas$ and $\to_{\SSS}$.

\begin{counterexample}\label{example:simultaneous-simultaneous-rewriting-full}
    Consider the PTRS $\overline{\SSS}_2$ with the three rules:
    
    \vspace*{-.5cm}
    \begin{minipage}[t]{4cm}
        \begin{align*}
            \tf(\tb,\tb) &\to \{1:\tf(\ta,\ta)\}\\
            \tf(\tc,\tc) &\to \{1:\tf(\ta,\ta)\}
        \end{align*}
    \end{minipage}
    \hspace*{.5cm}
    \begin{minipage}[t]{7cm}
        \vspace*{.2cm}
        \begin{align*}
            \ta &\to \{\nicefrac{1}{2}:\tb, \nicefrac{1}{2}:\tc\}
        \end{align*}
    \end{minipage}

\vspace*{.2cm}
    
    \noindent
    This PTRS is fAST.
    But as in \Cref{example:simultaneous-simultaneous-rewriting}, we have 
    $\{1:\tf(\ta,\ta)\}
    \iparaliftto_{\overline{\SSS}_2} \{\tfrac{1}{2}:\tf(\tb,\tb),\tfrac{1}{2}:\tf(\tc,\tc)\}
    \iparaliftto_{\overline{\SSS}_2} \{\nicefrac{1}{2}:\tf(\ta,\ta),
    \nicefrac{1}{2}:\tf(\ta,\ta)\}$,
    i.e., there are rewrite sequences with
    $\,\iparaliftto_{\overline{\SSS}_2}$ and thus, also with
    $\,\paraliftto_{\overline{\SSS}_2}$ that converge with probability 0. Hence,
    $\overline{\SSS}_2$ is not iAST or fAST w.r.t.\
 $\paraarrow_{\overline{\SSS}_2}$.
    Again, the same example also shows that fPAST and fPAST w.r.t.\ simultaneous rewriting
    are not equivalent either.
\end{counterexample}

Note that this kind of simultaneous rewriting is different from the ``ordinary'' parallelism used for non-probabilistic rewriting, which is typically denoted by $\to_{||}$. 
There, one may reduce multiple parallel redexes in a single rewrite step.
Here, we do not only allow reducing multiple redexes, but in addition
we ``merge'' the
corresponding terms in the multi-distributions that result from rewriting the different
redexes. Because of this merging, 
we only allow the simultaneous reduction of \emph{equal} redexes, whereas ``ordinary'' parallel rewriting allows the simultaneous reduction of arbitrary parallel redexes.
For example, for $\SSS_2$ from \Cref{example:diff-AST-vs-iAST-left-lin} we have
$\{1:\tf(\ta,\ta)\} \iparaliftto_{\SSS_2}
\{\tfrac{1}{2}:\tf(\tb,\tb),\tfrac{1}{2}:\tf(\tc,\tc)\}$, whereas
using ordinary parallel rewriting we would get $\{1:\tf(\ta,\ta)\} \iliftto_{|| \SSS_2}
\{\tfrac{1}{4}:\tf(\tb,\tb),\tfrac{1}{4}:\tf(\tb,\tc),\tfrac{1}{4}:\tf(\tc,\tb),\tfrac{1}{4}:\tf(\tc,\tc)\}$.

The following theorem shows that indeed, we do not need to require left-linearity when
moving from  iAST/iPAST w.r.t.\ $\paraarrow_{\SSS}$ to
fAST/fPAST w.r.t.\ $\to_{\SSS}$.

\setcounter{iASTtofASTTwo}{\value{theorem}}

\begin{restatable}[From iAST/iPAST to fAST/fPAST (2)]{theorem}{ASTAndIASTPropertyTwo}\label{properties-eq-AST-iAST-2}
    If a PTRS $\SSS$ is non-overlapping and right-linear, then 
    \begin{align*}
        \SSS \text{ is fAST} &\Longleftarrow \SSS \text{ is iAST w.r.t.\ } \paraarrow_{\SSS}\\
        \SSS \text{ is fPAST} &\Longleftarrow \SSS \text{ is iPAST w.r.t.\ } \paraarrow_{\SSS}\!
    \end{align*}
\end{restatable}

\begin{myproofsketch}
  We use an analogous construction as for the proof of
\Cref{properties-eq-AST-iAST-1}, but in addition, if we replace a non-innermost rewrite
step by an innermost one, then we check whether
in the original rewrite sequence,
the corresponding innermost redex is  ``inside'' the substitution used for
the non-innermost rewrite step. In that case, if this rewrite step applied a non-left-linear
rule, then we identify all other (equal) innermost redexes and use $\iparaarrow_{\SSS}$ to rewrite them simultaneously (as we did for the innermost redex $\ta$ in \Cref{example:simultaneous-simultaneous-rewriting}).
\end{myproofsketch}

Note that \Cref{example:simultaneous-simultaneous-rewriting-full} shows that
the  direction ``$\implies$'' does not hold in \Cref{properties-eq-AST-iAST-2}.
The following example shows that right-linearity in \Cref{properties-eq-AST-iAST-2} cannot be weakened to the
requirement that $\SSS$ is \emph{non-duplicating} (i.e., that no variable occurs more
often in a term on the right-hand side of a rule than on its left-hand side). 

\begin{counterexample}[Non-Duplicating Does Not Suffice]\label{example:example:diff-AST-vs-iAST-3-dup}
    Let $\td(\tf(\ta,\ta)^3)$ abbreviate $\td(\tf(\ta,\ta),\tf(\ta,\ta),\tf(\ta,\ta))$.
    Consider the PTRS $\SSS_5$ with the four rules:
    
    \vspace*{-.5cm}
    \begin{minipage}[t]{4cm}
        \begin{align*}
            \tf(x,x) &\to \{1:\tg(x,x)\}\\
            \ta &\to \{\nicefrac{1}{2}:\tb, \nicefrac{1}{2}:\tc\}
        \end{align*}
    \end{minipage}
    \hspace*{.5cm}
    \begin{minipage}[t]{7cm}
        \begin{align*}
            \tg(\tb,\tc) &\to \{1:\td(\tf(\ta,\ta)^3)\}\\
            \tg(\tc,\tb) &\to \{1:\td(\tf(\ta,\ta)^3)\}
        \end{align*}
    \end{minipage}

\medskip
    
    \noindent
    $\SSS_5$ is not fAST (and thus, also not fPAST), since the infinite rewrite sequence $\{1 : \tf(\ta,\ta)\}
    \liftto_{\SSS_5} \{1:\tg(\ta,\ta)\} \liftto_{\SSS_5}^2 \{\nicefrac{1}{4}:\tg(\tb,\tb),
    \nicefrac{1}{4}:\tg(\tb,\tc), \nicefrac{1}{4}:\tg(\tc,\tb),
    \nicefrac{1}{4}:\tg(\tc,\tc)\} \liftto_{\SSS_5} \{\nicefrac{1}{4}:\tg(\tb,\tb),
    \nicefrac{1}{4}:\td(\tf(\ta,\ta)^3),
    \nicefrac{1}{4}:\td(\tf(\ta,\ta)^3),
    \nicefrac{1}{4}:\tg(\tc,\tc)\}$ can be seen as a biased random walk on the number of
    $\tf(\ta,\ta)$-subterms that is not AST.   
    However, for every innermost evaluation with $\ito_{\SSS_5}$ or $\itopara_{\SSS_5}$ we
    have to rewrite the inner $\ta$-symbols first.
    Afterwards, the $\tf$-rule can only be used on redexes $\tf(t,t)$ where the resulting
    term $\tg(t,t)$ is a normal form. Thus,  $\SSS_5$ is iPAST (and hence, iAST) w.r.t.\
    $\paraarrow_{\SSS_5}$.
\end{counterexample}

Note that for wAST,
the direction of the implication in \Cref{rewrite-relation-relation-cor} is reversed, 
since wAST requires that for each start term, there \emph{exists} an infinite rewrite sequence that is almost-surely terminating, 
whereas fAST requires that \emph{all} infinite rewrite sequences are almost-surely terminating.
Thus, if there exists an infinite $\liftto_\SSS$-rewrite sequence that converges with probability $1$ 
(showing that $\SSS$ is wAST),
then this is also a valid $\paraliftto_\SSS$-rewrite sequence that converges with probability $1$
(showing that $\SSS$ is  wAST w.r.t.\ $\paraarrow_{\SSS}$).

\begin{restatable}[From $\to_{\SSS}$ to $\paraarrow_{\SSS}$ for wAST/wPAST]{corollary}{rewriteRelationRelationWeakAST}\label{rewrite-relation-relation-cor-weak-AST}
    If $\SSS$ is wAST (wPAST),
    then $\SSS$ is wAST (wPAST) w.r.t.\ $\paraarrow_{\SSS}$.
\end{restatable}

One may wonder whether simultaneous rewriting could also be used to improve
\Cref{properties-AST-vs-wAST} by removing the requirement of left-linearity,
but  \Cref{example:problem-sim-rewriting-wAST}
shows this is not possible.

\begin{counterexample}\label{example:problem-sim-rewriting-wAST}
    Consider the non-left-linear PTRS $\SSS_6$ with the two rules:

    \vspace*{-.5cm}
    \begin{minipage}[t]{5cm}
        \begin{align*}
            \tg &\to \{\nicefrac{3}{4}:\td(\tg,\tg), \nicefrac{1}{4}:\bot\}
        \end{align*}
    \end{minipage}
    \hspace*{.5cm}
    \begin{minipage}[t]{5cm}
        \begin{align*}
            \td(x,x) &\to \{1:x\}
        \end{align*}
    \end{minipage}
    \vspace*{.2cm}

    \noindent
    This PTRS is not fAST (and thus, also not fPAST), as we have
    $\{1:\tg\} \liftto_{\SSS_6} \{\nicefrac{3}{4}:\td(\tg,\tg), \nicefrac{1}{4}:\bot\}$,
    which corresponds to a random walk biased towards non-termination if we never use the $\td$-rule (since $\tfrac{3}{4} > \tfrac{1}{2}$).
    However, if we always use the $\td$-rule directly after the $\tg$-rule, then we
    essentially end up with a PTRS whose only rule is $\tg \to \{\nicefrac{3}{4}:\tc(\tg),
    \nicefrac{1}{4}:\bot\}$, which corresponds to flipping a biased coin until heads comes
    up.
       This proves that $\SSS_6$ is wPAST and hence, also wAST.
       As $\SSS_6$ is non-overlapping, right-linear, and non-erasing,
       this shows that a variant of \Cref{properties-AST-vs-wAST} without the requirement of left-linearity needs more than just moving to simultaneous rewriting.
\end{counterexample}

\subsection{Weakening Right-Linearity to Spareness}\label{sec:spareness}

To improve our results further, we introduce the notion of \emph{spareness}.
The idea of spareness is to require that variables which occur non-linear in right-hand
sides may only be instantiated by normal forms. We already used spareness for
non-probabilistic TRSs in 
\cite{frohn_analyzing_nodate} to
find classes of TRSs where
innermost and full runtime complexity coincide.
For a PTRS $\SSS$, we decompose its signature $\Sigma = \Sigma_{C} \uplus \Sigma_{D}$ such
that $f \in \Sigma_D$ iff $f = \rootsym(\ell)$ for some rule $\ell \to \mu \in \SSS$. 
The symbols in $\Sigma_{C}$ and $\Sigma_{D}$ are called \emph{constructors} and
\emph{defined symbols}, respectively. 

\begin{definition}[Spareness]\label{def:spareness}
    Let $\ell \to \mu \in \SSS$.
    A rewrite step $\ell\sigma \to_\SSS \mu\sigma$ is \emph{spare} if $\sigma(x)$ is in normal form w.r.t.\ $\SSS$ for every $x \in \Var$ that occurs more than once in some $r \in \Supp(\mu)$.
    A $\liftto_{\SSS}$-sequence is spare if each of its $\to_\SSS$-steps is spare.
    $\SSS$ is spare if each $\liftto_{\SSS}$-sequence that starts with $\{1 : t\}$
    for a basic term $t$ is spare.
    A term $t \in \TSet{\Sigma}{\VSet}$ is \emph{basic} if $t = f(t_1, \ldots, t_n)$ such that $f \in \Sigma_D$ and $t_i \in \TSet{\Sigma_C}{\VSet}$ for all $1 \leq i \leq n$.
\end{definition}

\begin{example}\label{example:spareness}
    Consider the PTRS $\SSS_7$ with the two rules:

    \vspace*{-.5cm}
    \begin{minipage}[t]{7cm}
        \begin{align*}
            \tg &\to \{\nicefrac{3}{4}:\td(\bot), \nicefrac{1}{4}:\tg\}
        \end{align*}
    \end{minipage}
    \hspace*{.5cm}
  \begin{minipage}[t]{4cm}
        \begin{align*}
            \td(x) &\to \{1:\tc(x,x)\}
        \end{align*}
    \end{minipage}

    \vspace*{.3cm}

    \noindent
    It is similar to the PTRS $\SSS_1$ from \Cref{example:diff-AST-vs-iAST-dup}, but we
    exchanged
    the symbols $\tg$ and $\bot$ in the right-hand side of the $\tg$-rule.
    This PTRS is orthogonal but duplicating due to the $\td$-rule.
    However, in any rewrite sequence that starts with $\{1:t\}$
    for a basic term $t$ we can only duplicate the constructor symbol $\bot$ but no defined symbol.
    Hence, $\SSS_7$ is spare.
\end{example}
  
In general, it is undecidable whether a PTRS is spare, since spareness is already
undecidable for non-probabilistic TRSs.
However, there exist computable sufficient conditions for
spareness, see \cite{frohn_analyzing_nodate}.

If a PTRS is spare, and we start with a basic term, then we will only duplicate normal
forms with our duplicating rules.
This means that the duplicating rules do not influence the (expected) runtime and, more importantly
for AST, the probability of termination. 
As in \cite{frohn_analyzing_nodate}, which analyzed runtime complexity, we have to
restrict ourselves to rewrite sequences that
start with basic terms.
So we only consider start terms
where a single algorithm is applied to data, i.e.,
we may not have any
nested defined symbols in our start terms.
This leads to the following theorem, where ``\emph{on basic terms}'' means that one only considers rewrite sequences that start with
$\{1:t\}$ for a basic term $t$. It can be proved by an analogous limit construction as in the proof of \Cref{properties-eq-AST-iAST-1}.

\setcounter{iASTtofASTThree}{\value{theorem}}

\begin{restatable}[From iAST/iPAST to fAST/fPAST (3)]{theorem}{ASTAndIASTPropertyThree}\label{properties-eq-AST-iAST-3}
    If a PTRS $\SSS$ is orthogonal and spare, then

    \vspace*{-.5cm}
    
    \begin{align*}
        \SSS \text{ is fAST on basic terms} &\Longleftrightarrow \SSS \text{ is iAST on basic terms}\\
        \SSS \text{ is fPAST on basic terms} &\Longleftrightarrow \SSS \text{ is iPAST on basic terms}\!
    \end{align*}
\end{restatable}

The requirement of
basic start terms is a real restriction for fAST, i.e., there exist PTRSs that are fAST on basic terms, but not fAST in general.

\begin{counterexample}\label{example:basic-terms-problem}
    Consider the PTRS $\SSS_8$ with the two rules:

    \vspace*{-.5cm}
    \begin{minipage}[t]{5cm}
        \begin{align*}
            \tg &\to \{\nicefrac{3}{4}:\ts(\tg), \nicefrac{1}{4}:\bot\}
        \end{align*}
    \end{minipage}
    \hspace*{.5cm}
  \begin{minipage}[t]{4cm}
        \begin{align*}
            \tf(\ts(x)) &\to \{1:\tc(\tf(x),\tf(x))\}
        \end{align*}
    \end{minipage}

    \vspace*{.3cm}

    \noindent
    This PTRS behaves similarly to $\SSS_1$ (see \Cref{example:diff-AST-vs-iAST-dup}).
    It is not fAST (and thus, also not fPAST), as we have
    $\{1:\tf(\tg)\} \liftto_{\SSS_8}^2 \{\nicefrac{3}{4}:\tc(\tf(\tg),\tf(\tg)), \nicefrac{1}{4}:\tf(\bot)\}$,
    which corresponds to a random walk biased towards non-termination (since $\tfrac{3}{4} > \tfrac{1}{2}$).

    However, the only basic terms for this PTRS are $\tg$ and $\tf(t)$ for terms $t$ that
    do not contain $\tg$ or $\tf$.
    A sequence starting with $\tg$ corresponds to flipping a biased coin and a sequence
    starting with $\tf(t)$ will clearly terminate. 
    Hence, $\SSS_8$ is fAST (and even fPAST) on basic terms.
    Furthermore, note that $\SSS_8$ is iAST analogous to $\SSS_1$, but not iPAST.
    This shows that \Cref{properties-eq-AST-iAST-3} cannot be extended to fAST in general. For fPAST, consider $\SSS_8'$ with the rules $\tf(\tf(\ta)) \to \{\nicefrac{3}{4}:\td(\tf(\tf(\ta))), \nicefrac{1}{4}:\bot\}$ and $\td(x) \to \{1:\tc(x,x)\}$, which is orthogonal, spare, and iPAST analogous to $\SSS_1$, but not fAST.
\end{counterexample}

One may wonder whether \Cref{properties-eq-AST-iAST-3} can nevertheless be used in order
to prove fAST of a PTRS $\SSS$ on all terms by using a suitable transformation from
$\SSS$ to another PTRS $\SSS'$ such that $\SSS$ is fAST on all terms iff $\SSS'$ is fAST
on basic terms.

There is an analogous difference in the complexity analysis of non-probabilistic
term rewrite systems.
There, the concept of \emph{runtime complexity}
is restricted to rewrite sequences that start with a basic term, whereas the concept of \emph{derivational complexity} allows arbitrary start terms.
In \cite{fuhs2019transformingdctorc}, a transformation was presented
that extends any (non-probabilistic) TRS $\R$ by so-called generator rules $\C{G}(\R)$ such that the
derivational complexity of $\R$ 
is the same as the runtime complexity of $\R \cup \C{G}(\R)$,
where $\C{G}(\R)$ are considered  to be \emph{relative} rules whose rewrite steps do not ``count'' for the complexity.
This transformation can indeed be reused to move from fAST \emph{on basic terms} to fAST
in general.

\begin{restatable}[]{lemma}{BasicASTToAST}\label{lemma:spareness-AST-proof-1}
    A PTRS $\SSS$ is fAST iff $\SSS \cup \C{G}(\SSS)$ is fAST on basic terms.
\end{restatable}

For every defined symbol $f$, the idea of the transformation is to introduce
a new constructor symbol $\tcons_f$ and for every function symbol $f$ it introduces a new defined
symbol $\tenc_{f}$.
As an example for $\SSS_8$ from \Cref{example:basic-terms-problem},
then instead of starting with the non-basic term $\tc(\tg,\tf(\tg))$,
we start with the basic term $\tenc_{\tc}(\tcons_{\tg},\tcons_{\tf}(\tcons_{\tg}))$, its
so-called \emph{basic variant}.
The new defined symbol $\tenc_{\tc}$ is used to first build the term $\tc(\tg,\tf(\tg))$
at the beginning of the rewrite sequence, i.e., it converts
all occurrences of $\tcons_{f}$ for $f \in \Sigma_D$ back into the defined symbol $f$, and then we can proceed as if we started with the term $\tc(\tg,\tf(\tg))$ directly.
For this conversion, we need another new defined symbol $\targenc$ that iterates through the term and
replaces all  new constructors $\tcons_{f}$ by the original defined symbol $f$. 
Thus, we define the generator rules as in
\cite{fuhs2019transformingdctorc} (just with trivial probabilities in the right-hand sides $\ell \to \{1:r\}$), since we do not need any probabilities during this initial construction of the original start term.

\begin{definition}[Generator Rules $\C{G}(\SSS)$]\label{def:generator-rules}
    Let $\SSS$ be a PTRS over the signature $\Sigma$. 
    Its \emph{generator rules}
    $\C{G}(\SSS)$ are the following set of rules
    {\small
    \allowdisplaybreaks
    \begin{align*}
       & \{\tenc_f(x_1, \ldots, x_n) \to \{1: f(\targenc(x_1), \ldots, \targenc(x_n))\} \mid f \in \Sigma\}\\
         \cup  \;&\{\targenc(\tcons_f(x_1, \ldots, x_n)) \to \{1: f(\targenc(x_1), \ldots, \targenc(x_n))\} \mid f \in \Sigma_D\}\\
         \cup \; & \{\targenc(f(x_1, \ldots, x_n)) \to \{1: f(\targenc(x_1), \ldots, \targenc(x_n))\} \mid f \in \Sigma_C\},
    \end{align*}}
    
    \noindent
    where $x_1, \ldots, x_n$ are pairwise different variables and where the function symbols $\targenc$, $\tcons_f$, and $\tenc_f$ are fresh (i.e., they do not occur in $\SSS$). 
    Moreover, we define $\Sigma_{\C{G}(\SSS)} = \{\tenc_f \mid f \in \Sigma\} \cup \{\targenc\} \cup
    \{\tcons_f \mid f \in \Sigma_D\}$. 
\end{definition}

\begin{example}
   For the PTRS $\SSS_8$ from \Cref{example:basic-terms-problem},
   we obtain the following generator rules $\C{G}(\SSS_8)$:

   \vspace*{-.2cm}

   \begin{longtable}{rcl}
        $\tenc_{\tg}$ & $\to$ &$\{1: \tg\}$\\
        $\tenc_{\tf}(x_1)$ & $\to$ &$\{1: \tf(\targenc(x_1))\}$\\
        $\tenc_{\tc}(x_1, x_2)$  &$\to$& $\{1: \tc(\targenc(x_1), \targenc(x_2))\}$\\
        $\tenc_{\ts}(x_1)$ & $\to$ &$\{1: \ts(\targenc(x_1))\}$\\
        $\tenc_{\bot}$ & $\to$& $\{1: \bot\}$\\
        $\targenc(\tcons_{\tg})$ & $\to$ &$\{1: \tg\}$\\
        $\targenc(\tcons_{\tf}(x_1))$ &$\to$ &$\{1: \tf(\targenc(x_1))\}$\\
        $\targenc(\tc(x_1, x_2))$ & $\to$ &$\{1: \tc(\targenc(x_1), \targenc(x_2))\}$\\
        $\targenc(\ts(x_1))$ & $\to$ &$\{1: \ts(\targenc(x_1))\}$\\
        $\targenc(\bot)$ & $\to$ &$\{1: \bot\}$
    \end{longtable}
 \end{example}

As mentioned, using the  symbols $\tcons_f$ and $\tenc_f$, as in
\cite{fuhs2019transformingdctorc} every term over 
$\Sigma$ can be transformed into a basic term over $\Sigma \cup \Sigma_{\C{G}(\SSS)}$.

However,  even if $\SSS$ is spare, the PTRS
$\SSS \cup \C{G}(\SSS)$ is not guaranteed to be spare, although the generator rules
themselves are right-linear. 
The problem is that the generator rules include  a rule like
$\tenc_{\tf}(x_1) \to \{1: \tf(\targenc(x_1))\}$ where
a defined symbol $\targenc$ occurs below the duplicating symbol $\tf$ on the right-hand
side.
Indeed, while $\SSS_8$ is spare,
$\SSS_8 \cup \C{G}(\SSS_8)$ is not. For example, when starting with the basic term
$\tenc_{\tf}(\ts(\tcons_{\tg}))$, we have
\[
\begin{array}{r@{\;\;}l@{\;\;}l}
\{1:\tenc_{\tf}(\ts(\tcons_{\tg}))\} &\liftto_{\C{G}(\SSS_8)}^2&
\{1:\tf(\ts(\targenc(\tcons_{\tg})))\}\\
&\liftto_{\SSS_8}&
\{1:\tc(\tf(\targenc(\tcons_{\tg})), \tf(\targenc(\tcons_{\tg})))\},
\end{array}
\]
where the last step is not spare.
In general, $\SSS \cup \C{G}(\SSS)$ is guaranteed to be spare if $\SSS$ is right-linear.
So we could modify \Cref{properties-eq-AST-iAST-3} into a theorem which states that $\SSS$
is fAST on all terms iff $\SSS \cup 
\C{G}(\SSS)$ is iAST on basic terms (which is equivalent to iAST of $\SSS$ on all terms, since every term can be generated from a basic term via $\C{G}(\SSS)$) for orthogonal and right-linear PTRSs
$\SSS$. However, this theorem would be subsumed by \Cref{properties-eq-AST-iAST-1}, where we 
already showed the equivalence of fAST and iAST if $\SSS$ is orthogonal and right-linear. Indeed,
our goal in \Cref{properties-eq-AST-iAST-3} was to find a weaker requirement than right-linearity.
Hence, such a transformational approach to move from fAST on all start terms to fAST
on basic terms does not seem viable for \Cref{properties-eq-AST-iAST-3}.

Finally, we can also combine our results on simultaneous rewriting and spareness to relax both left- and right-linearity in case of basic start terms.
The proof for the following theorem combines the proofs for \Cref{properties-eq-AST-iAST-2} and \Cref{properties-eq-AST-iAST-3}.

\setcounter{iASTtofASTFour}{\value{theorem}}

\begin{restatable}[From iAST/iPAST to fAST/fPAST (4)]{theorem}{ASTAndIASTPropertyFour}\label{properties-eq-AST-iAST-4}
  If $\SSS$ is non-overlapping and spare, then
  
  \vspace*{-.5cm}
  
    \begin{align*}
        \SSS \text{ is fAST on basic terms} &\Longleftarrow \SSS \text{ is iAST w.r.t.\ } \paraarrow_{\SSS} \text{ on basic terms}\\
        \SSS \text{ is fPAST on basic terms} &\Longleftarrow \SSS \text{ is iPAST w.r.t.\ } \paraarrow_{\SSS} \text{ on basic terms}\!
    \end{align*}
\end{restatable}

%% file: evaluation.tex
\section{Conclusion and Evaluation}\label{Evaluation}

In this paper, we
presented numerous new results on the relationship between full and restricted forms of
AST, including several criteria for PTRSs such that
innermost AST implies full AST. All of our results also hold for PAST, and  all of our criteria are suitable for automation (for spareness,
there exist sufficient conditions that can be checked  automatically).

We implemented our new criteria in our termination prover \textsf{AProVE}
\cite{JAR-AProVE2017}. For every PTRS, one can indicate
whether one wants to analyze its termination behavior for all start terms or only for basic start terms.
Up to now, \aprove's main technique for termination analysis of PTRSs was the probabilistic DP
framework from \cite{kassinggiesl2023iAST,FLOPS2024} which however can only prove iAST.
If one wants to analyze fAST for
a PTRS $\SSS$, then \textsf{AProVE} now first tries to prove that
the conditions of \Cref{properties-eq-AST-iAST-3} are satisfied if one is restricted to
basic start terms, or that the conditions of \Cref{properties-eq-AST-iAST-1} hold if one
wants to consider arbitrary start terms.
If this succeeds, then we can use the full probabilistic DP framework in order to prove
iAST, which then implies fAST. 
Otherwise, we try to prove all conditions of \Cref{properties-eq-AST-iAST-4} or \Cref{properties-eq-AST-iAST-2}, respectively.
If this succeeds, then we can use most of the processors from the
probabilistic DP framework to prove iAST, which again implies fAST.
If none of these theorems can be applied, then \textsf{AProVE}
tries to prove fAST using a direct application of polynomial orderings \cite{kassinggiesl2023iAST}.
Note that for AST w.r.t.\ basic start terms, \Cref{properties-eq-AST-iAST-3} generalizes \Cref{properties-eq-AST-iAST-1} and \Cref{properties-eq-AST-iAST-4} generalizes \Cref{properties-eq-AST-iAST-2}, since right-linearity implies spareness.

For our evaluation, we compare the \emph{old} \aprove{} without any of the new theorems
(which only uses direct applications of polynomial orderings to prove fAST), to variants of \aprove{}
where we activated each of the theorems individually, and finally to the \emph{new} \aprove{} strategy explained above.
The following diagram shows the theoretical subsumptions of each of these strategies for basic start terms, where an arrow from strategy A to strategy B means that B is  strictly better than A.

\begin{center}
    \begin{tikzpicture}
        \tikzstyle{adam}=[thick,draw=black!100,fill=white!100,minimum size=4mm, shape=rectangle split, rectangle split parts=2,rectangle split horizontal]
        \tikzstyle{empty}=[rectangle,thick,minimum size=4mm]

        \node[empty] at (0, 0)  (a) {\emph{old} \aprove{}};
        
        \node[empty] at (3, 0.5)  (b1) {\Cref{properties-eq-AST-iAST-1}};
        \node[empty] at (3, -0.5)  (b2) {\Cref{properties-eq-AST-iAST-2}};
        
        \node[empty] at (6, 0.5)  (c1) {\Cref{properties-eq-AST-iAST-3}};
        \node[empty] at (6, -0.5)  (c2) {\Cref{properties-eq-AST-iAST-4}};
    
        \node[empty] at (9, 0)  (d) {\emph{new} \aprove{}};

        \draw (a) edge[->] (b1);
        \draw (a) edge[->] (b2);
        \draw (b1) edge[->] (c1);
        \draw (b2) edge[->] (c2);
        \draw (c1) edge[->] (d);
        \draw (c2) edge[->] (d);
    \end{tikzpicture}
\end{center}

We used the benchmark set of 100 PTRSs from
\cite{FLOPS2024}, and extended it by 15 new PTRSs that contain all the examples
presented in this paper and some additional examples which illustrate the power of each strategy.
\aprove{} can prove iAST for 93 of these 118 PTRSs.
The following table shows for how many of these 93 PTRSs the respective strategy 
allows us to conclude fAST for basic start terms from \aprove{}'s proof of iAST.

\smallskip

\begin{center}
    \begin{tabular}{||c | c | c | c | c | c||}
     \hline
     \emph{old} \aprove{} & \Cref{properties-eq-AST-iAST-1} & \Cref{properties-eq-AST-iAST-2} & \Cref{properties-eq-AST-iAST-3} & \Cref{properties-eq-AST-iAST-4} & \emph{new} \aprove{} \\ [0.5ex] 
     \hline
     36 & 48 & 44 & 58 & 56 & 61 \\ 
     \hline
    \end{tabular}
\end{center}

\smallskip

From the 61 examples that we can solve by using both \Cref{properties-eq-AST-iAST-3} and
\Cref{properties-eq-AST-iAST-4} in ``\emph{new} \textsf{AProVE}'', 5 examples  (that are
all right-linear) can only be solved by \Cref{properties-eq-AST-iAST-3}, 3 examples (where one is right-linear and the others only spare) can only be solved by \Cref{properties-eq-AST-iAST-4}, and 53 can be solved by both.
If one considers arbitrary start terms, then the \emph{new} \aprove{} can conclude fAST
(using only \Cref{properties-eq-AST-iAST-1} and \Cref{properties-eq-AST-iAST-2}) for 49 examples.

Currently, we only use the switch from full to innermost rewriting as a preprocessing step
before applying the DP framework.
As future work, we want to develop a processor within the DP framework that can perform
this switch in a modular way. 
Then, the criteria of our theorems do not have to be required for the whole
PTRS anymore, but just for specific sub-problems
within the termination proof. 
This, however, requires developing a DP framework for fAST directly, which we will
investigate in future work.

For details on our experiments, our collection of examples, and for instructions on how to run our implementation
in \textsf{AProVE} via its \emph{web interface} or locally, we refer to:
\begin{center}
  \url{https://aprove-developers.github.io/InnermostToFullAST/}
\end{center}

\noindent
In addition, an artifact is available at \cite{aprove_artifact}.

\medskip

\noindent
\textbf{Acknowledgements.} We thank Stefan Dollase for pointing us to \cite{fuhs2019transformingdctorc}.

%% file: appendix.tex
\vspace*{-.4cm}

\section*{Appendix}

\vspace*{-.1cm}

\Cref{appendix} contains all proofs for our new contributions and observations regarding AST.
Afterwards, in \Cref{PASTandSAST}, we introduce the notion of \emph{strong} almost-sure termination (SAST) and
prove our results for PAST and SAST as well.

\section{Proofs}\label{appendix}

In this section, we prove all our new contributions and observations concerning AST.
Most of our proofs do not handle rewrite sequences with $\liftto$, but instead we work with rewrite sequence trees that were first described in \cite{reportkg2023iAST}.
We already gave an example for such a tree in \Cref{example:diff-AST-vs-iAST-dup}.
We start by recapitulating this notion.

\begin{definition}[Rewrite Sequence Tree (RST)] \label{def:rewrite-sequence-tree}
    Let $\SSS$ be a PTRS\@.
    $\F{T}\!=\!(V,E,L)$ {\normalsize is an} $\SSS${\normalsize\emph{-rewrite sequence tree (RST)}} if

    \begin{enumerate}
    \item[(1)] $V \neq \emptyset$ is a possibly infinite set of nodes and $E \subseteq V \times V$ is a set of directed edges, such that $(V, E)$ is a (possibly infinite) directed tree where $vE = \{ w \mid (v,w) \in E \}$ is finite for every $v \in V$.
    \item[(2)]  $L : V \rightarrow (0,1] \times \TSet{\Sigma}{\VSet}$ labels every node $v$ by a probability $p_v$ and a term $t_v$.
    For the root $v \in V$ of the tree, we have $p_v = 1$.
    \item[(3)] For all $v \in V$: If $vE = \{w_1, \ldots, w_k\}$, then $t_v \to_{\SSS} \{\tfrac{p_{w_1}}{p_v}:t_{w_1}, \ldots, \tfrac{p_{w_k}}{p_v}:t_{w_k}\}$.
    \end{enumerate}
    If we restrict condition (3) to innermost / leftmost-innermost steps, then $\F{T}$ is an innermost / leftmost-innermost $\SSS$-rewrite sequence tree, respectively.
\end{definition}

When it is not clear about which RST we are talking, we will always explicitly indicate the tree. 
For instance, for the probability $p_v$ of the node $v \in V$ of some RST $\F{T} =
(V,E,L)$, we may also write $p_v^{\F{T}}$. 
$\ctleaf^{\F{T}}$ denotes the set of leaves of the 
RST $\F{T}$ and for a node $x$ in $\F{T}$, $d^{\F{T}}(x)$ 
denotes the depth of node $x$ in $\F{T}$.

\begin{definition}[$|\F{T}|_{\ctleaf}$, Convergence Probability] \label{def:rewrite-sequence-tree-convergence-notation}
    Let $\SSS$ be a PTRS.
    For any $\SSS$-RST $\F{T}$ we define $|\F{T}|_{\ctleaf} = \sum_{v \in \ctleaf} p_v$.
    We say that an RST $\F{T}$ \emph{converges with
    probability} $p \in \IR$ if we have $|\F{T}|_{\ctleaf} = p$.
\end{definition}

It is now easy to observe that a PTRS $\SSS$ is AST (i.e., for all $\liftto_{\SSS}$-rewrite
sequences $(\mu_n)_{n \in \IN}$ we have $\lim_{n \to \infty} |\mu_n|_\SSS = 1$) iff for all
$\SSS$-RSTs $\F{T}$ we have $|\F{T}|_{\ctleaf} = 1$. 
To see this, note that every infinite $\liftto_{\SSS}$-rewrite sequence $(\mu_n)_{n
\in \IN}$ that begins
with a single start term (i.e., $\mu_0 = \{1:t\}$) can be represented by an infinite
$\SSS$-RST $\F{T}$ that is fully evaluated (i.e., for every leaf
$v$, $t_v$ is an  $\SSS$-normal form) such that $\lim_{n \to \infty} |\mu_n|_\SSS =
|\F{T}|_{\ctleaf}$ and vice versa. So $\SSS$ is AST iff all fully evaluated 
$\SSS$-RSTs with a single start term
converge with probability 1.

\begin{example}\label{example:random-walk-rst}
    Let $\SSS_{\trw 2}$ be the PTRS with the rule $\tg(x) \to \{\nicefrac{1}{2}:\tg^2(x), \nicefrac{1}{2}:x\}$.
    Here, we write $\tg^2(x)$ to abbreviate $\tg(\tg(x))$, etc.
    Consider the infinite rewrite sequence:
    \[\begin{array}{ll}
    &\{1:\tg(\O)\}\\
    \liftto_{\SSS_{\trw 2}}&\{\nicefrac{1}{2}:\tg^2(\O),\nicefrac{1}{2}:\O\}\\
    \liftto_{\SSS_{\trw 2}}&\{\nicefrac{1}{4}:\tg^3(\O), \nicefrac{1}{4}:\tg(\O), \nicefrac{1}{2}:\O\}\\
    \liftto_{\SSS_{\trw 2}}&\{\nicefrac{1}{8}:\tg^4(\O),\nicefrac{1}{8}:\tg^2(\O), \nicefrac{1}{8}:\tg^2(\O), \nicefrac{1}{8}:\O, \nicefrac{1}{2}:\O\}\\
    \liftto_{\SSS_{\trw 2}}& \ldots \!
    \end{array}\]
    This sequence can be represented by the following $\SSS_{\trw 2}$-RST:
    \begin{center}
        \small
        \begin{tikzpicture}
            \tikzstyle{adam}=[thick,draw=black!100,fill=white!100,minimum size=4mm, shape=rectangle split, rectangle split parts=2,rectangle split horizontal]
            \tikzstyle{empty}=[rectangle,thick,minimum size=4mm]
        
            \node[adam] at (-4, 0)  (a) {$1$\nodepart{two}$\tg(\O)$};
            \node[adam] at (-6, -1)  (b) {$\nicefrac{1}{2}$\nodepart{two}$\tg^2(\O)$};
            \node[adam] at (-2, -1)  (c) {$\nicefrac{1}{2}$\nodepart{two}$\O$};
            \node[adam] at (-8, -2)  (d) {$\nicefrac{1}{4}$\nodepart{two}$\tg^3(\O)$};
            \node[adam] at (-4, -2)  (e) {$\nicefrac{1}{4}$\nodepart{two}$\tg(\O)$};
            \node[adam] at (-9, -3)  (f) {$\nicefrac{1}{8}$\nodepart{two}$\tg^4(\O)$};
            \node[adam] at (-7, -3)  (g) {$\nicefrac{1}{8}$\nodepart{two}$\tg^2(\O)$};
            \node[adam] at (-5, -3)  (h) {$\nicefrac{1}{8}$\nodepart{two}$\tg^2(\O)$};
            \node[adam] at (-3, -3)  (i) {$\nicefrac{1}{8}$\nodepart{two}$\O$};
            \node[empty] at (-9, -4)  (j) {$\ldots$};
            \node[empty] at (-7, -4)  (k) {$\ldots$};
            \node[empty] at (-5, -4)  (l) {$\ldots$};
            
            \draw (a) edge[->] (b);
            \draw (a) edge[->] (c);
            \draw (b) edge[->] (d);
            \draw (b) edge[->] (e);
            \draw (d) edge[->] (f);
            \draw (d) edge[->] (g);
            \draw (e) edge[->] (h);
            \draw (e) edge[->] (i);
            \draw (f) edge[->] (j);
            \draw (g) edge[->] (k);
            \draw (h) edge[->] (l);
        \end{tikzpicture}
    \end{center}
\end{example}

Furthermore, for every $\SSS$-RST $\F{T}$ that is not fully evaluated, there exists an $\SSS$-RST $\F{T}'$ that is fully evaluated such that $|\F{T}|_{\ctleaf} \geq |\F{T}'|_{\ctleaf}$.
To get from $\F{T}$ to $\F{T}'$ we can simply perform arbitrary (possibly infinitely many)
rewrite steps at the leaves that are not in normal form to fully evaluate the tree.
So $\SSS$ is AST iff all
$\SSS$-RSTs with a single start term converge with probability 1.

It remains to prove that it suffices to only regard $\liftto$-rewrite sequences that start
with a single start term.

\begin{lemma}[Single Start Terms Suffice for AST] \label{lemma:PTRS-AST-single-start-term}
    Suppose that there exists an infinite $\liftto_{\SSS}$-rewrite sequence 
    $(\mu_n)_{n \in \mathbb{N}}$ that converges with probability $<1$.
    Then there is also an infinite $\liftto_{\SSS}$-rewrite sequence 
    $(\mu'_n)_{n \in \mathbb{N}}$ with a single start term, 
    i.e., $\mu'_0 = \{1:t\}$, that converges with probability $<1$.
\end{lemma}

\begin{myproof}
    Let $(\mu_n)_{n \in \mathbb{N}}$ be an $\liftto_{\SSS}$-rewrite sequence that converges with probability $<1$.
    Suppose that we have $\mu_0 = \{p_1:t_1, \ldots, p_k:t_k\}$.
    Let $(\mu^j_n)_{n \in \mathbb{N}}$ with $\mu^j_0 =
    \{1 : t_j\}$ denote the infinite $\liftto_{\SSS}$-rewrite sequence that uses
    the same rules
    as $(\mu_n)_{n \in \mathbb{N}}$ does on the term $t_j$ for every $1 \leq j \leq k$.
    Assume for a contradiction that for every $1 \leq j \leq k$ the
    $\liftto_{\SSS}$-rewrite sequence  $(\mu^j_n)_{n \in \mathbb{N}}$
    converges with probability $1$.
    Then we would have
    \[
    \begin{array}{rl}
    &\lim_{n \to \infty}|\mu_{n}|_{\SSS}\\
        =&\lim_{n \to \infty} \sum_{(p:t) \in \mu_{n}, t \in \mathtt{NF}_{\SSS}} p\\
        =&\lim_{n \to \infty} \sum_{1 \leq j \leq k} p_j \cdot \sum_{(p:t) \in \mu^j_n, t \in \mathtt{NF}_{\SSS}} p\\
        =&\sum_{1 \leq j \leq k} p_j \cdot \lim_{n \to \infty} \sum_{(p:t) \in \mu^j_n, t \in \mathtt{NF}_{\SSS}} p\\
        =&\sum_{1 \leq j \leq k} p_j \cdot \lim_{n \to \infty}|\mu^j_n|_{\SSS}\\
        =&\sum_{1 \leq j \leq k} p_j \cdot 1 \\
        =&\sum_{1 \leq j \leq k} p_j \\
        =&1
        \end{array}
    \]
    which is a contradiction to our assumption that we have $\lim_{n \to \infty}|\mu_{n}|_{\SSS} < 1$.
    Therefore, we have at least one $1 \leq j \leq k$ such that $(\mu^j_n)_{n \in
        \mathbb{N}}$
        converges with probability $<1$.
\end{myproof}

Obviously, all of these observations also hold for iAST and for liAST.
We now obtain the following corollary.

\begin{corollary}[Characterizing AST with RSTs]\label{cor:Characterizing AST with RSTs}
    Let $\SSS$ be a PTRS. Then $\SSS$ is fAST (iAST/liAST) iff for all (innermost/leftmost-innermost) $\SSS$-RSTs
    $\F{T}$ we have $|\F{T}|_{\ctleaf} = 1$.
\end{corollary}

Next, we prove all new results from \cref{Relating AST and its Restricted Forms}.
For this, we first recapitulate the notion of a \emph{sub-rewrite sequence tree} (sub-RST) from \cite{reportkg2023iAST}.

\begin{definition}[Sub-RST] \label{def:chain-tree-induced-sub}
	Let $\SSS$ be a PTRS and $\F{T} = (V,E,L)$ be an $\SSS$-RST.
	Let $W \subseteq V$ be non-empty, weakly connected, and for all $x \in W$ we have $xE \cap W = \emptyset$ or $xE \cap W = xE$.
	Then, we define the \emph{sub-rewrite sequence tree} (or \emph{sub-RST}) $\F{T}[W]$ by $\F{T}[W] = (W,E \cap (W \times W),L^W)$.
	Let $w \in W$ be the root of $\F{T}[W]$.
	To ensure that the root of our sub-RST has the probability $1$ again,
	we use the labeling $L^W(x) = (\frac{p_{x}^{\F{T}}}{p_w^{\F{T}}}: t_{x}^{\F{T}})$ for all nodes $x \in W$.
\end{definition}

The property of being non-empty and weakly connected ensures that the resulting 
graph $G^{\F{T}[W]} = (W,E \cap (W \times W))$ is a tree again.
The property that we either have $xE \cap W = \emptyset$ or $xE \cap W = xE$ ensures that the sum of 
probabilities for the successors of a node $x$ is equal to the probability for the node $x$ itself.

\setcounter{auxctr}{\value{theorem}}
\setcounter{theorem}{\value{iASTtofASTOne}}

\begin{theorem}[From iAST to fAST (1)]
    If a PTRS $\SSS$ is orthogonal and right-linear (i.e., non-overlapping and linear), then:
    \begin{align*}
        \SSS \text{ is fAST} &\Longleftrightarrow \SSS \text{ is iAST}\!
    \end{align*}
\end{theorem}

\setcounter{theorem}{\value{auxctr}}

\begin{myproof}
    We only have to prove the non-trivial direction ``$\Longleftarrow$''.
    Let $\SSS$ be a PTRS that is non-overlapping and linear.
    Furthermore, assume that $\SSS$ is not fAST.
    This means that there exists an RST $\F{T}$ such that $|\F{T}|_{\ctleaf} = c$ for some $0 \leq c < 1$.
    We create a new \emph{innermost} RST $\F{T}^{(\infty)}$ such that $|\F{T}^{(\infty)}|_{\ctleaf} \leq |\F{T}|_{\ctleaf} = c < 1$, which shows that $\SSS$ is not iAST either.
    W.l.o.G.,  at least one rewrite step in $\F{T}$ is performed at some node $x$ with a
    redex that is not an innermost redex (otherwise we can use $\F{T}^{(\infty)}
    = \F{T}$). 
    The core steps of the proof are the following:
    \begin{enumerate}
        \item[1.] We iteratively move innermost rewrite steps to a higher position in the tree using a construction $\Phi(\circ)$. 
        The limit of this iteration, namely $\F{T}^{(\infty)}$, is an \emph{innermost} $\SSS$-RST that converges with probability at most $c < 1$.
        \begin{itemize}
            \item[1.1.] We formally define the construction $\Phi(\circ)$ that replaces a certain subtree $\F{T}_x$ by a new subtree $\Phi(\F{T}_x)$, by moving an innermost rewrite step to the root.
            \item[1.2.] We show $|\Phi(\F{T}_x)|_{\ctleaf} \leq |\F{T}_x|_{\ctleaf}$.
            \item[1.3.] We show that $\Phi(\F{T}_x)$ is indeed a valid RST\@.
        \end{itemize}
    \end{enumerate}

    \smallskip
          
    \noindent 
    \textbf{\underline{1. We iteratively move innermost rewrite steps to a higher position.}}

    \noindent 
    In $\F{T}$ there exists at least one rewrite step performed at some node $x$, which is not an innermost rewrite step.
    Furthermore, we can assume that this is the first such rewrite step in the path from the root to the node $x$ and that $x$ is a node of minimum depth with this property.
    Let $\F{T}_x$ be the sub-RST that starts at node $x$, i.e., $\F{T}_x = \F{T}[xE^*]$.
    We then construct a new tree $\Phi(\F{T}_x)$ such that $|\Phi(\F{T}_x)|_{\ctleaf} = |\F{T}_x|_{\ctleaf}$,
    and where we use an innermost rewrite step at the root
    node $x$ instead of the old one (i.e., we pushed the first non-innermost rewrite step
    deeper into the tree).
    This construction only works because $\SSS$ is non-overlapping and
    linear.\footnote{For the construction,
    non-overlappingness is essential (while
    a related construction could also be defined without linearity).
    However, linearity is needed to ensure that the probability of termination in the new tree
    is not larger than in the original one.}
    Then, by replacing the subtree $\F{T}_x$ with the new tree $\Phi(\F{T}_x)$ in $\F{T}$, we get
    an $\SSS$-RST $\F{T}^{(1)}$, with $|\F{T}^{(1)}|_{\ctleaf} = |\F{T}|_{\ctleaf}$,
    and where we use an innermost rewrite step at node $x$ instead of the old rewrite step, as desired.
    We can then do such a replacement iteratively for every use of a non-innermost rewrite step, i.e., we again replace the first non-innermost rewrite step in $\F{T}^{(1)}$ to get $\F{T}^{(2)}$ with $|\F{T}^{(2)}|_{\ctleaf} = |\F{T}^{(1)}|_{\ctleaf}$, and so on.
    In the end, the limit of all these RSTs $\lim_{i \to \infty} \F{T}^{(i)}$ is an innermost $\SSS$-RST, 
    that we denote by $\F{T}^{(\infty)}$ and that converges with probability at most $c <
    1$. 
    So while the termination probability remains the same in each step, it can
    decrease in the limit.\footnote{As an example, consider a tree $\F{T}$ which is just
    a finite path and its path length increases in each iteration by one. Then the limit $\F{T}^{(\infty)}$ is an infinite path and converges with probability $0$, while all the $\F{T}, \F{T}^{(1)}, \ldots$ converge with probability $1$.}       
    Hence, $\SSS$ is not iAST either.

    To see that $\F{T}^{(\infty)}$ is indeed a valid innermost $\SSS$-RST, note that in
    every iteration of the construction we turn a non-innermost rewrite step at minimum depth
    into an innermost one.
    Hence, for every depth $H$ of the tree, we eventually turned every non-innermost rewrite step up to
    depth $H$ into an innermost one so that the construction will not change the tree above depth $H$ anymore, i.e.,
    there exists an $m_H$ such that $\F{T}^{(\infty)}$ and $\F{T}^{(i)}$ are the same trees up to depth $H$ for all $i \geq m_H$.
    This means that the sequence $\lim_{i \to \infty} \F{T}^{(i)}$ really converges into an innermost $\SSS$-RST.

    Next, we want to prove that we have $|\F{T}^{(\infty)}|_{\ctleaf} \leq c$.
    By induction on $n$ one can prove that $|\F{T}^{(i)}|_{\ctleaf} = c$ for all $1 \leq i \leq
    n$, since we have $|\F{T}^{(i)}|_{\ctleaf} = |\F{T}^{(i-1)}|_{\ctleaf}$ for all $i \geq 2$ and $|\F{T}^{(1)}|_{\ctleaf} = |\F{T}|_{\ctleaf} = c$.
    Assume for a contradiction that $\F{T}^{(\infty)}$ converges with probability greater
    than $c$, i.e., that $|\F{T}^{(\infty)}|_{\ctleaf} > c$.
    Then there exists a depth $H \in \IN$ such that
    $\sum_{x \in \ctleaf^{\F{T}^{(\infty)}}\!, \, d^{\F{T}^{(\infty)}}\!(x) \leq H} p_x > c$.
    Again, let $m_{H+1} \in \IN$ such that $\F{T}^{(\infty)}$ and $\F{T}^{(m_{H+1})}$ are the same trees up to depth $H+1$, so that they have the same leaves up to depth $H$.
    But this would mean that
    $|\F{T}^{(m_{H+1})}|_{\ctleaf} \geq \sum_{x \in \ctleaf^{\F{T}^{(m_{H+1})}}\!, \,
    d^{\F{T}^{(m_{H+1})}}\!(x) \leq H} p_x = \sum_{x \in \ctleaf^{\F{T}^{(\infty)}}\!,\,
    d^{\F{T}^{(\infty)}}\!(x) \leq H} p_x > c$, which is a contradiction to
    $|\F{T}^{(m_{H+1})}|_{\ctleaf} = c$.
  
    \smallskip
          
    \noindent 
    \textbf{\underline{1.1 Construction of $\Phi(\circ)$}}

    \noindent 
    It remains to define
    the mentioned construction $\Phi(\circ)$.
    Let $\F{T}_x$ be an $\SSS$-RST that performs a non-innermost rewrite step at the root node
    $x$. This step has the form $t_x^{\F{T}_x} \to_{\SSS} \{p_{y_1}^{\F{T}_x}:t_{y_1}^{\F{T}_x}, \ldots,
    p_{y_k}^{\F{T}_x}:t_{y_k}^{\F{T}_x}\}$ using the rule
    $\bar{\ell} \to \{ \bar{p}_1:\bar{r}_1, \ldots, \bar{p}_k:\bar{r}_k\}$, the
    substitution $\bar{\sigma}$, and the position $\bar{\pi}$ such that $t_x^{\F{T}_x}|_{\bar{\pi}} = \bar{\ell} \bar{\sigma}$.
    Then we have $t_{y_j}^{\F{T}_x} = t_x^{\F{T}_x}[\bar{r}_j \bar{\sigma}]_{\bar{\pi}}$ for all $1 \leq j \leq k$.
    Instead of applying a non-innermost rewrite step at the root $x$
    we want to directly apply an innermost rewrite step.
    Let $\tau$ be the position of some innermost redex in $t_x^{\F{T}_x}$ below $\bar{\pi}$.

    \begin{wrapfigure}[6]{r}{0.37\textwidth}
        \vspace*{-1.3cm}
        \begin{center}
            \begin{tikzpicture}[scale=0.5]
                \begin{pgfonlayer}{nodelayer}
                    \node [style=target,pin={[pin distance=0.05cm, pin edge={,-}] 140:\tiny \textcolor{blue}{$x$}}] (3) at (0, 3) {};
                    \node [style=none] (6) at (1.5, 0) {};
                    \node [style=none] (7) at (-1.5, 0) {};
                    \node [style=none] (9) at (-2, -1) {};
                    \node [style=none] (10) at (2, -1) {};
                    \node [style=moveBlock] (12) at (-0.75, 0.75) {};
                    \node [style=moveBlock] (13) at (0, 0) {};
                    \node [style=moveBlock] (14) at (0.75, 0.25) {};
                    \node [style=moveBlock] (15) at (-0.75, 0.75) {};
                    \node [style=none] (16) at (-0.5, 1.5) {};
                    \node [style=none] (17) at (0, 1.25) {};
                    \node [style=none] (18) at (0.5, 1.5) {};
                    \node [style=none] (19) at (-0.75, 0) {};
                    \node [style=none] (20) at (0, -0.75) {};
                    \node [style=none] (21) at (0.75, -0.5) {};
                \end{pgfonlayer}
                \begin{pgfonlayer}{edgelayer}
                    \draw (3) to (6.center);
                    \draw (3) to (7.center);
                    \draw [style=dotWithoutHead] (7.center) to (9.center);
                    \draw [style=dotWithoutHead] (6.center) to (10.center);
                    \draw [style=dotWithoutHead, in=15, out=-105, looseness=0.50] (3) to (16.center);
                    \draw [style=dotWithoutHead, in=120, out=-90, looseness=0.75] (3) to (17.center);
                    \draw [style=dotWithoutHead, in=135, out=-75] (3) to (18.center);
                    \draw [style=dotHead, in=90, out=-30, looseness=0.75] (18.center) to (14);
                    \draw [style=dotHead, in=90, out=-150, looseness=0.75] (16.center) to (15);
                    \draw [style=dotHead, in=90, out=-45] (17.center) to (13);
                    \draw [style=dashHead, bend right=75, looseness=2.00] (14) to (3);
                    \draw [style=dashHead, bend left=75, looseness=1.75] (15) to (3);
                    \draw [style=dashHead, bend right=105, looseness=2.75] (13) to (3);
                    \draw [style=dotWithoutHead] (15) to (19.center);
                    \draw [style=dotWithoutHead] (13) to (20.center);
                    \draw [style=dotWithoutHead] (14) to (21.center);
                \end{pgfonlayer}
            \end{tikzpicture}
        \end{center}
    \end{wrapfigure}
    The construction creates a new $\SSS$-RST $\Phi(\F{T}_x) = (V',E',L')$ whose root is
    labeled with $(1:t_x^{\F{T}_x})$ such that $|\Phi(\F{T}_x)|_{\ctleaf} =
    |\F{T}_x|_{\ctleaf}$, and that directly performs the first rewrite step at position
    $\tau$ in the original tree $\F{T}_x$
    (which is an innermost rewrite step) at the root of the tree, by pushing it
    from the original nodes in the tree $\F{T}_x$ to the root of the new tree  $\Phi(\F{T}_x)$. 
    (It could also be that this innermost redex was never reduced in $\F{T}_x$.)
    This can be seen in the diagram above.
    This push only results in the same termination probability due to our
    restriction that $\SSS$ is linear.

    In the figure below, we illustrate the effect of
    $\Phi$, where the original tree $\F{T}_x$ with $x = v_0$
    is on the left and $\Phi(\F{T}_x)$ is on the right.
    \begin{center}
        \centering
        \scriptsize
		\begin{tikzpicture}
			\tikzstyle{adam}=[rectangle,thick,draw=black!100,fill=white!100,minimum size=3mm]
			\tikzstyle{empty}=[shape=circle,thick,minimum size=8mm]
			\tikzstyle{circle}=[shape=circle,draw=black!100,fill=white!100,thick,minimum size=3mm]
			
			\node[empty] at (-3.5, 0.5)  (name) {$T_{v_0}$};
			\node[adam] at (-3.5, 0)  (la) {$v_0$};

			\node[circle] at (-4.5, -1)  (lb1) {$v_1$};
			\node[adam] at (-3.5, -1)  (lb2) {$v_2$};
			\node[adam] at (-2.5, -1)  (lb3) {$v_3$};

			\node[adam] at (-4.5, -2)  (lc1) {$v_4$};

			\node[circle] at (-3, -2)  (ld1) {$v_5$};
			\node[circle] at (-2, -2)  (ld3) {$v_6$};
		
			\node[adam] at (-3, -3)  (lf1) {$v_7$};
			\node[adam] at (-2, -3)  (lf3) {$v_8$};

			\node[adam] at (-3, -4)  (lg1) {$v_9$};

			\draw (la) edge[->] (lb1);
			\draw (la) edge[->] (lb2);
			\draw (la) edge[->] (lb3);
			\draw (lb1) edge[->] (lc1);
			\draw (lb3) edge[->] (ld1);
			\draw (lb3) edge[->] (ld3);
			\draw (ld1) edge[->] (lf1);
			\draw (ld3) edge[->] (lf3);
			\draw (lf1) edge[->] (lg1);

			\node[empty] at (-5.1,-1.3)  (Z) {$Z$};

            \draw [dashed] (-5,-1.5) -- (-4,-1.5) -- (-4,-2.5) -- (-1,-2.5);

			\node[empty] at (-0.5, -1.5)  (lead) {\huge $\leadsto$};
			
			\node[empty] at (2.5, 0.5)  (name2) {$\Phi(T_{v_0})$};
			\node[circle] at (2.5, 0)  (a) {$\hat{v}$};

			\node[adam] at (2.5, -1)  (b1) {$1.v_0$};

			\node[adam] at (1.5, -2)  (c1) {$1.v_1$};
			\node[adam] at (2.5, -2)  (c2) {$1.v_2$};
			\node[adam] at (3.5, -2)  (c3) {$1.v_3$};

			\node[adam] at (3, -3)  (d1) {$1.v_5$};
			\node[adam] at (4, -3)  (d2) {$1.v_6$};

			\node[adam] at (3, -4)  (f1) {$v_9$};

			\draw (a) edge[->] (b1);
			\draw (b1) edge[->] (c1);
			\draw (b1) edge[->] (c2);
			\draw (b1) edge[->] (c3);
			\draw (c3) edge[->] (d1);
			\draw (c3) edge[->] (d2);
			\draw (d1) edge[->] (f1);
		\end{tikzpicture}
    \end{center}
    However, since we are allowed to rewrite above $\tau$ in the original tree $\F{T}_x$, the actual position of the innermost redex that was originally at position $\tau$ might change during the application of a rewrite step.
    Hence, we recursively define the position $\varphi_{\tau}(v)$ that contains precisely this redex
    for each node $v$ in $\F{T}_x$ until we rewrite at this position.
    Initially, we have $\varphi_{\tau}(x) = \tau$.
    Whenever we have defined $\varphi_{\tau}(v)$ for some node $v$, and we have
    $t_v^{\F{T}_x} \to_{\SSS} \{p_{w_1}^{\F{T}_x}:t_{w_1}^{\F{T}_x}, \ldots,
    p_{w_m}^{\F{T}_x}:t_{w_m}^{\F{T}_x}\}$ for the direct successors $vE
    = \{w_1, \ldots, w_m\}$, using the rule $\ell \to \{
    p_1:r_1, \ldots, p_m:r_m\}$, the substitution $\sigma$, and the position $\pi$, we do the following:
    If $\varphi_{\tau}(v) = \pi$, meaning that we rewrite this innermost redex, then we set $\varphi_\tau(w_j) = \bot$ for all $1 \leq j \leq m$ to indicate that we have rewritten the innermost redex.
    If we have $\varphi_{\tau}(v) \bot \pi$, meaning that the rewrite step takes place on a
    position that is parallel to $\varphi_{\tau}(v)$, then we set
    $\varphi_\tau(w_j) = \varphi_{\tau}(v)$ for all $1 \leq j \leq
    m$, as the position of the innermost redex did not change.
    Otherwise, we have $\pi < \varphi_{\tau}(v)$ (since we cannot rewrite below $\varphi_{\tau}(v)$ as it is an innermost redex), and thus there exists a $\chi \in \IN^+$ such that $\pi.\chi = \varphi_{\tau}(v)$.
    Since the rules of $\SSS$ are non-overlapping, the redex must be completely ``inside''
    the used substitution $\sigma$, and we can find a position $\alpha_q$ of a variable $q$ in $\ell$ and another position $\beta$ such that $\chi = \alpha_q.\beta$.
    Furthermore, since the rule is linear, $q$ only occurs once in $\ell$ and at most once in $r_j$ for all $1 \leq j \leq m$.
    If $q$ occurs in $r_j$ at a position $\rho_q^j$, then we set $\varphi_\tau(w_j) = \pi.\rho_q^j.\beta$.
    Otherwise, we set $\varphi_\tau(w_j) = \top$ to indicate that the innermost redex was erased during the computation.
    Finally, if $\varphi_{\tau}(v) \in \{\bot, \top\}$, then we set $\varphi_\tau(w_j)
    = \varphi_{\tau}(v)$ for all $1 \leq j \leq m$ as well.
    So to summarize, $\varphi_{\tau}(v)$ is
    now either the position of the innermost redex in $t_{v}$, $\top$ to indicate that the redex was erased, or $\bot$ to indicate that we have rewritten the redex.

    In the figure above, the circled nodes represent the nodes where we perform a rewrite
    step at position $\varphi_{\tau}(v)$.
    We now define the $\SSS$-RST $\Phi(\F{T}_x)$
    whose root $\hat{v}$ is labeled with $(1:t_x^{\F{T}_x})$ and 
    that directly performs the rewrite step $t_x^{\F{T}_x} = t_{\hat{v}}^{\Phi(\F{T}_x)} \ito_{\SSS, \tau} \{\hat{p}_{1}:t_{1.x}^{\Phi(\F{T}_x)}, \ldots, \hat{p}_{h}:t_{h.x}^{\Phi(\F{T}_x)}\}$, 
    with the rule $\hat{\ell} \to \{ \hat{p}_1:\hat{r}_1, \ldots, \hat{p}_h:\hat{r}_h\} \in \SSS$, a substitution $\hat{\sigma}$, 
    and the position $\tau$, at the new root $\hat{v}$.
    Here, we have $t_{\hat{v}}^{\Phi(\F{T}_x)}|_{\tau} = \hat{\ell} \hat{\sigma}$.
    Let $Z$ be the set of all nodes $z$ such that $\varphi_{\tau}(z) \neq \bot$, 
    i.e., at node $z$ we did not yet perform the rewrite step with the innermost redex.
    In the example we have $Z = \{v_0, \ldots, v_6\} \setminus \{v_4\}$.
    For each of these nodes $z \in Z$ and each $1 \leq e \leq h$,
    we create a new node $e.z \in V'$ with edges as in $\F{T}_x$ for the nodes in $Z$, e.g., for the node $1.v_3$ we create edges to $1.v_5$ and $1.v_6$.
    Furthermore, we add the edges from the new root $\hat{v}$ to the nodes $e.{x}$ for all $1 \leq e \leq h$.
    Note that $x$ was the root in the tree $\F{T}_x$ and has to be contained in $Z$.
    For example, for the node $\hat{v}$ we create an edge to $1.v_0$.
    We define the labeling of the nodes in $\Phi(\F{T}_x)$ as follows
    for all nodes $z$ in $Z$:        
    \begin{itemize}
        \item[] (T-1) $t_{e.z}^{\Phi(\F{T}_x)} = t_{z}^{\F{T}_x}[\hat{r}_e \hat{\sigma}]_{\varphi_{\tau}(z)}$ if $\varphi_{\tau}(z) \in \IN^*$ and $t_{e.z}^{\Phi(\F{T}_x)} = t_{z}^{\F{T}_x}$ if $\varphi_{\tau}(z) = \top$.
        \item[] (T-2) $p_{e.z}^{\Phi(\F{T}_x)} = p_z^{\F{T}_x} \cdot \hat{p}_e$ 
    \end{itemize}
    Now, for a leaf $e.z' \in V'$ either $z' \in V$ is also a leaf (e.g., node $v_2$ in our example) 
    or we rewrite the innermost redex at
    position $\varphi_{\tau}(z')$ at node $z'$ in $\F{T}_x$ (e.g., node $v_1$ in our example).
    In the latter case, due to non-overlappingness, the same rule
    $\hat{\ell} \to \{ \hat{p}_1:\hat{r}_1, \ldots, \hat{p}_h:\hat{r}_h\}$ and
    the same substitution $\hat{\sigma}$ were used.
    If we rewrite $t_{z'}^{\F{T}_x} \ito_{\SSS, \varphi_{\tau}(z')} \{\hat{p}_{1}:t_{w'_1}^{\F{T}_x}, \ldots, \hat{p}_{h}:t_{w'_h}^{\F{T}_x}\}$, 
    then we have $t_{w'_e}^{\F{T}_x} =
    t_{z'}^{\F{T}_x}[\hat{r}_e \hat{\sigma}]_{\varphi_{\tau}(z')} \stackrel{\mbox{\scriptsize (T-1)}}{=} t_{e.z'}^{\Phi(\F{T}_x)}$ 
    and $p_{w'_e}^{\F{T}_x} \stackrel{\mbox{\scriptsize (T-2)}}{=} p_{z'}^{\F{T}_x} \cdot \hat{p}_e =
    p_{e.z'}^{\Phi(\F{T}_x)}$.
    Thus, we can copy the rest of this subtree of $\F{T}_x$ to
    our newly generated tree $\Phi(\F{T}_x)$.
    In our example, $v_1$ has the only successor $v_4$, hence we can copy the subtree starting at node $v_4$, which is only the node itself, to the node $1.v_1$ in $\Phi(\F{T}_x)$.
    For $v_5$, we have the only successor $v_7$, hence we can copy the subtree starting at node $v_7$, which is the node itself together with its successor $v_9$, to the node $1.v_5$ in $\Phi(\F{T}_x)$.
    So essentially, we just had to define how to construct
    $\Phi(\F{T}_x)$ for
    the part of the tree before we reach nodes $v$ with $\varphi_{\tau}(v) = \bot$ in $\F{T}_x$.
    Now we have to show that $|\Phi(\F{T}_x)|_{\ctleaf} = |\F{T}_x|_{\ctleaf}$
    and that $\Phi(\F{T}_x)$ is indeed a valid $\SSS$-RST 
    (i.e., that the edges between nodes $e.z$ with $z \in Z$ and its successors correspond to rewrite steps with $\SSS$).

    \noindent
    \underline{\textbf{1.2 We show $|\Phi(\F{T}_x)|_{\ctleaf} = |\F{T}_x|_{\ctleaf}$.}}

    \noindent
    Let $v$ be a leaf in $\Phi(\F{T}_x)$.
    If $v = e'.z$ for some node $z \in Z$ that is a leaf in $\F{T}_x$ (e.g., node $1.v_2$), then also $e.z$ must be a leaf in $\Phi(\F{T}_x)$ for every $1 \leq e \leq h$.
    Here, we get $\sum_{1 \leq e \leq h} p_{e.z}^{\Phi(\F{T}_x)} \stackrel{\text{(T-2)}}{=} \sum_{1 \leq e \leq h} p_z^{\F{T}_x} \cdot \hat{p}_e = p_z^{\F{T}_x} \cdot \sum_{1 \leq e \leq h} \hat{p}_e = p_z^{\F{T}_x} \cdot 1 = p_z^{\F{T}_x}$, and thus 
    \[\sum_{\substack{e.z \in \ctleaf^{\Phi(\F{T}_x)}\\ z \in Z,  z \in \ctleaf^{\F{T}_x}}}
    p_{e.z}^{\Phi(\F{T}_x)} = \sum_{\substack{z \in \ctleaf^{\F{T}_x}\\ z \in Z}} \left( \sum_{1 \leq e \leq h}
    p_{e.z}^{\Phi(\F{T}_x)} \right) = \sum_{\substack{z \in \ctleaf^{\F{T}_x}\\ z \in Z}} \!\! p_z^{\F{T}_x}\]

    If $v = e.z$ for some node $z \in Z$ that is not a leaf in $\F{T}_x$ (e.g., node $1.v_1$), then we know by construction 
    that the $e$-th successor $w_e$ of $z$ in $\F{T}_x$ is not contained in $Z$ and is a leaf of $\F{T}_x$.
    Here, we get $p_{e.z}^{\Phi(\F{T}_x)} \stackrel{\text{(T-2)}}{=} p_z^{\F{T}_x} \cdot \hat{p}_e = p_{w_e}^{\F{T}_x}$, and thus 
    \[\sum_{\substack{e.z \in \ctleaf^{\Phi(\F{T}_x)}\\z \in Z,  z \notin \ctleaf^{\F{T}_x}}} p_{e.z}^{\Phi(\F{T}_x)} = \sum_{\substack{w_e \in \ctleaf^{\F{T}_x}\\ w_e = (zE)_e, w_e \not\in Z, z \in Z}}
    p_{e.z}^{\Phi(\F{T}_x)} = \sum_{\substack{w_e \in \ctleaf^{\F{T}_x}\\ w_e = (zE)_e, w_e \not\in Z, z \in Z}} p_{w_e}^{\F{T}_x}\]
    Here, $w_e = (zE)_e$ denotes that $w_e$ is the $e$-th successor of $z$.

    Finally, if $v$ does not have the form $v = e.z$, then $v$ is also a leaf in $\F{T}_x$
    with $p_{v}^{\Phi(\F{T}_x)} = p_{v}^{\F{T}_x}$ and for both $v$ and its predecessor $u$ we have $v,u \not\in Z$, and thus 
    \[\sum_{\substack{v \in \ctleaf^{\Phi(\F{T}_x)}\\v \in \ctleaf^{\F{T}_x}}} p_v^{\Phi(\F{T}_x)} = \sum_{\substack{v \in \ctleaf^{\F{T}_x}\\ v \in uE, u \not\in Z}} p_v^{\Phi(\F{T}_x)} = \sum_{\substack{v \in \ctleaf^{\F{T}_x}\\ v \in uE, u \not\in Z}} p_v^{\F{T}_x}\]
    Note that these cases cover each leaf of $\F{T}_x$ exactly once.
    These three equations imply:
    {\small
    \allowdisplaybreaks
    \begin{align*}
        & |\Phi(\F{T}_x)|_{\ctleaf}\\
        = & \sum_{z \in \ctleaf^{\Phi(\F{T}_x)}} p_{z}^{\Phi(\F{T}_x)}\\
        = & \sum_{\substack{e.z \in \ctleaf^{\Phi(\F{T}_x)}\\ z \in Z,  z \in \ctleaf^{\F{T}_x}}}
        p_{e.z}^{\Phi(\F{T}_x)} + \sum_{\substack{e.z \in \ctleaf^{\Phi(\F{T}_x)}\\
        z \in Z,  z \notin \ctleaf^{\F{T}_x}}} p_{e.z}^{\Phi(\F{T}_x)}
        + \sum_{\substack{v \in \ctleaf^{\Phi(\F{T}_x)}\\v \in \ctleaf^{\F{T}_x}}} p_v^{\Phi(\F{T}_x)} \\ 
        = & \sum_{\substack{z \in \ctleaf^{\F{T}_x}\\ z \in Z}} \!\! p_z^{\F{T}_x}
        + \sum_{\substack{w_e \in \ctleaf^{\F{T}_x}\\ w_e = (zE)_e, w_e \not\in Z, z \in Z}} p_{w_e}^{\F{T}_x}
        + \sum_{\substack{v \in \ctleaf^{\F{T}_x}\\ v \in uE, u \not\in Z}} p_v^{\F{T}_x} \\
        = & \sum_{z \in \ctleaf^{\F{T}_x}} p_z^{\F{T}_x} \\
        = & |\F{T}_x|_{\ctleaf} 
    \end{align*}
    }

    \noindent
    \underline{\textbf{1.3 We show that $\Phi(\F{T}_x)$ is indeed a valid RST\@.}}

    \noindent
    Finally, we prove that $\Phi(\F{T}_x)$ is a valid $\SSS$-RST.
    Here, we only need to show that $t_{e.z}^{\Phi(\F{T}_x)} \to_{\SSS} \{p_{e.{w_1}}^{\Phi(\F{T}_x)}:t_{e.{w_1}}^{\Phi(\F{T}_x)}, \ldots, p_{e.{w_m}}^{\Phi(\F{T}_x)}:t_{e.{w_m}}^{\Phi(\F{T}_x)}\}$ for all $z \in Z$ as all the other edges and labelings were already present in $\F{T}_x$, which is a valid $\SSS$-RST, and we have already seen that we have a valid innermost rewrite step at the new root $\hat{v}$.

    Let $z \in Z$.
    In the following, we distinguish between two different cases for a rewrite step at a
    node $z$ of $\F{T}_x$:
    \begin{enumerate}
        \item[(A)] We rewrite at a position
        parallel to $\varphi_{\tau}(z)$ or $\varphi_{\tau}(z) = \top$.
        \item[(B)] We rewrite at a position above $\varphi_{\tau}(z)$.
        Note that this is the most interesting case, where we need  the properties L
    (i.e., \underline{l}inearity) and NO.
    \end{enumerate}
    Let $t_z^{\F{T}_x} \to_{\SSS} \{p_z^{\F{T}_x} \cdot p_1:t_{w_1}^{\F{T}_x}, \ldots, p_z^{\F{T}_x} \cdot p_m:t_{w_m}^{\F{T}_x}\}$, 
    with a rule $\ell \to \{ p_1:r_1, \ldots, p_{m}:r_{m}\} \in \SSS$, 
    a substitution $\sigma$, and a position $\pi$ such that $t_z^{\F{T}_x}|_{\pi} = \ell \sigma$. 
    We have $t_{w_j}^{\F{T}_x} = t_z^{\F{T}_x}[r_j \sigma]_{\pi}$ for all $1 \leq j \leq m$.

    \smallskip
    \noindent
    \textbf{(A)} We start with the case where
    we have $\pi \bot \varphi_{\tau}(z)$ or $\varphi_{\tau}(z) = \top$.
    By (T-1), we get $t_{e.z}^{\Phi(\F{T}_x)} = t_{z}^{\F{T}_x}[\hat{r}_e \hat{\sigma}]_{\varphi_{\tau}(z)}$ if $\varphi_{\tau}(z) \in \IN^*$ 
    and $t_{e.z}^{\Phi(\F{T}_x)} = t_{z}^{\F{T}_x}$ if $\varphi_{\tau}(z) = \top$.
    In both cases, we can rewrite $t_{e.z}^{\Phi(\F{T}_x)}$ using the same rule, 
    the same substitution, and the same position, 
    as we have $t_{e.z}^{\Phi(\F{T}_x)}|_{\pi} = t_{z}^{\F{T}_x}[\hat{r}_e \hat{\sigma}]_{\varphi_{\tau}(z)}|_{\pi} = t_z^{\F{T}_x}|_{\pi} = \ell \sigma$ 
    or directly $t_{e.z}^{\Phi(\F{T}_x)}|_{\pi} = t_z^{\F{T}_x}|_{\pi} = \ell \sigma$.

It remains to show that $t_{e.{w_j}}^{\Phi(\F{T}_x)} = t_{e.z}^{\Phi(\F{T}_x)}[r_j \sigma]_{\pi}$ for all $1 \leq j \leq m$, i.e., that the labeling we defined for $\Phi(\F{T}_x)$ corresponds to this rewrite step.
    Let $1 \leq j \leq m$.
    If $\varphi_{\tau}(z) \in \IN^*$, then we have 
    $t_{e.z}^{\Phi(\F{T}_x)}[r_j \sigma]_{\pi} = t_{z}^{\F{T}_x}[\hat{r}_e \hat{\sigma}]_{\varphi_{\tau}(z)}[r_j \sigma]_{\pi} \stackrel{\varphi_{\tau}(z) \bot \pi}{=} t_{z}^{\F{T}_x}[r_j \sigma]_{\pi}[\hat{r}_e \hat{\sigma}]_{\varphi_{\tau}(z)} = t_{w_j}^{\F{T}_x}[\hat{r}_e \hat{\sigma}]_{\varphi_{\tau}(z)} = t_{e.{w_j}}^{\Phi(\F{T}_x)}$.
    If $\varphi_{\tau}(z) = \top$, then
    $t_{e.z}^{\Phi(\F{T}_x)}[r_j \sigma]_{\pi} = t_{z}^{\F{T}_x}[r_j \sigma]_{\pi} = t_{w_j}^{\F{T}_x} = t_{e.{w_j}}^{\Phi(\F{T}_x)}$.
    Finally, note that the probabilities of the labeling are correct, as we are using the same rule with the same probabilities in both trees.

    \smallskip
    \noindent
    \textbf{(B)} If we have $\pi < \varphi_{\tau}(z)$, then there exists a $\chi \in \IN^+$ such that $\pi.\chi = \varphi_{\tau}(z)$.
    Since the rules of $\SSS$ are non-overlapping, the redex must be completely ``inside''
    the used substitution $\sigma$, and we can find a position $\alpha_q$ of a variable $q$ in $\ell$ and another position $\beta$ such that $\chi = \alpha_q.\beta$.
    Furthermore, since the rule is linear,
    $q$ only occurs once in $\ell$ and at most once in $r_j$ for all $1 \leq j \leq m$.
    Let $\rho_q^j$ be the position of $q$ in $r_j$ if it exists.
    By (T-1), we get $t_{z}^{\F{T}_x}[\hat{r}_e \hat{\sigma}]_{\varphi_{\tau}(z)} = t_{e.z}^{\Phi(\F{T}_x)}$.
    We can rewrite $t_{e.z}^{\Phi(\F{T}_x)}$ using the same rule, 
    the same substitution, and the same position, 
    as we have $t_{e.z}^{\Phi(\F{T}_x)}|_{\pi} = t_{z}^{\F{T}_x}[\hat{r}_e \hat{\sigma}]_{\varphi_{\tau}(z)}|_{\pi} = t_{z}^{\F{T}_x}|_{\pi}[\hat{r}_e \hat{\sigma}]_{\chi} = \ell \sigma'$, where $\sigma'(q) = \sigma(q)[\hat{r}_e \hat{\sigma}]_{\beta}$ and $\sigma'(q') = \sigma(q')$ for all other variables $q' \neq q$.

It remains to show that $t_{e.{w_j}}^{\Phi(\F{T}_x)} = t_{e.z}^{\Phi(\F{T}_x)}[r_j \sigma']_{\pi}$ for all $1 \leq j \leq m$, i.e., that the labeling we defined for $\Phi(\F{T}_x)$ corresponds to this rewrite step.
    Let $1 \leq j \leq m$.
    If $\rho_q^j$ exists, then we have
    \begin{align*}
    t_{e.z}^{\Phi(\F{T}_x)}[r_j \sigma']_{\pi} &=
    t_{z}^{\F{T}_x}[\hat{r}_e \hat{\sigma}]_{\varphi_{\tau}(z)}[r_j \sigma']_{\pi} =
    t_{z}^{\F{T}_x}[\hat{r}_e \hat{\sigma}]_{\pi.\alpha_q.\beta}[r_j \sigma']_{\pi}\\
    &= t_{z}^{\F{T}_x}[r_j \sigma]_{\pi}[\hat{r}_e \hat{\sigma}]_{\pi.\rho_q^j.\beta} =
    t_{w_j}^{\F{T}_x}[\hat{r}_e \hat{\sigma}]_{\varphi_{\tau}(w_j)} = t_{e.{w_j}}^{\Phi(\F{T}_x)}.
    \end{align*}
    Otherwise, we erase the precise redex and get
    \begin{align*}
    t_{e.z}^{\Phi(\F{T}_x)}[r_j \sigma]_{\pi} &=
    t_{z}^{\F{T}_x}[\hat{r}_e \hat{\sigma}]_{\varphi_{\tau}(z)}[r_j \sigma']_{\pi} =
    t_{z}^{\F{T}_x}[\hat{r}_e \hat{\sigma}]_{\pi.\alpha_q.\beta}[r_j \sigma']_{\pi}\\
    &= t_{z}^{\F{T}_x}[r_j \sigma]_{\pi} = t_{w_j}^{\F{T}_x} = t_{e.{w_j}}^{\Phi(\F{T}_x)}.
    \end{align*}
    Finally, note that the probabilities of the labeling are correct, as we are using the same rule with the same probabilities in both trees.
\end{myproof}

\setcounter{auxctr}{\value{theorem}}
\setcounter{theorem}{\value{wASTtofAST}}

\begin{theorem}[From wAST to fAST]
    If a PTRS $\SSS$ is non-overlapping, linear, and non-erasing, then 
    \begin{align*}
        \SSS \text{ is fAST} &\Longleftrightarrow \SSS \text{ is wAST}\!
    \end{align*}
\end{theorem}

\setcounter{theorem}{\value{auxctr}}

\begin{myproof}
    We only have to prove the non-trivial direction ``$\Longleftarrow$''.
    Let $\SSS$ be a PTRS that is non-overlapping, linear, and non-erasing.
    Furthermore, assume for a contradiction that $\SSS$ is wAST but not fAST.
    This means that there exists an RST $\F{T}$ such that $|\F{T}|_{\ctleaf} = c$ for some $0 \leq c < 1$, and we can assume that $\F{T}$ is fully evaluated (since extending an RST to a fully evaluated one never increases its convergence probability).
    Let $t \in \TSet{\Sigma}{\VSet}$ such that the root of $\F{T}$ is labeled with
    $(1:t)$.
    Since $\SSS$ is wAST, there exists another fully evaluated RST $\tilde{\F{T}} = (\tilde{V}, \tilde{E},
    \tilde{L})$ such that $|\tilde{\F{T}}|_{\ctleaf} = 1$ and the root of $\tilde{\F{T}}$
    is also
    labeled with $(1:t)$.
    Hence, in $\F{T}$ there is at least one rewrite step performed at some node $x$ that is different to the rewrite step performed in $\tilde{\F{T}}$, and we choose such a node $x$ of minimal depth.
    The core steps of the proof are the same as for the proof of \Cref{properties-eq-AST-iAST-1}.
    We iteratively push the rewrite steps that would be performed in $\tilde{\F{T}}$ at node $x$ to this node in $\F{T}$.
    Then, the limit of this construction is exactly $\tilde{\F{T}}$, which would mean that
    $|\tilde{\F{T}}|_{\ctleaf} \leq c < 1$, which is the desired contradiction.
    For this, we have to adjust the construction $\Phi(\circ)$.
    The rest of the proof remains completely the same.
    \smallskip
          
    \noindent 
    \textbf{\underline{1.1 Construction of $\Phi(\circ)$}}

    \noindent
    Let $\F{T}_x = \F{T}[xE^*]$ be an $\SSS$-RST that performs a rewrite step at position $\zeta$ at the root node $x$, i.e., $t_x^{\F{T}_x} \to_{\SSS, \zeta} \{p_{y_1}^{\F{T}_x}:t_{y_1}^{\F{T}_x}, \ldots, p_{y_k}^{\F{T}_x}:t_{y_k}^{\F{T}_x}\}$ using the rule $\bar{\ell} \to \{ \bar{p}_1:\bar{r}_1, \ldots, \bar{p}_k:\bar{r}_k\}$, and the substitution $\bar{\sigma}$ such that $t_x^{\F{T}_x}|_{\zeta} = \bar{\ell} \bar{\sigma}$.
    Then $t_{y_j}^{\F{T}_x} = t_x^{\F{T}_x}[\bar{r}_j \bar{\sigma}]_{\zeta}$ for all $1 \leq j \leq k$.
    Furthermore, assume that
    $\tilde{T}_x = \tilde{\F{T}}[x\tilde{E}^*]$ rewrites at position $\tau$ (using a rule
    $\hat{\ell} \to \{ \hat{p}_1:\hat{r}_1, \ldots, \hat{p}_h:\hat{r}_h\} \in \SSS$ and  
    the substitution $\hat{\sigma}$)
    with $\tau \neq \zeta$. (Note that
    if $\tau = \zeta$, then by non-overlappingness the rewrite step would be the same.)
    Instead of applying the rewrite step at position $\zeta$ at the root $x$
    we want to directly apply the rewrite step at position $\tau$.

    \begin{wrapfigure}[6]{r}{0.37\textwidth}
        \vspace*{-1.3cm}
        \begin{center}
            \begin{tikzpicture}[scale=0.5]
                \begin{pgfonlayer}{nodelayer}
                    \node [style=target,pin={[pin distance=0.05cm, pin edge={,-}] 140:\tiny \textcolor{blue}{$x$}}] (3) at (0, 3) {};
                    \node [style=none] (6) at (1.5, 0) {};
                    \node [style=none] (7) at (-1.5, 0) {};
                    \node [style=none] (9) at (-2, -1) {};
                    \node [style=none] (10) at (2, -1) {};
                    \node [style=moveBlock] (12) at (-0.75, 0.75) {};
                    \node [style=moveBlock] (13) at (0, 0) {};
                    \node [style=moveBlock] (14) at (0.75, 0.25) {};
                    \node [style=moveBlock] (15) at (-0.75, 0.75) {};
                    \node [style=none] (16) at (-0.5, 1.5) {};
                    \node [style=none] (17) at (0, 1.25) {};
                    \node [style=none] (18) at (0.5, 1.5) {};
                    \node [style=none] (19) at (-0.75, 0) {};
                    \node [style=none] (20) at (0, -0.75) {};
                    \node [style=none] (21) at (0.75, -0.5) {};
                \end{pgfonlayer}
                \begin{pgfonlayer}{edgelayer}
                    \draw (3) to (6.center);
                    \draw (3) to (7.center);
                    \draw [style=dotWithoutHead] (7.center) to (9.center);
                    \draw [style=dotWithoutHead] (6.center) to (10.center);
                    \draw [style=dotWithoutHead, in=15, out=-105, looseness=0.50] (3) to (16.center);
                    \draw [style=dotWithoutHead, in=120, out=-90, looseness=0.75] (3) to (17.center);
                    \draw [style=dotWithoutHead, in=135, out=-75] (3) to (18.center);
                    \draw [style=dotHead, in=90, out=-30, looseness=0.75] (18.center) to (14);
                    \draw [style=dotHead, in=90, out=-150, looseness=0.75] (16.center) to (15);
                    \draw [style=dotHead, in=90, out=-45] (17.center) to (13);
                    \draw [style=dashHead, bend right=75, looseness=2.00] (14) to (3);
                    \draw [style=dashHead, bend left=75, looseness=1.75] (15) to (3);
                    \draw [style=dashHead, bend right=105, looseness=2.75] (13) to (3);
                    \draw [style=dotWithoutHead] (15) to (19.center);
                    \draw [style=dotWithoutHead] (13) to (20.center);
                    \draw [style=dotWithoutHead] (14) to (21.center);
                \end{pgfonlayer}
            \end{tikzpicture}
        \end{center}
    \end{wrapfigure}
    The construction creates a new RST $\Phi(\F{T}_x) = (V',E',L')$ such that
    $|\Phi(\F{T}_x)|_{\ctleaf} = |\F{T}_x|_{\ctleaf}$,
    and that directly performs a rewrite step at position $\tau$ at the root of the tree, by pushing it from the original nodes in the tree $\F{T}_x$ to the root.
    This push only results in the same termination probability due to our
    restriction that $\SSS$ is linear and non-erasing.
    We need the non-erasing property now, because $\tau$ may be above $\zeta$, which was not possible in the proof of \Cref{properties-eq-AST-iAST-1}.

    Again, since we are allowed to rewrite above $\tau$ in the original tree $\F{T}_x$, the
    actual position of the redex that was originally at position $\tau$ might change
    during the application of a rewrite step.
    Hence, we recursively define the position $\varphi_{\tau}(v)$
    that contains precisely this redex for each node $v$ in $\F{T}_x$ until we rewrite at this position.
    Compared to the proof of \Cref{properties-eq-AST-iAST-1}, 
    since the rules in $\SSS$ are non-erasing, 
    we only have $\varphi_{\tau}(v) \in \IN^*$ or $\varphi_{\tau}(v) = \bot$. 
    The option $\varphi_{\tau}(v) = \top$ is not possible anymore.
    Initially, we have $\varphi_{\tau}(x) = \tau$.
    Whenever we have defined $\varphi_{\tau}(v)$ for some node $v$, and we have
    $t_v^{\F{T}_x} \to_{\SSS} \{p_{w_1}^{\F{T}_x}:t_{w_1}^{\F{T}_x}, \ldots, p_{w_m}^{\F{T}_x}:t_{w_m}^{\F{T}_x}\}$ 
    for the direct successors $vE = \{w_1, \ldots, w_m\}$, using the rule $\ell \to \{ p_1:r_1, \ldots, p_m:r_m\}$, 
    the substitution $\sigma$, and position $\pi$, we do the following:
    If $\varphi_{\tau}(v) = \pi$, then we set $\varphi_\tau(w_j) = \bot$ for all $1 \leq j \leq m$ to indicate that we have rewritten the redex.
    If we have $\varphi_{\tau}(v) \bot \pi$, meaning that the rewrite step takes place parallel to $\varphi_{\tau}(v)$, then we set $\varphi_\tau(w_j) = \varphi_{\tau}(v)$ for all $1 \leq j \leq m$, as the position did not change.
    If we have $\varphi_{\tau}(v) < \pi$, then we set $\varphi_\tau(w_j)
    = \varphi_{\tau}(v)$ for all $1 \leq j \leq m$ as well, as the position did not change
    either.
    If we have $\pi < \varphi_{\tau}(v)$, then there exists a $\chi \in \IN^+$ such that $\pi.\chi = \varphi_{\tau}(v)$.
    Since the rules of $\SSS$ are non-overlapping, the redex must be completely inside the
    used substitution $\sigma$, and we can find a position $\alpha_q$ of a variable $q$ in $\ell$ and another position $\beta$ such that $\chi = \alpha_q.\beta$.
    Furthermore, since the rule is linear and non-erasing,
    $q$ only occurs once in $\ell$ and once in $r_j$ for all $1 \leq j \leq m$.
    Let $\rho_q^j$ be the position of $q$ in $r_j$.
    Here, we set $\varphi_\tau(w_j) = \pi.\rho_q^j.\beta$.
    Finally, if $\varphi_{\tau}(v) = \bot$, then we set $\varphi_\tau(w_j)
    = \varphi_{\tau}(v) = \bot$ for all $1 \leq j \leq m$ as well.

    Again, we now define the $\SSS$-RST $\Phi(\F{T}_x)$ whose root is labeled with $(1:t_x^{\F{T}_x})$ such that $|\Phi(\F{T}_x)|_{\ctleaf} = |\F{T}_x|_{\ctleaf}$, 
    and that directly performs the rewrite step $t_{x}^{\F{T}_x} \to_{\SSS, \tau} \{\hat{p}_{1}:t_{1.x}^{\Phi(\F{T}_x)}, \ldots, \hat{p}_{h}:t_{h.x}^{\Phi(\F{T}_x)}\}$, 
    with the rule $\hat{\ell} \to \{ \hat{p}_1:\hat{r}_1, \ldots, \hat{p}_h:\hat{r}_h\} \in \SSS$, 
    the substitution $\hat{\sigma}$, 
    and the position $\tau$, at the new root $\hat{v}$.
    Here, we have $t_x^{\F{T}_x}|_{\tau} = \hat{\ell} \hat{\sigma}$.
    Let $Z$ be the set of all nodes $v$ such that $\varphi_{\tau}(v) \neq \bot$.
    For each of these nodes $z \in Z$ and each $1 \leq e \leq h$,
    we create a new node $e.z \in V'$ with edges as in $\F{T}_x$ for the nodes in $Z$.
    Furthermore, we add the edges from the new root $\hat{v}$ to the nodes $e.{x}$ for all $1 \leq e \leq h$.
    Remember that $x$ was the root in the tree $\F{T}_x$ and has to be contained in $Z$.
    We define the labeling of the nodes in  $\Phi(\F{T}_x)$ as follows
    for all nodes $z$ in
    $Z$:
    \begin{itemize}
        \item[] (T-1) $t_{e.z}^{\Phi(\F{T}_x)} = t_{z}^{\F{T}_x}[\hat{r}_e \delta]_{\varphi_{\tau}(z)}$ for the substitution $\delta$ such that $t_{z}^{\F{T}_x}|_{\varphi_{\tau}(z)} = \hat{\ell} \delta$
        \item[] (T-2) $p_{e.z}^{\Phi(\F{T}_x)} = p_z^{\F{T}_x} \cdot \hat{p}_e$ 
    \end{itemize}
    Now, for a leaf $e.z' \in V'$ either $z' \in V$ is also a leaf
    or we rewrite at the position $\varphi_{\tau}(z')$ in node $z'$ in $\F{T}_x$.
    If we rewrite
    $t_{z'}^{\F{T}_x} \ito_{\SSS, \varphi_{\tau}(z')} \{p_{w'_1}^{\F{T}_x}:t_{w'_1}^{\F{T}_x}, \ldots, p_{w'_h}^{\F{T}_x}:t_{w'_h}^{\F{T}_x}\}$,
    then we have $t_{w'_e}^{\F{T}_x} = t_{z'}^{\F{T}_x}[\hat{r}_e \delta]_{\varphi_{\tau}(z')} \stackrel{\mbox{\scriptsize (T-1)}}{=} t_{e.z'}^{\Phi(\F{T}_x)}$ for the substitution $\delta$ such that $t_{z'}^{\F{T}_x}|_{\varphi_{\tau}(z')} = \hat{\ell} \delta$
    and $p_{w'_e}^{\F{T}_x} \stackrel{\mbox{\scriptsize (T-2)}}{=} p_{z'}^{\F{T}_x} \cdot \hat{p}_e = p_{e.z'}^{\Phi(\F{T}_x)}$.
    Thus, we can copy the rest of this subtree of $\F{T}_x$ in
    our newly generated tree $\Phi(\F{T}_x)$.
    So essentially, we just had to define how to construct
    $\Phi(\F{T}_x)$ for
    the part of the tree before we reach the
    nodes $v$ with $\varphi_{\tau}(v) = \bot$ in $\F{T}_x$.
    As in the proof of \Cref{properties-eq-AST-iAST-1}, we get $|\Phi(\F{T}_x)|_{\ctleaf} =
    |\F{T}_x|_{\ctleaf}$.
    We only need to show that $\Phi(\F{T}_x)$ is a valid $\SSS$-RST, i.e., that
    $t_{e.z}^{\Phi(\F{T}_x)} \to_{\SSS} \{p_{e.{w_1}}^{\Phi(\F{T}_x)}:t_{e.{w_1}}^{\Phi(\F{T}_x)}, \ldots, p_{e.{w_m}}^{\Phi(\F{T}_x)}:t_{e.{w_m}}^{\Phi(\F{T}_x)}\}$ 
    for all $z \in Z$ as in the proof of \Cref{properties-eq-AST-iAST-1}.

\smallskip

    \noindent
    \underline{\textbf{1.3 We show that $\Phi(\F{T}_x)$ is indeed a valid RST\@.}}

    \noindent
    Let $z \in Z$.
    In the following, we distinguish between three different cases for a rewrite step at node $z$ of $\F{T}_x$:
    \begin{enumerate}
        \item[(A)] We rewrite at a position parallel to $\varphi_{\tau}(z)$.
        \item[(B)] We rewrite at a position above $\varphi_{\tau}(z)$.
        \item[(C)] We rewrite at a position below $\varphi_{\tau}(z)$.
    \end{enumerate}
    The cases (A) and (B) are analogous as before but in Case (A) we cannot erase the redex anymore.
    We only need to look at the new Case (C).

    \noindent
    \textbf{(C) If we have}
    $t_z^{\F{T}_x} \to_{\SSS} \{p_z^{\F{T}_x} \cdot p_1:t_{w_1}^{\F{T}_x}, \ldots, p_z^{\F{T}_x} \cdot p_m:t_{w_m}^{\F{T}_x}\}$, 
    then there is a rule $\ell \to \{ p_1:r_1, \ldots, p_{m}:r_{m}\} \in \SSS$, 
    a substitution $\sigma$, and a position $\pi$ with $t_z^{\F{T}_x}|_{\pi} = \ell \sigma$. 
    Then $t_{w_j}^{\F{T}_x} = t_z^{\F{T}_x}[r_j \sigma]_{\pi}$ for all $1 \leq j \leq m$.
    Additionally, we assume that $\varphi_{\tau}(z) < \pi$, and thus there exists a $\chi \in \IN^+$ such that $\pi = \varphi_{\tau}(z).\chi$.
    By (T-1), we get $t_{z}^{\F{T}_x}[\hat{r}_e \delta]_{\varphi_{\tau}(z)} = t_{e.z}^{\Phi(\F{T}_x)}$ for the substitution $\delta$ such that $t_{z}^{\F{T}_x}|_{\varphi_{\tau}(z)} = \hat{\ell} \delta$.
    Since the rules of $\SSS$ are non-overlapping, including the rule $\hat{\ell} \to \{ \hat{p}_1:\hat{r}_1, \ldots, \hat{p}_h:\hat{r}_h\}$ that we use at the root of $\Phi(\F{T}_x)$, 
    the redex for the current rewrite step must be completely ``inside'' the substitution $\delta$, and we can find a variable position $\alpha_q$ of a variable $q$ in $\hat{\ell}$ and another position $\beta$ such that $\chi = \alpha_q.\beta$.
    Furthermore, since the rule is also linear
    and non-erasing, $q$ occurs exactly once in $\hat{\ell}$ and exactly once in $\hat{r}_j$ for all $1 \leq j \leq h$.
    Let $\rho_q^e$ be the position of $q$ in $\hat{r}_e$.
    We can rewrite $t_{e.z}^{\Phi(\F{T}_x)}$ using the same rule, 
    the same substitution, and the position $\pi' = \varphi_{\tau}(z).\rho_q^e.\beta$, 
    as we have $t_{e.z}^{\Phi(\F{T}_x)}|_{\pi'} =
    t_{z}^{\F{T}_x}[\hat{r}_e \delta]_{\varphi_{\tau}(z)}|_{\pi'} = \hat{r}_e \delta|_{\rho_q^e.\beta}
    = \delta(q)|_{\beta} = t_{z}^{\F{T}_x}|_{\varphi_{\tau}(z)}|_{\alpha_q}|_{\beta} =
    t_{z}^{\F{T}_x}|_{\varphi_{\tau}(z).\alpha_q.\beta} = t_{z}^{\F{T}_x}|_{\varphi_{\tau}(z).\chi} =
    t_{z}^{\F{T}_x}|_{\pi} = \ell \sigma$.
    
    It remains to show that $t_{e.{w_j}}^{\Phi(\F{T}_x)} = t_{e.z}^{\Phi(\F{T}_x)}[r_j \sigma]_{\pi'}$ for all $1 \leq j \leq m$, i.e., that the labeling we defined for $\Phi(\F{T}_x)$ corresponds to this rewrite step.
    Let $1 \leq j \leq m$.
    We have
    $t_{e.z}^{\Phi(\F{T}_x)}[r_j \sigma]_{\pi'} = t_{z}^{\F{T}_x}[\hat{r}_e \delta]_{\varphi_{\tau}(z)}[r_j \sigma]_{\pi'} = t_{z}^{\F{T}_x}[\hat{r}_e \delta']_{\varphi_{\tau}(z)} = t_{e.{w_j}}^{\Phi(\F{T}_x)}$ for the substitution $\delta'$ with $\delta'(q) = \delta(q)[r_j \sigma]_{\beta}$ and $\delta'(q') = \delta(q')$ for all other variables $q' \neq q$.
    With this new substitution, we get $t_{w_j}^{\F{T}_x}|_{\varphi_{\tau}(w_j)} = t_{z}^{\F{T}_x}[r_j \sigma]_{\varphi_{\tau}(z).\alpha_q.\beta}|_{\varphi_{\tau}(z)} = t_{z}^{\F{T}_x}|_{\varphi_{\tau}(z)}[r_j \sigma]_{\alpha_q.\beta} = (\hat{\ell} \delta)[r_j \sigma]_{\alpha_q.\beta} = \hat{\ell} \delta'$.
    Finally, note that the probabilities of the labeling are correct, as we are using the same rule with the same probabilities in both trees.
\end{myproof}

\setcounter{auxctr}{\value{theorem}}
\setcounter{theorem}{\value{liASTtoiAST}}

\begin{theorem}[From liAST to iAST]
    If a PTRS $\SSS$ is non-overlapping, then
    \begin{align*}
        \SSS \text{ is iAST} &\Longleftrightarrow \SSS \text{ is liAST}\!
    \end{align*}
\end{theorem}

\setcounter{theorem}{\value{auxctr}}

\begin{myproof}
    The idea and the construction of this proof are completely analogous
    to the one of \Cref{properties-eq-AST-iAST-1}.
    We iteratively move leftmost innermost rewrite steps to a higher position in the innermost RST.
    Hence, the resulting tree is a leftmost innermost $\SSS$-RST.

    The construction of $\Phi(\circ)$ is also analogous to the one
    in \Cref{properties-eq-AST-iAST-1}.
    The only difference to the proof of \Cref{properties-eq-AST-iAST-1} is that the original tree $\F{T}_x$ is already an innermost $\SSS$-RST.
    This means that during our construction only Case (A) can occur, as we cannot rewrite above a redex in an innermost $\SSS$-RST.
    And for Case (A), we only need the property of being non-overlapping.
\end{myproof}

Next, we prove the new results from \cref{Improving Applicability} regarding AST.

\setcounter{auxctr}{\value{theorem}}
\setcounter{theorem}{\value{iASTtofASTTwo}}

\begin{theorem}[From iAST to fAST (2)]
    If a PTRS $\SSS$ is non-overlapping and right-linear, then 
    \begin{align*}
        \SSS \text{ is fAST} &\Longleftarrow \SSS \text{ is iAST w.r.t.\ } \paraarrow_{\SSS}\!
    \end{align*}
\end{theorem}

\setcounter{theorem}{\value{auxctr}}

\begin{myproof}
    Once again, we use the idea and the construction of the proof for \Cref{properties-eq-AST-iAST-1}.
    We iteratively move innermost rewrite steps to a higher position in the tree.
    But in this case, for moving these innermost rewrite steps, we also allow using rewrite steps with $\itopara_{\SSS}$ instead of $\ito_{\SSS}$.
    Hence, the resulting tree is an innermost $\SSS$-RST w.r.t.\ $\itopara_{\SSS}$.
    The construction of $\Phi(\circ)$ also remains similar to the one in \Cref{properties-eq-AST-iAST-1}.
    \smallskip
          
    \noindent 
    \textbf{\underline{1.1 Construction of $\Phi(\circ)$}}

    \noindent
    Let $\F{T}_x$ be an $\SSS$-RST that performs a non-innermost rewrite step at the root node
    $x$, i.e., $t_x^{\F{T}_x} \to_{\SSS} \{p_{y_1}^{\F{T}_x}:t_{y_1}^{\F{T}_x}, \ldots,
    p_{y_k}^{\F{T}_x}:t_{y_k}^{\F{T}_x}\}$ using the rule
    $\bar{\ell} \to \{ \bar{p}_1:\bar{r}_1, \ldots, \bar{p}_k:\bar{r}_k\}$, the substitution
    $\bar{\sigma}$,
    and the position $\bar{\pi}$ such that $t_x^{\F{T}_x}|_{\bar{\pi}} = \bar{\ell} \bar{\sigma}$.
    Then we have $t_{y_j}^{\F{T}_x} = t_x^{\F{T}_x}[\bar{r}_j \bar{\sigma}]_{\bar{\pi}}$ for all $1 \leq j \leq k$.
    Instead of applying a non-innermost rewrite step at the root $x$
    we want to directly apply an innermost rewrite step.
    Let $\tau$ be the position of some innermost redex in $t_x$ that is below $\bar{\pi}$.
    
    The construction creates a new RST $\Phi(\F{T}_x) = (V',E',L')$ such that
    $|\Phi(\F{T}_x)|_{\ctleaf} = |\F{T}_x|_{\ctleaf}$. In $\Phi(\F{T}_x)$,
    directly at the root one
    performs the first rewrite step at position $\tau$ 
    (which is an innermost rewrite step)
    and possibly simultaneously at some other innermost positions using the
    relation $\itopara_{\SSS}$, by pushing it from the original nodes
    in the tree $\F{T}_x$ to the root.
    This push only works due to our
    restriction that $\SSS$ is right-linear.
    We can now remove the left-linearity requirement by using $\itopara_{\SSS}$ at the root instead of just $\ito_{\SSS}$.

    Again, we recursively define the position $\varphi_{\tau}(v)$
    that contains precisely this redex for each node $v$ in $\F{T}_x$ until we rewrite at this position.
    This works exactly as for \Cref{properties-eq-AST-iAST-1}, again, due to our restriction of right-linearity.
    So to recapitulate, $\varphi_{\tau}(v)$
    for a node $v$ is either the position of the
    redex in $t_v$, $\top$ to indicate that the redex was erased, or $\bot$ to indicate
    that we have rewritten the redex.
    Also, the construction of the tree is similar to the one for \Cref{properties-eq-AST-iAST-1}.
    Let $Z$ be the set of all nodes $z$ such that $\varphi_{\tau}(z) \neq \bot$.
    For each of these nodes $z \in Z$ and each $1 \leq e \leq h$,
    we create a new node $e.z \in V'$ with edges as in $\F{T}_x$ for the nodes in $Z$.
    Furthermore, we add the edges from the new root $\hat{v}$ to the nodes $e.{x}$ for all $1 \leq e \leq h$.
    
    Next, we define the labeling for each of the new nodes.
    This is the part that differs compared to \Cref{properties-eq-AST-iAST-1}.
    Since $\SSS$ is non-overlapping, the position $\tau$ must be completely ``inside'' the 
    substitution $\bar{\sigma}$, 
    and we can find a variable $q$ of $\bar{\ell}$ such that $\tau = \bar{\pi}.\alpha_q.\beta$ 
    for some variable position $\alpha_q$ of $q$ in $\bar{\ell}$ and some other position $\beta$.
    However, since the left-hand side $\bar{\ell}$ may contain the same
    variable several times (as $\SSS$ does not have to be left-linear), there may exist multiple occurrences of $q$ in $\bar{\ell}$.
    Let $\{\alpha_1, \ldots, \alpha_n\}$ be the set of all positions $\alpha$ of $\bar{\ell}$ such that $\bar{\ell}|_{\alpha} = q$.
    The root $\hat{v}$ of $\Phi(\F{T}_x)$ is labeled with $(1 : t_x^{\F{T}_x})$, 
    and we perform the rewrite step 
    $t_{x}^{\F{T}_x} \iparaarrow_{\SSS, \Gamma} \{\hat{p}_{1}:t_{1.x}^{\Phi(\F{T}_x)}, \ldots, \hat{p}_{h}:t_{h.x}^{\Phi(\F{T}_x)}\}$, 
    with the rule $\hat{\ell} \to \{ \hat{p}_1:\hat{r}_1, \ldots, \hat{p}_h:\hat{r}_h\} \in \SSS$, 
    a substitution $\hat{\sigma}$, 
    and the set of positions $\Gamma = \{\bar{\pi}.\alpha_1.\beta, \ldots, \bar{\pi}.\alpha_n.\beta\}$, at the new root $\hat{v}$.
    Here, we have $t_x^{\Phi(\F{T}_x)}|_{\gamma} = \hat{\ell} \hat{\sigma}$ for all $\gamma \in \Gamma$ and get $t_{e.x}^{\Phi(\F{T}_x)} =
    t_{x}[\hat{r}_e \hat{\sigma}]_{\bar{\pi}.\alpha_1.\beta}\ldots[\hat{r}_e \hat{\sigma}]_{\bar{\pi}.\alpha_n.\beta}$
    for all $1 \leq e \leq h$.

    After this rewrite step, we mirror the rewrite step from the root $x$ at each node $e.x$ for all $1 \leq e \leq h$.
    This is possible, since we have $t_{e.x}^{\Phi(\F{T}_x)}|_{\bar{\pi}} = \bar{\ell} \bar{\sigma}'$ using the substitution $\bar{\sigma}'$ with $\bar{\sigma}'(q) = \bar{\sigma}(q)[\hat{r}_e \hat{\sigma}]_{\beta}$ and $\bar{\sigma}'(q') = \bar{\sigma}(q')$ for all other variables $q' \neq q$.
    Note that this is only possible because we have rewritten all occurrences of the same redex at a position $\gamma \in \Gamma$ simultaneously.
    Otherwise, we would not be able to define $\bar{\sigma}'$ like this, because the matching might fail in some cases (see \Cref{example:diff-AST-vs-iAST-left-lin}).
    With this definition of $\bar{\sigma}'$ we really have $\bar{\ell} \bar{\sigma}' = \bar{\ell} \bar{\sigma}[\hat{r}_e \hat{\sigma}]_{\alpha_1.\beta}\ldots[\hat{r}_e \hat{\sigma}]_{\alpha_n.\beta} = t_{x}^{\F{T}_x}|_{\bar{\pi}}[\hat{r}_e \hat{\sigma}]_{\alpha_1.\beta}\ldots[\hat{r}_e \hat{\sigma}]_{\alpha_n.\beta} = t_{x}^{\F{T}_x}[\hat{r}_e \hat{\sigma}]_{\bar{\pi}.\alpha_1.\beta}\ldots[\hat{r}_e \hat{\sigma}]_{\bar{\pi}.\alpha_n.\beta}|_{\bar{\pi}} = t_{e.x}^{\Phi(\F{T}_x)}|_{\bar{\pi}}$.
    Since $\SSS$ is right-linear, we know that $q$ can occur at most once in every $\bar{r}_j$.
    In this case, let $\psi_j$ be the position of $q$ in $\bar{r}_j$ for all $1 \leq j \leq k$.
    Then initially, we have $\varphi_{y_j}(\tau) = \bar{\pi}.\psi_j.\beta$ for all $1 \leq j \leq k$.
    Now, for all $1 \leq e \leq h$ and $1 \leq j \leq k$, if $q$ exists in $\bar{r}_j$, 
    then we get $t_{e.{y_j}}^{\Phi(\F{T}_x)} = t_{e.x}^{\Phi(\F{T}_x)}[\bar{r}_j \bar{\sigma}']_{\bar{\pi}} 
    = t_{e.x}^{\Phi(\F{T}_x)}[\bar{r}_j \bar{\sigma}]_{\bar{\pi}}[\hat{r}_e \hat{\sigma}]_{\bar{\pi}.\psi_j.\beta} 
    = t_{y_j}^{\F{T}_x}[\hat{r}_e \hat{\sigma}]_{\varphi_{y_j}(\tau)}$, 
    and if $q$ does not exist in $\bar{r}_j$, 
    then we get $t_{e.{y_j}}^{\Phi(\F{T}_x)} = t_{e.x}^{\Phi(\F{T}_x)}[\bar{r}_j \bar{\sigma}']_{\bar{\pi}} 
    = t_{e.x}^{\Phi(\F{T}_x)}[\bar{r}_j \bar{\sigma}]_{\bar{\pi}} 
    = t_{y_j}^{\F{T}_x}$ and $\varphi_{y_j}(\tau) = \top$.

    The rest of this construction is now completely analogous to the one for \Cref{properties-eq-AST-iAST-1}.
    The labeling of the nodes $e.z$ for $z \in Z$ with $z \neq x$ is defined by:
    \begin{itemize}
        \item[] (T-1) $t_{e.z}^{\Phi(\F{T}_x)} = t_{z}^{\F{T}_x}[\hat{r}_e \hat{\sigma}]_{\varphi_{\tau}(z)}$ if $\varphi_{\tau}(z) \in \IN^*$ and $t_{e.z}^{\Phi(\F{T}_x)} = t_z^{\F{T}_x}$ if $\varphi_{\tau}(z) = \top$.
        \item[] (T-2) $p_{e.z}^{\Phi(\F{T}_x)} = p_z^{\F{T}_x} \cdot \hat{p}_e$ 
    \end{itemize}
    Now, for a leaf $e.z' \in V'$ either $z' \in V$ is also a leaf
    or we rewrite the innermost redex at position $\varphi_{\tau}(z')$ at node $z'$ in $\F{T}_x$.
    If we rewrite $t_{z'}^{\F{T}_x} \ito_{\SSS, \varphi_{\tau}(z')} \{\hat{p}_{1}:t_{w'_1}^{\F{T}_x}, \ldots, \hat{p}_{h}:t_{w'_h}^{\F{T}_x}\}$, 
    then we have $t_{w'_e}^{\F{T}_x} =
    t_{z'}^{\F{T}_x}[\hat{r}_e \hat{\sigma}]_{\varphi_{\tau}(z')} \stackrel{\mbox{\scriptsize (T-1)}}{=} t_{e.z'}^{\Phi(\F{T}_x)}$ 
    and $p_{w'_e}^{\F{T}_x} =  p_{z'}^{\F{T}_x} \cdot \hat{p}_e \stackrel{\mbox{\scriptsize (T-2)}}{=}
    p_{e.z'}^{\Phi(\F{T}_x)}$. 
    Thus, we can again copy the rest of this subtree of $\F{T}_x$ to
    our newly generated tree $\Phi(\F{T}_x)$.
    As for \Cref{properties-eq-AST-iAST-1}, one can now prove
    that $|\Phi(\F{T}_x)|_{\ctleaf} = |\F{T}_x|_{\ctleaf}$, 
    and that all other rewrite steps between a node $e.z$ and its successors are valid $\to_\SSS$-rewrite steps.
\end{myproof}

\setcounter{auxctr}{\value{theorem}}
\setcounter{theorem}{\value{iASTtofASTThree}}

\begin{theorem}[From iAST to fAST (3)]
    If a PTRS $\SSS$ is orthogonal and spare, then 
    \begin{align*}
        \SSS \text{ is fAST on basic terms} &\Longleftrightarrow \SSS \text{ is iAST on basic terms}\!
    \end{align*}
\end{theorem}

\setcounter{theorem}{\value{auxctr}}

\begin{myproof}
    We only consider the non-trivial direction ``$\Longleftarrow$''. Let $\SSS$ be iAST on basic terms. Then
    we have to 
    show that all $\SSS$-RSTs that start with $(1:t)$  converge with probability $1$
    if $t$ is a basic term.
    The proof for this part is completely analogous to the one of \Cref{properties-eq-AST-iAST-1}.
    We iteratively move the innermost rewrite steps to a higher position using the construction $\Phi(\circ)$.
    Note that since $\SSS$ is spare and left-linear,  in
    the construction of $\Phi(\circ)$ and in the proof of
    Case (B), if the innermost redex is below a redex $\ell\sigma$ that is reduced next via a
    rule $\ell \to \{p_1:r_1, \ldots, p_m:r_m \}$, then the innermost redex is completely
    ``inside'' the used substitution $\sigma$, and it corresponds to a variable $q$ which 
    occurs only once in $\ell$ and at most once in $r_j$ for all $1 \leq j \leq m$,
    due to spareness of $\SSS$.
    Hence, we can use the same construction as in \Cref{properties-eq-AST-iAST-1}.
\end{myproof}

For \Cref{lemma:spareness-AST-proof-1} we need some more auxiliary functions from \cite{fuhs2019transformingdctorc} to
decode a basic term over $\Sigma \cup \Sigma_{\C{G}(\SSS)}$ into the original term over $\Sigma$.

\begin{definition}[Constructor Variant, Basic Variant, Decoded Variant]
    Let $\SSS$ be a PTRS over the signature $\Sigma$. 
    For a term $t \in \TSet{\Sigma}{\VSet}$, we define its \emph{constructor variant} $\cv(t)$ inductively as follows: 
    \begin{itemize}
        \item[$\bullet$] $\cv(x) = x$ for $x \in \VSet$
        \item[$\bullet$] $\cv(f(t_1, \ldots, t_n)) = f(\cv(t_1), \ldots, \cv(t_n))$ for $f \in \Sigma_C$
        \item[$\bullet$] $\cv(f(t_1, \ldots, t_n)) = \tcons_f(\cv(t_1), \ldots, \cv(t_n))$ for $f \in \Sigma_D$
    \end{itemize}
    For a term $t \in \TSet{\Sigma}{\VSet}$ with $t = f(t_1, \ldots, t_n)$, we define its \emph{basic variant} $\bv(f(t_1, \ldots, t_n)) = \tenc_f(\cv(t_1), \ldots, \cv(t_n))$.
    For a term $t \in \TSet{\Sigma \cup \Sigma_{\C{G}(\SSS)}}{\VSet}$, we define its \emph{decoded variant} $\dv(t) \in \TSet{\Sigma}{\VSet}$ as follows: 
    \begin{itemize}
        \item[$\bullet$] $\dv(x) = x$ for $x \in \VSet$
        \item[$\bullet$] $\dv(\targenc(t)) = \dv(t)$
        \item[$\bullet$] $\dv(f(t_1, \ldots, t_n)) = g(\dv(t_1), \ldots, \dv(t_n))$ for $f \in \{g, \tcons_g, \tenc_g\}$ with $g \in \Sigma_D$
        \item[$\bullet$] $\dv(f(t_1, \ldots, t_n)) = g(\dv(t_1), \ldots, \dv(t_n))$ for $f \in \{g, \tenc_g\}$ with $g \in \Sigma_C$
    \end{itemize}
\end{definition}

The only difference to the auxiliary functions from \cite{fuhs2019transformingdctorc} is that $\dv$ now also removes the $\targenc$ symbols from a term.
This was handled differently in \cite{fuhs2019transformingdctorc} in order to ensure
$|\!\dv(t)| = |t|$, but this is irrelevant for our proofs regarding AST.
Here, $|t|$ denotes the size of a term $t \in \TSet{\Sigma}{\VSet}$ and is recursively defined by $|t| = 1$ if $t \in \VSet$ and $|t| = 1 + |t_1| + \ldots + |t_n|$ if $t = f(t_1, \ldots, t_n)$ with $f \in \Sigma \cup \Sigma_{\C{G}(\SSS)}$.

\BasicASTToAST*

\begin{myproof}
    
    \noindent   
    ``$\Longleftarrow$''

    \noindent   
    Assume that $\SSS$ is not fAST.
    Then there exists an $\SSS$-RST $\F{T} = (V,E,L)$ that converges with probability $<1$
    whose root is labeled with $(1:t)$ for some $t \in \TSet{\Sigma}{\VSet}$.
    We construct an $(\SSS \cup \C{G}(\SSS))$-RST $\F{T}' = (V',E',L')$ whose root is labeled with $(1:\bv(t))$ and with $|\F{T}'|_{\ctleaf} = |\F{T}|_{\ctleaf}$.
    Therefore, there exists an $(\SSS \cup \C{G}(\SSS))$-RST $\F{T}'$ that converges with
    probability $<1$ and starts with the basic term $\bv(t)$, i.e., $\SSS \cup \C{G}(\SSS)$ is not fAST on basic terms.
    
    In \cite{fuhs2019transformingdctorc} it was shown that 
    $\{1 : \bv(t)\} \iliftto_{\C{G}(\SSS)}^* \{1: t \delta_t\}$
    (\cite{fuhs2019transformingdctorc} did not consider PTRSs but since all the
    probabilities in the rules of $\C{G}(\SSS)$ are trivial, the proof
    in \cite{fuhs2019transformingdctorc}
    directly translates to the probabilistic setting).
    Here, for any term $t \in \TSet{\Sigma}{\VSet}$ the substitution $\delta_t$ is defined by $\delta_t(x) = \targenc(x)$ if $x \in \Var(t)$ and $\delta_t(x) = x$ otherwise.
    The $(\SSS \cup \C{G}(\SSS))$-RST $\F{T}'$ first performs these innermost rewrite steps to get from $\bv(t)$ to $t \delta_t$, and then we can mirror the rewrite steps from $\F{T}$.
    To be precise, we use the same underlying tree structure and an adjusted labeling such that $p_x^{\F{T}} = p_x^{\F{T}'}$ and $t_x^{\F{T}} \delta_t = t_x^{\F{T}'}$ for all $x \in V$.
    Since the tree structure and the probabilities are the same, we then get $|\F{T}|_{\ctleaf} = |\F{T}'|_{\ctleaf}$.
    To be precise, the set of leaves in $\F{T}$ is equal to the set of leaves in $\F{T}'$, and they have the same probabilities.
    Since $|\F{T}|_{\ctleaf} < 1$, we thus have $|\F{T}'|_{\ctleaf} < 1$.
    \begin{center}
        \scriptsize
        \begin{tikzpicture}
            \tikzstyle{adam}=[thick,draw=black!100,fill=white!100,minimum size=4mm, shape=rectangle split, rectangle split parts=2,rectangle split
            horizontal]
            \tikzstyle{empty}=[rectangle,thick,minimum size=4mm]
            
            \node[adam] at (-3.5, 0)  (a) {$1$ \nodepart{two} $t$};
            \node[adam] at (-5, -0.8)  (b) {$p_1$ \nodepart{two} $t_{1}$};
            \node[adam] at (-2, -0.8)  (c) {$p_2$ \nodepart{two} $t_{2}$};
            \node[adam] at (-6, -1.6)  (d) {$p_3$ \nodepart{two} $t_3$};
            \node[adam] at (-4, -1.6)  (e) {$p_4$ \nodepart{two} $t_4$};
            \node[adam] at (-2, -1.6)  (f) {$p_5$ \nodepart{two} $t_5$};
            \node[empty] at (-6, -2.4)  (g) {$\ldots$};
            \node[empty] at (-4, -2.4)  (h) {$\ldots$};
            \node[empty] at (-2, -2.4)  (i) {$\ldots$};

            \node[empty] at (-0.5, -1.2)  (arrow) {\Huge $\leadsto$};
            
            \node[adam] at (3.5, 1.6)  (init) {$1$ \nodepart{two} $\bv(t)$};
            \node[empty] at (3.5, 0.8)  (init2) {$\ldots$};
            \node[adam] at (3.5, 0)  (a2) {$1$ \nodepart{two} $t \delta_t$};
            \node[adam] at (2, -0.8)  (b2) {$p_1$ \nodepart{two} $t_1 \delta_t$};
            \node[adam] at (5, -0.8)  (c2) {$p_2$ \nodepart{two} $t_2 \delta_t$};
            \node[adam] at (1, -1.6)  (d2) {$p_3$ \nodepart{two} $t_3 \delta_t$};
            \node[adam] at (3, -1.6)  (e2) {$p_4$ \nodepart{two} $t_4 \delta_t$};
            \node[adam] at (5, -1.6)  (f2) {$p_5$ \nodepart{two} $t_5 \delta_t$};
            \node[empty] at (1, -2.4)  (g2) {$\ldots$};
            \node[empty] at (3, -2.4)  (h2) {$\ldots$};
            \node[empty] at (5, -2.4)  (i2) {$\ldots$};
        
            \draw (a) edge[->] (b);
            \draw (a) edge[->] (c);
            \draw (b) edge[->] (d);
            \draw (b) edge[->] (e);
            \draw (c) edge[->] (f);
            \draw (d) edge[->] (g);
            \draw (e) edge[->] (h);
            \draw (f) edge[->] (i);

            \draw (init) edge[->] (init2);
            \draw (init2) edge[->] (a2);
            \draw (a2) edge[->] (b2);
            \draw (a2) edge[->] (c2);
            \draw (b2) edge[->] (d2);
            \draw (b2) edge[->] (e2);
            \draw (c2) edge[->] (f2);
            \draw (d2) edge[->] (g2);
            \draw (e2) edge[->] (h2);
            \draw (f2) edge[->] (i2);
        \end{tikzpicture}
    \end{center}
   
    \noindent
    ``$\implies$''

    \noindent 
    Let $\SSS \cup \C{G}(\SSS)$ be not fAST on basic terms.
    Then there exists an $(\SSS \cup \C{G}(\SSS))$-RST $\F{T} = (V,E,L)$ that converges with probability
    $<1$ whose root is labeled with $(1:t)$ for some term $t \in \TSet{\Sigma \cup \Sigma_{\C{G}(\SSS)}}{\VSet}$.
    We construct an $\SSS$-RST $\F{T}' = (V',E',L')$ inductively such that for all leaves
    $x$ of $\F{T}'$ during the construction there exists a node $\varphi(x)$ of
    $\F{T}$ such that
    \begin{equation}
    \label{claimBasic} t_{x}^{\F{T}'} = \dv(t_{\varphi(x)}^{\F{T}}) \; \text{ and } \;
    p_{x}^{\F{T}'} =
    p_{\varphi(x)}^{\F{T}}.
    \end{equation}
    Here, $\varphi$ is injective, i.e., every
    leaf $x$ of $\F{T}'$ is mapped to a (unique) node $\varphi(x)$ of
    $\F{T}$.
    Furthermore, after this construction, if $x$ is still a leaf in $\F{T}'$, then $\varphi(x)$ is also a leaf in $\F{T}$ (since $\C{G}(\SSS)$ is terminating, no path of $\F{T}$ consists only of $\C{G}(\SSS)$-steps from some node on).
    Hence, we get
    $|\F{T}|_{\ctleaf} = \sum_{x \in \ctleaf^{\F{T}}}p_x^{\F{T}} \geq \sum_{x \in \ctleaf^{\F{T}'}}p_x^{\F{T}'} = |\F{T}'|_{\ctleaf}$.
    Since $|\F{T}|_{\ctleaf} < 1$, we thus have $|\F{T}'|_{\ctleaf} < 1$.
    This means that $\SSS$ is also not fAST.

    We label the root of $\F{T}'$ with $(1:\dv(t))$.
    By letting $\varphi$ map the root of  $\F{T}'$  to the root of $\F{T}$, the
    claim \eqref{claimBasic} is 
    clearly satisfied.
    As long as there is still a node $x$ in $\F{T}'$ such that $\varphi(x)$ is not a leaf in $\F{T}$, we do the following.
    If we perform a rewrite step with $\C{G}(\SSS)$ at node $\varphi(x)$, i.e.,
    $t_{\varphi(x)}^{\F{T}} \to_{\C{G}(\SSS)} \{1: t_{y}^{\F{T}}\}$ for the only successor
    $y$ of $\varphi(x)$, then we have $\dv(t_{\varphi(x)}^{\F{T}}) = \dv(t_{y}^{\F{T}})$,
    i.e., we do nothing in $\F{T}'$ but simply
    change the definition of $\varphi$ such that 
    $\varphi(x)$ is now $y$.
    To see why $\dv(t_{\varphi(x)}^{\F{T}}) = \dv(t_{y}^{\F{T}})$ holds, note that we have $\dv(\ell) = \dv(r)$ for all rules $\ell \to \{1:r\} \in \C{G}(\SSS)$, since $\dv(\tenc_f(x_1, \ldots, x_n)) = f(x_1, \ldots, x_n) = \dv(f(\targenc(x_1), \ldots, \targenc(x_n)))$, and analogously $\dv(\targenc(\tcons_f(x_1, \ldots, x_n))) = f(x_1, \ldots, x_n) = \dv(f(\targenc(x_1),\linebreak \ldots, \targenc(x_n)))$.
    This can then be lifted to arbitrary rewrite steps.

    Otherwise, let $\varphi(x)E = \{y_1, \ldots, y_k\}$ be the set of successors of
    $\varphi(x)$ in $\F{T}$, and we have $t_{\varphi(x)}^{\F{T}} \to_{\SSS} \{\tfrac{p_{y_1}^{\F{T}}}{p_{\varphi(x)}^{\F{T}}}:t_{y_1}^{\F{T}}, \ldots, \tfrac{p_{y_k}^{\F{T}}}{p_{\varphi(x)}^{\F{T}}}:t_{y_k}^{\F{T}}\}$.
    Since $t_{x}^{\F{T}'} = \dv(t_{\varphi(x)}^{\F{T}})$ and $p_{x}^{\F{T}'} =
    p_{\varphi(x)}^{\F{T}}$ by the induction hypothesis, 
    then we also have
    $t_x^{\F{T}'} \to_{\SSS} \{\tfrac{p_{y_1}^{\F{T}'}}{p_{x}^{\F{T}'}}:t_{y_1}^{\F{T}'}, \ldots, \tfrac{p_{y_k}^{\F{T}'}}{p_{x}^{\F{T}'}}:t_{y_k}^{\F{T}'}\}$
    for terms $t_{y_j}^{\F{T}'}$ with  $t_{y_j}^{\F{T}'} = \dv(t_{y_j}^{\F{T}})$ and for
    $p_{y_j}^{\F{T}'} = p_{y_j}^{\F{T}}$. Thus, when defining $\varphi(y_j) = y_j$ for all $1 \leq j \leq k$, 
    \eqref{claimBasic} is satisfied for the new leaves $y_1,\ldots,y_k$ of $\F{T}'$.
\end{myproof}

\setcounter{auxctr}{\value{theorem}}
\setcounter{theorem}{\value{iASTtofASTFour}}

\begin{theorem}[From iAST to fAST (4)]
    If a PTRS $\SSS$ is non-overlapping and spare, then 
    \begin{align*}
        \SSS \text{ is fAST on basic terms} &\Longleftarrow \SSS \text{ is iAST w.r.t.\ } \paraarrow_{\SSS} \text{ on basic terms}\!
    \end{align*}
\end{theorem}

\setcounter{theorem}{\value{auxctr}}

\begin{myproof}
    The proof is completely analogous to the one of \Cref{properties-eq-AST-iAST-2}.
    We iteratively move the innermost rewrite steps to a higher position using the construction $\Phi(\circ)$.
    Note that since $\SSS$ is spare, in the construction of $\Phi(\circ)$ and in the proof of
    Case (B) (i.e., rewriting at a position above $\varphi_{\tau}(z)$, see the proof of \Cref{properties-eq-AST-iAST-1})
    and the construction of the label at the nodes $e.y_1, \ldots, e.y_k$ (the second rewrite step after the root), 
    if the innermost redex is below a redex $\ell\sigma$ that is reduced next via a
    rule $\ell \to \{p_1:r_1, \ldots, p_m:r_m \}$, then the innermost redex is completely
    ``inside'' the used substitution $\sigma$, and it corresponds to a variable $q$ which 
    occurs only once in $\ell$ and at most once in $r_j$ for all $1 \leq j \leq m$,
    due to spareness of $\SSS$.
    Hence, we can use the same construction as in \Cref{properties-eq-AST-iAST-2}.
\end{myproof}

%% file: pastandsast.tex
\section{PAST \& SAST}\label{PASTandSAST}

In this section, we prove all our new contributions and observations regarding PAST.
But first, we define
a stricter notion than PAST, namely \emph{strong almost-sure termination}
\cite{dblp:conf/vmcai/fuc19,avanzini2020probabilistic}, 
which requires a finite bound on the expected derivation lengths of all rewrite
sequences with the same starting term in case of non-determinism.
Thus, for a term $t$, the \emph{expected derivation height} $\edh(t)$ 
considers all possible rewrite sequences that start with $\{1:t\}$.

\begin{definition}[$\edh$, SAST]\label{def:expected-derivation-height-SAST}
    Let $\SSS$ be a PTRS, $t \in \TSet{\Sigma}{\VSet}$ be a term, and let
    $s \in \{ \mathsf{f}, \mathsf{i}, \mathsf{li} \}$ be a rewrite strategy where
    $\mathsf{f}$, $\mathsf{i}$, and $\mathsf{li}$ stand for full, innermost, and
    leftmost-innermost rewriting, respectively.
    By $\edh_s(t) = \sup\{\edl(\vec{\mu}) \mid \vec{\mu} \text{ is a
    $\xliftto{s}_{\SSS}$-rewrite sequence with } \mu_0 = \{1:t\}\}$ we denote the \emph{expected derivation height} of $t$.
	$\SSS$ is \emph{strongly almost-surely terminating (SAST)} 
    (\emph{innermost SAST (iSAST)} / \emph{leftmost-innermost SAST (liSAST)}) 
    if $\edh_{\mathsf{f}}(t)$ ($\edh_{\mathsf{i}}(t)$ / $\edh_{\mathsf{li}}(t)$) is finite    
    for all $t \in \TSet{\Sigma}{\VSet}$.
\end{definition}

Note that in contrast to wAST and wPAST, we did not define any notion of weak SAST. The
reason is that the definition of weak forms of termination  always only requires the
\emph{existence} of some suitable rewrite sequence. But the definition of SAST
imposes a requirement on \emph{all} rewrite sequences. Thus, it is not clear how to
obtain a useful definition for ``weak SAST'' that differs from wPAST.
   
It is well known that SAST implies PAST \cite{avanzini2020probabilistic}.
However, even for PTRSs with \emph{finitely} many rules, SAST and PAST are \emph{not} equivalent. 
A counterexample is the PTRS with the rules
\[
\begin{array}{rcl@{\qquad}rcl}
\tf(x) &\to& \{\nicefrac{1}{2}:\tf(\ts(x)), \nicefrac{1}{2}:\tb\} & \th(\ts(x)) &\to& \{1:\mathsf{q}(\th(x))\}\\
\tf(x) &\to& \{1:\tg(x)\} & \th(\O) &\to& \{1:\ta(\ta(\ta(\ta(\O))))\}\\
\tg(\O) &\to& \{1:\O\} & \mathsf{q}(\ta(x)) &\to& \{1:\mathsf{q}_1(x)\}\\
\tg(\ts(x)) &\to& \{1:\tg_1(\ts(x))\} & \mathsf{q}_1(x) &\to& \{1:\mathsf{q}_2(x)\}\\
\tg_1(\ts(x)) &\to& \{1:\th(\ts(x))\} & \mathsf{q}_2(x) &\to& \{1:\ta(\ta(\ta(\ta(\mathsf{q}(x)))))\}
\end{array}
\]
over a signature with only unary symbols and constants.
For $n > 0$, the term $\tg(\ts^n(\O))$ needs $4^{n+1}-1+n$ steps:
after $2+n$ steps it is rewritten to $\mathsf{q}^n(\th(\O))$ and then to $\mathsf{q}^n(\ta^4(\O))$, and
each $\mathsf{q}$ replaces every $\ta$ below it by four $\ta$s using $3$ steps per $\ta$, so that
we reach the normal form $\ta^{4^{n+1}}(\mathsf{q}^n(\O))$ after
$2 + n + 1 + 3 \cdot \sum_{i=1}^n 4^i = 4^{n+1}-1+n$ steps.
The PTRS is PAST. To see this, consider the possible derivations of $\tf(\O)$. If
we never use the rule from $\tf$ to $\tg$, then the expected derivation length is 
$\tfrac{1}{2} \cdot 1 + \tfrac{1}{4} \cdot 2 + \tfrac{1}{8} \cdot 3 + \ldots =
\sum_{i=1}^\infty \tfrac{i}{2^i} = 2$
and if we never use the probabilistic $\tf$-rule, then the derivation length is $2$ as well.
Otherwise, if we use the probabilistic
$\tf$-rule in the first $k \geq 1$ steps and the rule from $\tf$ to $\tg$ in the $(k+1)$-th step, then 
the expected derivation length is 
$(\sum_{i=1}^k \tfrac{i}{2^i}) + \tfrac{k + 1 + 4^{k+1} - 1 + k}{2^k} = (\sum_{i=1}^k \tfrac{i}{2^i}) +
2^{k+2} + \tfrac{2 \cdot k}{2^k}$.

Thus, for every rewrite sequence starting with $\tf(\O)$, the expected derivation length is
finite, and a similar argument holds for every other start term. However,
since $k$ can be any number, for any $k \geq 1$, there is a rewrite sequence starting with 
$\tf(\O)$ whose expected derivation 
length is  $(\sum_{i=1}^k \tfrac{i}{2^i}) +
2^{k+2} + \tfrac{2 \cdot k}{2^k} \geq 
2^{k+2}$. In other words, although every rewrite sequence starting with 
$\tf(\O)$ has finite expected derivation length, the supremum over the lengths
for all such rewrite sequences is infinite. Hence, this PTRS is not SAST.
 
Now we present our new results when regarding PAST and SAST instead of AST.
So now we investigate restricted forms like iPAST, liPAST, etc.\ (and similar for SAST). 
While the theorems are the same as for AST, except for \Cref{properties-eq-AST-iAST-3,properties-eq-AST-iAST-4},
they need a slight change in their proofs as we now need to argue about expected derivation length/height and
not just about the termination probability.

We first show that PAST and SAST can also be formulated
in terms of rewrite sequence trees.

\begin{theorem}[PAST via RSTs]\label{thm:PAST-via-RSTs}
	Let $\SSS$ be a PTRS and $\F{T}$ be an $\SSS$-RST.  
    By
    \[
        \edl(\F{T}) = \sum_{x \in V \setminus \ctleaf} p_x^{\F{T}} = \sum_{n = 0}^{\infty}
        \;\; \sum_{\substack{x \in V \setminus \ctleaf \\ d^{\F{T}}(x) = n}} p_x^{\F{T}}
    \]
    we define the \emph{expected derivation length}
    of $\F{T}$, where $d^{\F{T}}(x)$ again denotes the depth of a node $x$ in the tree $\F{T}$.
    Here, the root has depth $0$.
    Then $\SSS$ is PAST (iPAST / liPAST) 
    iff $\edl(\F{T})$ is finite
    for every 
    (innermost / leftmost-innermost) $\SSS$-RST $\F{T}$.
    Similarly,  $\SSS$ is wPAST iff for every term $t$ there exists a fully evaluated $\SSS$-RST $\F{T}$
    whose root is labeled $(1:t)$
    such that $\edl(\F{T})$ is finite.
\end{theorem}

\begin{theorem}[SAST via RSTs]\label{thm:SAST-via-RSTs}
    Let $\SSS$ be a PTRS, $\F{T}$ be an $\SSS$-RST, and let
    $s \in \{ \mathsf{f}, \mathsf{i}, \mathsf{li} \}$ be a rewrite strategy.
 Then $\SSS$ is SAST (iSAST / liSAST) iff 
$\sup\{\edl(\F{T}) \mid \F{T}$ is an
    $\SSS$-RST w.r.t.\ the strategy $s$ whose root is labeled with  $(1:t)\}$
    is finite    
    for all $t \in \TSet{\Sigma}{\VSet}$
where $s = \mathsf{f}$ ($s = \mathsf{i}$ / $s = \mathsf{li}$).
\end{theorem}

Both of these theorems are again easy to prove, similar to the characterization of AST
with RSTs in \Cref{cor:Characterizing AST with RSTs}. 
First note that every infinite $\liftto_{\SSS}$-rewrite sequence $\vec{\mu} = (\mu_n)_{n \in \IN}$ that begins
with a single start term (i.e., $\mu_0 = \{1:t\}$) 
can be represented by an infinite
$\SSS$-RST $\F{T}$ that is fully evaluated such that $|\F{T}|_{\ctleaf} = \lim_{n \to \infty} |\mu_n|_{\SSS}$ and
$\edl(\vec{\mu}) = \sum_{n = 0}^{\infty} (1 - |\mu_n|_{\SSS}) =
\sum_{n = 0}^{\infty} \sum_{\substack{x \in V \setminus \ctleaf \\ d^{\F{T}}(x) = n}} p_x^{\F{T}} = \edl(\F{T})$ and vice versa,
since whenever we reach a normal form after $n$ steps, this is both a normal form in
$\mu_n$ and a leaf at depth $n$ of the RST with the same probability, and otherwise,
if we have a term $t$ in $\Supp(\mu_n)$ that is not in normal form, 
then there exists a (unique) node in $\F{T}$ at depth $n$
with the same probability that is also not a leaf.
So $\SSS$ is PAST for rewrite sequences that begin with a single start term
iff all fully evaluated $\SSS$-RSTs have finite expected derivation lengths.

To show that this is equivalent to PAST,
it remains to show that it suffices to only regard $\liftto$-rewrite sequences that begin
with a single start term. Hence, the following lemma adapts \Cref{lemma:PTRS-AST-single-start-term} from AST to PAST.

\begin{lemma}[Single Start Terms Suffice for PAST] \label{lemma:PTRS-PAST-single-start-term}
    Suppose that there exists an infinite $\liftto_{\SSS}$-rewrite sequence 
    $(\mu_n)_{n \in \mathbb{N}}$ with infinite expected derivation length.
    Then there is also an infinite $\liftto_{\SSS}$-rewrite sequence 
    $(\mu'_n)_{n \in \mathbb{N}}$ with a single start term, 
    i.e., $\mu'_0 = \{1:t\}$, that has infinite expected derivation length.
\end{lemma}

\begin{myproof}
    Let $\vec{\mu} = (\mu_n)_{n \in \mathbb{N}}$ be an $\liftto_{\SSS}$-rewrite sequence 
    that has infinite expected derivation length.
    Suppose that we have $\mu_0 = \{p_1:t_1, \ldots, p_k:t_k\}$.
    Let $\vec{\mu}^j = (\mu^j_n)_{n \in \mathbb{N}}$ with $\mu^j_0 =
    \{1 : t_j\}$ denote the infinite $\liftto_{\SSS}$-rewrite sequence that uses
    the same rules
    as $(\mu_n)_{n \in \mathbb{N}}$ does on the term $t_j$ for every $1 \leq j \leq k$.
    Assume for a contradiction that for every $1 \leq j \leq k$ the
    $\liftto_{\SSS}$-rewrite sequence  $(\mu^j_n)_{n \in \mathbb{N}}$
    has finite expected derivation length.
    Then we would have
    \[
    \begin{array}{rl}
    &\edl(\vec{\mu})\\
        =&\sum_{n = 0}^{\infty} (1 - |\mu_n|_{\SSS})\\
        =&\sum_{n = 0}^{\infty} \sum_{1 \leq j \leq k} p_j \cdot (1 - |\mu_n^j|_{\SSS})\\
        =&\sum_{1 \leq j \leq k} p_j \cdot \sum_{n = 0}^{\infty} (1 - |\mu_n^j|_{\SSS})\\
        =&\sum_{1 \leq j \leq k} p_j \cdot \edl(\vec{\mu}^j)\\
        <&\infty
        \end{array}
    \]
    which is a contradiction to our assumption that we have $\edl(\vec{\mu}) = \infty$.
    Therefore, we have at least one $1 \leq j \leq k$ such that $\edl(\vec{\mu}^j) = \infty$.
\end{myproof}

\Cref{thm:SAST-via-RSTs} is then a direct consequence of \Cref{thm:PAST-via-RSTs}.
Note that we do not need a version of \Cref{lemma:PTRS-PAST-single-start-term} for SAST,
since SAST already considers single start terms only.

Now we show that \Cref{properties-eq-AST-iAST-1} also holds for PAST and SAST.

\setcounter{auxctr}{\value{theorem}}
\setcounter{theorem}{\value{iASTtofASTOne}}

\begin{theorem}[From iPAST/iSAST to fPAST/fSAST (1)]
    If a PTRS $\SSS$ is orthogonal and right-linear (i.e., non-overlapping and linear), then:
    \begin{align*}
        \SSS \text{ is fPAST} &\Longleftrightarrow \SSS \text{ is iPAST}\\
        \SSS \text{ is fSAST} &\Longleftrightarrow \SSS \text{ is iSAST}\!
    \end{align*}
\end{theorem}

\setcounter{theorem}{\value{auxctr}}

\begin{myproof}
  Again, we only prove the non-trivial direction ``$\Longleftarrow$''.

  \smallskip

  \noindent
  \textbf{\underline{1. From iPAST to fPAST:}}

    \noindent   
    Let $\SSS$ be a PTRS that is non-overlapping and linear.
    Furthermore, assume that $\SSS$ is not fPAST.
    This means that there exists an RST $\F{T}$ such that $\edl(\F{T}) = \infty$.
    We create a new \emph{innermost} RST $\F{T}^{(\infty)}$, using the same construction
    as in the proof of \Cref{properties-eq-AST-iAST-1} for AST.
    If we can show that this construction yields $\edl(\F{T}^{(\infty)}) \geq \edl(\F{T}) = \infty$, then this shows that $\SSS$ is also not iPAST.

Let $x$ again be a topmost node in $\F{T}$ where a non-innermost step is performed and let
$\F{T}^{(1)}, \F{T}^{(2)}, \ldots$ be defined via the transformation $\Phi$ as in the
proof of \Cref{properties-eq-AST-iAST-1} for AST. So $\F{T}^{(1)}$ results from replacing the
subtree $\F{T}_x$ by $\Phi(\F{T}_x) = (V', E', L')$.
Our goal is to show that in the construction of the trees $\F{T}^{(1)}, \F{T}^{(2)},
\ldots$, every leaf $v$ of $\F{T}$ is turned into leaves of 
$\F{T}^{(1)}, \F{T}^{(2)},
\ldots$ whose probabilities sum up to $p^{\F{T}}(v)$, and whose depths are
greater or equal
than the original depth $d^{\F{T}}(v)$ of $v$ in $\F{T}$. This then implies $\infty =
\edl(\F{T}) \leq \edl(\F{T}^{(1)})  \leq \edl(\F{T}^{(2)}) \leq \ldots$\
From this observation, we then conclude 
     $\infty =
\edl(\F{T}) \leq
\edl(\F{T}^{(\infty)})$.

    \smallskip
          
    \noindent 
    \textbf{\underline{1.1 We prove $\infty =
        \edl(\F{T}) \leq \edl(\F{T}^{(1)})  \leq \edl(\F{T}^{(2)}) \leq \ldots$}}

      \noindent
      We start by considering how the leaves of $\F{T}$ correspond to
      the leaves of 
      $\F{T}^{(1)}$.
      Let $v$ be a leaf in $\F{T}$.
    If $v \not\in xE^*$, then $v$ is also a leaf in $\F{T}^{(1)}$, labeled with the same probability, and at the same depth.
    Otherwise, $v \in xE^*$, which means that either $v \in Z$, $v \not\in Z$ but its
    predecessor is in $Z$, or neither $v$ nor its predecessor is in $Z$.

    If $v \in Z$ (like, e.g., the node $v_2$ in the example of Step 1.1 in the proof of
    \Cref{properties-eq-AST-iAST-1} for AST), then also $e.v$ must be a leaf in $\Phi(\F{T}_x)$ for every $1 \leq e \leq h$.
    Here, we get $\sum_{1 \leq e \leq h} p_{e.v}^{\Phi(\F{T}_x)} = p_v^{\F{T}_x}$ as in the proof
    of \Cref{properties-eq-AST-iAST-1} for AST.
    In addition, we also know that if $e.v$ is at depth $m$ in $\Phi(\F{T}_x)$, then $v$ is at depth $m-1$ in $\F{T}_x$.

    If $v \not\in Z$ but $v$ is the $e$-th successor of a node $z \in Z$
(like the node $v_8$ in the example in the proof of \Cref{properties-eq-AST-iAST-1} for AST), then $e.z$ is a leaf in $\Phi(\F{T}_x)$.
    Here, we get $p_{e.z}^{\Phi(\F{T}_x)} = p_{v}^{\F{T}_x}$ as in the proof of
    \Cref{properties-eq-AST-iAST-1} for AST.
    In addition, we also know that if $e.z$ is at depth $m$ in $\Phi(\F{T}_x)$, then $v$ is at depth $m$ in $\F{T}_x$.

    Finally, if neither $v$ nor its predecessor is in $Z$
    (like the node $v_9$ in the example in the proof of \Cref{properties-eq-AST-iAST-1}
    for AST), then $v$ is also a leaf in $\Phi(\F{T}_x)$ with $p_{v}^{\Phi(\F{T}_x)} = p_{v}^{\F{T}_x}$ and at the same depth.

    These cases cover each leaf of $\F{T}^{(1)}$ exactly once and all leaves in
    $\Phi(\F{T}_x)$ are
    at a depth greater or equal than the corresponding leaves in $\F{T}_x$, 
    implying that for each leaf $u$ in $\F{T}$ we can find a set of leaves $\Omega_u^1$ in $\F{T}^{(1)}$ 
    such that $d^{\F{T}}(u) \leq d^{\F{T}^{(1)}}(w)$ for each $w \in \Omega_u^1$,
$\sum_{w \in \Omega_u^1} p^{\F{T}^{(1)}}_w = p^{\F{T}}_u$, 
    and $\ctleaf^{\F{T}^{(1)}} = \biguplus_{u \in \ctleaf^{\F{T}}} \Omega_u^1$.

    We can now use the same observation for each leaf $w$ in the tree $\F{T}^{(1)}$ in
    order to get a new set $\Xi_w^2$ of leaves in $\F{T}^{(2)}$ and define $\Omega_u^2 = \bigcup_{w \in \Omega_u^1} \Xi_w^2$.
    Again, we get $d^{\F{T}}(u) \leq d^{\F{T}^{(2)}}(w)$ for each $w \in \Omega_u^2$,
$\sum_{w \in \Omega_u^2} p^{\F{T}^{(2)}}_w = p^{\F{T}}_u$, 
    and $\ctleaf^{\F{T}^{(2)}} = \biguplus_{u \in \ctleaf^{\F{T}}} \Omega_u^2$.
    We now do this for each $i \in \IN$ to define the set $\Omega_u^i$ for each leaf $u$ in the original tree $\F{T}$.
Overall, this implies  $\infty =
\edl(\F{T}) \leq \edl(\F{T}^{(1)})  \leq \edl(\F{T}^{(2)}) \leq \ldots$.

  \smallskip
          
    \noindent 
    \textbf{\underline{1.2 We prove   $\infty =
\edl(\F{T}) \leq
        \edl(\F{T}^{(\infty)})$.}}

      \noindent
From the construction of the $\Omega_u^{i}$ above, in
the end, we get
\[\ctleaf^{\F{T}^{(\infty)}} = \biguplus_{u \in \ctleaf^{\F{T}}} \limsup_{i \to \infty} \Omega_u^{i},\] where $\limsup_{i \to \infty} \Omega_u^{i} = \bigcap_{i' \in \IN} \bigcup_{i > i'} \Omega_u^{i} = \{w \mid w$ is contained in infinitely many $\Omega_u^{i}\}$.
    To see this, let $v \in \ctleaf^{\F{T}^{(\infty)}}$.
    Remember that for every depth $H$ of the tree, 
    there exists an $m_H$ such that $\F{T}^{(\infty)}$ and $\F{T}^{(i)}$ are the same trees up to depth $H$ for all $i \geq m_H$.
    This means that the node $v$ must be a leaf in all trees $\F{T}^{(m)}$ with $m \geq m_{d^{\F{T}^{(\infty)}}(v)+1}$, 
    i.e., it is contained in $\biguplus_{u \in \ctleaf^{\F{T}}} \limsup_{i \to \infty} \Omega_u^{i}$.
    The other direction of the equality follows in the same manner.

Since the probabilities of the leaves in $\F{T}^{(i)}$ always add up to the probability of
the corresponding leaf in $\F{T}$, we have $\sum_{v \in \limsup_{i \to \infty}
  \Omega_u^{i}} p_v^{\F{T}^{(\infty)}} \leq p_u^{\F{T}}$ for all $u \in \ctleaf^{\F{T}}$.
We now show $\infty =
\edl(\F{T}) \leq
        \edl(\F{T}^{(\infty)})$ by considering the two cases where 
$\sum_{v \in \limsup_{i \to \infty}
          \Omega_u^{i}} p_v^{\F{T}^{(\infty)}} < p_u^{\F{T}}$ for some $u \in \ctleaf^{\F{T}}$
        and where $\sum_{v \in \limsup_{i \to \infty}
          \Omega_u^{i}} p_v^{\F{T}^{(\infty)}} = p_u^{\F{T}}$ for all $u \in
        \ctleaf^{\F{T}}$.

 If we have $\sum_{v \in \limsup_{i \to \infty} \Omega_u^{i}} p_v^{\F{T}^{(\infty)}} < p_u^{\F{T}}$ for some $u \in \ctleaf^{\F{T}}$, then we obtain
    \[ 
        \begin{array}{rclcl}
            1 &\geq& |\F{T}|_{\ctleaf}
            &=&
            \sum_{u \in \ctleaf^{\F{T}}} p_u^{\F{T}} \\
            &>&
            \sum_{u \in \ctleaf^{\F{T}}} \sum_{v \in \limsup_{i \to \infty} \Omega_u^{i}} p_v^{\F{T}^{(\infty)}}
            &=&
            \sum_{v \in \biguplus_{u \in \ctleaf^{\F{T}}} \limsup_{i \to \infty} \Omega_u^{i}} p_v^{\F{T}^{(\infty)}}\\
            &=&
            \sum_{v \in \ctleaf^{\F{T}^{(\infty)}}} p_v^{\F{T}^{(\infty)}}
            &=& 
            |\F{T}^{(\infty)}|_{\ctleaf}
        \end{array}
    \]
    and from $|\F{T}^{(\infty)}|_{\ctleaf} < 1$ we directly get $\edl(\F{T}^{(\infty)}) =
    \infty$.

    Otherwise, we have $\sum_{v \in \limsup_{i \to \infty} \Omega_u^{i}}
    p_v^{\F{T}^{(\infty)}} = p_u^{\F{T}}$ for all $u \in
        \ctleaf^{\F{T}}$, and thus we obtain
    \begin{align*}
            &\textstyle \edl(\F{T})\\
            ={}&\textstyle \sum_{n = 0}^{\infty} \sum_{\substack{u \in V \setminus \ctleaf^{\F{T}} \\ d^{\F{T}}(u) = n}} p_u^{\F{T}}\\
            ={}&\textstyle \sum_{n = 0}^{\infty} (1 - \sum_{\substack{u \in \ctleaf^{\F{T}} \\ d^{\F{T}}(u) \leq n}} p_u^{\F{T}})\\
            &\qquad \textstyle(\textcolor{blue}{\sum_{\substack{u \in V \setminus \ctleaf \\ d^{\F{T}}(u) = n}} p_u^{\F{T}} = 1 - \sum_{\substack{u \in \ctleaf \\ d^{\F{T}}(u) \leq n}} p_u^{\F{T}}})\\
            ={}&\textstyle \sum_{n = 0}^{\infty} (1 - \sum_{\substack{u \in \ctleaf^{\F{T}} \\ d^{\F{T}}(u) \leq n}} \sum_{v \in \limsup_{i \to \infty} \Omega_u^{i}} p_v^{\F{T}^{(\infty)}})\\
            &\qquad \textstyle(\textcolor{blue}{p_u^{\F{T}} = \sum_{v \in \limsup_{i \to \infty} \Omega_u^{i}} p_v^{\F{T}^{(\infty)}}})\\
            \leq{}&\textstyle \sum_{n = 0}^{\infty} (1 - \sum_{u \in \ctleaf^{\F{T}} } \sum_{\substack{v \in \limsup_{i \to \infty} \Omega_u^{i}\\ d^{\F{T}^{(\infty)}}(v) \leq n}} p_v^{\F{T}^{(\infty)}})\\
            &\qquad \textstyle(\textcolor{blue}{\forall v \in \limsup_{i \to \infty} \Omega_u^{i}: d^{\F{T}}(u) \leq d^{\F{T}^{(\infty)}}(v)})\\
            ={}&\textstyle \sum_{n = 0}^{\infty} (1 - \sum_{\substack{v \in \biguplus_{u \in \ctleaf^{\F{T}}} \limsup_{i \to \infty} \Omega_u^{i}\\ d^{\F{T}^{(\infty)}}(v) \leq n}} p_v^{\F{T}^{(\infty)}})\\
            &\qquad \textstyle(\textcolor{blue}{\Omega_u^i \text{ all disjoint}})\\
            ={}&\textstyle \sum_{n = 0}^{\infty} (1 - \sum_{\substack{v \in \ctleaf^{\F{T}^{(\infty)}}\\ d^{\F{T}^{(\infty)}}(v) \leq n}} p_v^{\F{T}^{(\infty)}})\\
            &\qquad \textstyle(\textcolor{blue}{\ctleaf^{\F{T}^{(\infty)}} = \biguplus_{u \in \ctleaf^{\F{T}}} \limsup_{i \to \infty} \Omega_u^{i}})\\
            ={}&\textstyle \sum_{n = 0}^{\infty} \sum_{\substack{u \in V \setminus \ctleaf^{\F{T}^{(\infty)}} \\ d^{\F{T}^{(\infty)}}(u) = n}} p_u^{\F{T}^{(\infty)}}\\
            &\qquad \textstyle(\textcolor{blue}{\sum_{\substack{u \in V \setminus \ctleaf^{\F{T}^{(\infty)}} \\ d^{\F{T}^{(\infty)}}(u) = n}} p_u^{\F{T}^{(\infty)}} = 1 - \sum_{\substack{v \in \ctleaf^{\F{T}^{(\infty)}} \\ d^{\F{T}^{(\infty)}}(v) \leq n}} p_v^{\F{T}^{(\infty)}}})\\
            ={}&\textstyle \edl(\F{T}^{(\infty)})
    \end{align*}
    and therefore, $\edl(\F{T}^{(\infty)}) \geq \edl(\F{T}) = \infty$.

    \smallskip
    \noindent 
    \textbf{\underline{2. From iSAST to fSAST:}}

    \noindent 
    The proof for SAST is now a direct consequence of the proof for PAST.
    Let $\SSS$ be a PTRS that is non-overlapping, linear, and not fSAST.
    If there exists a single RST $\F{T}$ such that $\edl(\F{T}) = \infty$, then we do the same as for PAST.
    Otherwise, there exists an infinite sequence of RSTs $(\F{T}_i)_{i \in \IN}$ such that $\lim\limits_{i \to \infty} \edl(\F{T}_i) = \infty$.
    For each of these infinite RSTs we apply the construction as before such that we get an \emph{innermost} RST $\F{T}_i^{(\infty)}$ with $\edl(\F{T}_i^{(\infty)}) \geq \edl(\F{T}_i)$ for all $i \in \IN$.
    Thus, $(\F{T}_i^{(\infty)})_{i \in \IN}$ is an infinite sequence of innermost RSTs such that $\lim\limits_{i \to \infty} \edl(\F{T}_i^{(\infty)}) = \infty$, which proves that $\SSS$ is not iSAST either.
\end{myproof}

\setcounter{auxctr}{\value{theorem}}
\setcounter{theorem}{\value{wASTtofAST}}

\begin{theorem}[From wPAST to fPAST]
    If a PTRS $\SSS$ is non-overlapping, linear, and non-erasing, then 
    \begin{align*}
        \SSS \text{ is fPAST} &\Longleftrightarrow \SSS \text{ is wPAST}\!
    \end{align*}
\end{theorem}

\setcounter{theorem}{\value{auxctr}}

\begin{myproof}
    Same construction as in the proof of \Cref{properties-AST-vs-wAST} for AST.
    Again, we also get $\edl(\F{T}^{(\infty)}) \geq \edl(\F{T})$, which we then use as in the proof of \Cref{properties-eq-AST-iAST-1} regarding PAST.
\end{myproof}

\setcounter{auxctr}{\value{theorem}}
\setcounter{theorem}{\value{liASTtoiAST}}

\begin{theorem}[From liPAST/liSAST to iPAST/iSAST]
    If a PTRS $\SSS$ is non-overlapping, then
    \begin{align*}
        \SSS \text{ is iPAST} &\Longleftrightarrow \SSS \text{ is liPAST}\\
        \SSS \text{ is iSAST} &\Longleftrightarrow \SSS \text{ is liSAST}\!
    \end{align*}
\end{theorem}

\setcounter{theorem}{\value{auxctr}}

\begin{myproof}
    Same construction as in the proof of \Cref{properties-iAST-vs-liAST} for AST.
    Again, we also get $\edl(\F{T}^{(\infty)}) \geq \edl(\F{T})$, which we then use as in the proof of \Cref{properties-eq-AST-iAST-1} regarding PAST/SAST.
\end{myproof}

\setcounter{auxctr}{\value{theorem}}
\setcounter{theorem}{\value{iASTtofASTTwo}}

\begin{theorem}[From iPAST/iSAST to fPAST/fSAST (2)]
    If a PTRS $\SSS$ is non-overlapping and right-linear, then 
    \begin{align*}
        \SSS \text{ is fPAST} &\Longleftarrow \SSS \text{ is iPAST w.r.t.\ } \paraarrow_{\SSS}\\
        \SSS \text{ is fSAST} &\Longleftarrow \SSS \text{ is iSAST w.r.t.\ } \paraarrow_{\SSS}\!
    \end{align*}
\end{theorem}

\setcounter{theorem}{\value{auxctr}}

\begin{myproof}
    Same construction as in the proof of \Cref{properties-eq-AST-iAST-2} for AST.
    Again, we also get $\edl(\F{T}^{(\infty)}) \geq \edl(\F{T})$, which we then use as in the proof of \Cref{properties-eq-AST-iAST-1} regarding PAST/SAST.
\end{myproof}

\setcounter{auxctr}{\value{theorem}}
\setcounter{theorem}{\value{iASTtofASTThree}}

\begin{theorem}[From  iPAST/iSAST to fPAST/fSAST (3)]
    If a PTRS $\SSS$ is orthogonal and spare, then 
    \begin{align*}
        \SSS \text{ is fPAST on basic terms} &\Longleftrightarrow \SSS \text{ is iPAST on basic terms}\\
        \SSS \text{ is fSAST on basic terms} &\Longleftrightarrow \SSS \text{ is iSAST on basic terms}\!
    \end{align*}
\end{theorem}

\setcounter{theorem}{\value{auxctr}}

\begin{myproof}
    Same construction as in the proof of \Cref{properties-eq-AST-iAST-3} for AST.
    Again, we also get $\edl(\F{T}^{(\infty)}) \geq \edl(\F{T})$, which we then use as in the proof of \Cref{properties-eq-AST-iAST-1} regarding PAST/SAST.
\end{myproof}

\setcounter{auxctr}{\value{theorem}}
\setcounter{theorem}{\value{iASTtofASTFour}}

\begin{theorem}[From  iPAST/iSAST to fPAST/fSAST (4)]
    If a PTRS $\SSS$ is non-overlapping and spare, then 
    \begin{align*}
        \SSS \text{ is fPAST on basic terms} &\Longleftarrow \SSS \text{ is iPAST w.r.t.\ } \paraarrow_{\SSS} \text{ on basic terms}\\
        \SSS \text{ is fSAST on basic terms} &\Longleftarrow \SSS \text{ is iSAST w.r.t.\ } \paraarrow_{\SSS} \text{ on basic terms}\!
    \end{align*}
\end{theorem}

\setcounter{theorem}{\value{auxctr}}

\begin{myproof}
    Same construction as in the proof of \Cref{properties-eq-AST-iAST-4} for AST.
    Again, we also get $\edl(\F{T}^{(\infty)}) \geq \edl(\F{T})$, which we then use as in the proof of \Cref{properties-eq-AST-iAST-1} regarding PAST/SAST.
\end{myproof}